\documentclass[aps,pra,notitlepage,onecolumn,superscriptaddress,showpacs]{revtex4-1}
\usepackage[dvips]{graphics}
\usepackage{graphicx}
\usepackage{amsfonts}
\usepackage{amssymb}
\usepackage{amsmath}
\usepackage[utf8]{inputenc}
\usepackage[T1]{fontenc}
\usepackage{lmodern}
\usepackage{subfigure}
\usepackage{color}
\usepackage{hyperref}

\begin{document}

\title{Nonadiabatic couplings and gauge-theoretical structure of curved quantum waveguides}

\date{\today}

\author{J. Stockhofe}
\email{jstockho@physnet.uni-hamburg.de}
\affiliation{Zentrum f\"ur Optische Quantentechnologien, Universit\"at Hamburg, Luruper Chaussee 149, 22761 Hamburg, Germany}

\author{P. Schmelcher}
\affiliation{Zentrum f\"ur Optische Quantentechnologien, Universit\"at Hamburg, Luruper Chaussee 149, 22761 Hamburg, Germany}
\affiliation{The Hamburg Centre for Ultrafast Imaging, Universit\"at Hamburg, Luruper Chaussee 149, 22761 Hamburg, Germany}


\begin{abstract}
We investigate the quantum mechanics of a single particle constrained to move along an arbitrary smooth reference curve by a confinement that is allowed to vary along the waveguide.
The Schrödinger equation is evaluated in the adapted coordinate frame and a transverse mode decomposition is performed, taking into account both curvature and torsion effects and the possibility of a cross-section potential that changes along the curve in an arbitrary way. We discuss the adiabatic structure of the problem, and examine nonadiabatic couplings that arise due to the curved geometry, the varying transverse profile and their interplay.
The exact multi-mode matrix Hamiltonian is taken as the natural starting point for few-mode approximations. Such approximate equations are provided, and it is worked out how these recover known results for twisting waveguides and can be applied to other types of waveguide designs.
The quantum waveguide Hamiltonian is recast into a form that clearly illustrates how it generalizes the Born-Oppenheimer Hamiltonian encountered in molecular physics. 
In analogy to the latter, we explore the local gauge structure inherent to the quantum waveguide problem and suggest the usefulness of diabatic states, giving an explicit construction of the adiabatic-to-diabatic basis transformation.
\end{abstract}

\maketitle

\section{Introduction}
Structures that are designed to controllably guide waves along certain directions in space are abundant in physics and essential for a large number of technological applications. Waveguides are routinely used to channel electromagnetic, optical, acoustic and nowadays also quantum mechanical waves. 
Quantum waveguide models arise in the description of electrons in nanowires (cf. \cite{Londergan1999i} and references therein) or guided on a chip \cite{Hoffrogge2011}, neutrons propagating along fibres \cite{Klein1983,Alvarez-Estrada1984}, undulated optical slab waveguides \cite{Szameit2010,Kartashov2011}, 
but on a more abstract level can also provide insights into the dynamics of chemical, see e.g. \cite{Marcus1966,Torrontegui2011}, or nuclear \cite{Bittner2013} reactions.
Experiments dedicated to the investigation of quantum waveguide systems include magnetoresistance measurements \cite{Timp1988} that have been interpreted as indications of resonances due to bound states in the curved quantum wire \cite{Exner1989a} and studies of microwave resonators simulating the Schrödinger equation \cite{Carini1992,Carini1993,Carini1997,Carini1997a,Bittner2013}.
With the advent of flexible and highly controllable technology for trapping and manipulating ultracold atoms and degenerate quantum gases, new experimental realizations of quantum waveguides were put forward, in particular making use of highly miniatuarized atom chip magnetic traps \cite{Reichel2002,Folman2002}.
Ultracold bosons have been explored in the context of atom optics \cite{Adams1994}, and experiments have demonstrated the possibility of controllably guiding them in elongated potentials \cite{Bongs2001,Leanhardt2002,Guerin2006}, through beam splitters \cite{Cassettari2000,Gattobigio2012}, trapping them in ring-shaped geometries \cite{Sauer2001,Gupta2005,Arnold2006,Henderson2009}, or in the evanescent light field around an ultrathin optical nanofibre \cite{Sagu'e2007,Sagu'e2008,Reitz2012}.
Promising applications making use of bosonic quantum waveguides range from sensitive interferometry \cite{Kreutzmann2004,Cronin2009} to quantum information processing \cite{Schmiedmayer2002} and atomtronic devices \cite{Wright2013}, such that their structural and dynamical properties have attracted much attention \cite{Leboeuf2001,*Jaaskelainen2002a,*Bromley2003,*Koehler2005,*Bromley2004a,*Gaididei2005,*Schwartz2006,*Paul2007,*Gattobigio2010,*Ernst2010,*Tacla2011,*Conti2013,*Campo2013}. A point of particular recent interest is probing and tailoring the transverse (``vibrational'') state of a guided matter wave \cite{Loiko2011,Buecker2013,Martinez-Garaot2013}.

Beyond the context of cold atoms, there is a substantial literature on theoretical approaches to the quantum waveguide problem. 
Circumventing conceptual problems in the quantization procedure of ideally constrained systems \cite{Kaplan1997}, 
a single quantum particle in the presence of a potential that strongly grows when leaving a given curve was studied in \cite{Costa1981,Costa1982}, taking the limit of the confining potential being uniform along the curve and becoming infinitely strong and thin. It was shown that in this limit the wavefunction (after a suitable rescaling) factorizes into a longitudinal and a transverse part, where the longitudinal factor experiences an effective attractive geometric potential term proportional to the square of the local curvature. This finding stimulated a large number of mathematically rigorous studies considering this ultrathin waveguide limit in more detail and proving the existence of geometrically induced bound states due to the curvature, see the reviews \cite{Duclos1995,Hurt2000,Exner2008,Krejcirik2012} and references therein. Furthermore, it was shown that local deformations in the transverse shape of a hard-wall waveguide potentially also lead to the existence of bound states \cite{Bulla1997,Borisov2001}. 
When going beyond the ultrathin waveguide limit with slowly (if at all) varying transverse profile, the adiabatic factorization of the wavefunction breaks down and different transverse modes are nonadiabatically coupled to each other, both due to curvature and torsion and local variations in the cross-section profile of the waveguide. Individually, these effects have been addressed theoretically using the transverse mode decomposition technique, mapping the problem to
a set of coupled differential equations. 
On the one hand, in \cite{Duclos1995} this was worked out for the special case of a circular hard-wall transverse potential that does not change along the waveguide, serving as the starting point for a perturbative expansion of the eigenvalue spectrum, see also \cite{Grushin2008,Grushin2009}. 
On the other hand, for a straight waveguide with a spatially varying cross-section, the transverse mode decomposition has been performed in \cite{Nakazato1991,Jaaskelainen2002}, while only a lowest order correction to this including weak curvature was suggested in \cite{Jaaskelainen2002a}. The decomposition method is also applied in recent works \cite{Baek2008,Gravesen2008}. 

The transverse mode decomposition for the quantum waveguide is reminiscent of the Born-Oppenheimer expansion \cite{Born1927,Born1954,Domcke2004} in molecular physics. 
Making use of the formal similarity between the problems as discussed e.g. in \cite{Belov2006}, some of the mathematical techniques developed for the Born-Oppenheimer problem (recently summarized in \cite{Jecko2013}) have been carried over to the realm of the quantum waveguide. This goes in particular for the so-called space-adiabatic perturbation theory reviewed in \cite{Teufel2003,Panati2003}, which allows the construction of effective Hamiltonians governing the time-evolution inside almost invariant subspaces that to lowest order in the perturbation parameter coincide with the adiabatically decoupled single mode spaces. Going beyond this lowest order, the almost invariant subspace is modified by a prescribed ``tilt'' admixing other modes \cite{Panati2007}. 
Assuming a particular scaling behaviour of the various length scales, the first few orders of this adiabatic perturbation theory expansion for the quantum waveguide problem 
have been worked out in \cite{Wachsmuth2010,Wachsmuth2013} (see also the recent work \cite{Haag2014}). While this perturbation scheme is mathematically rigorous and insightful from a formal point of view, it can be expected that in many situations the plain transverse eigenstates will persist to play a crucial role as immediate, intuitive points of reference, also since they are accessible to direct measurements \cite{Jaaskelainen2002b,Dall2010,Dall2011}.
Knowledge of the nonadiabatic coupling matrix elements between them determined by the interplay of curved geometry and cross-section deformations provides a natural starting point for understanding, and ultimately engineering, the longitudinal dynamics and transverse profiles of guided waves.

In the present work, we put forward a theory that details these nonadiabatic couplings between transverse modes for a quantum waveguide of essentially arbitrary curvature and torsion and an arbitrary cross-section potential which is allowed to vary along the longitudinal direction. This is achieved by means of a transverse mode decomposition. 
The known results of either the cross-section being constant or the waveguide being straight are recovered by this comprehensive approach in a natural way. In particular, in the absence of curvature the molecular Born-Oppenheimer Hamiltonian with one nuclear degree of freedom is reobtained as a limiting case, and we provide its generalization due to the curved geometry.
Having set up the exact theory with infinitely many transverse modes, we explore the conditions for adiabatic decoupling using suitable series expansions of the matrix elements. Beyond the adiabatic limit, the nonadiabatic couplings will not become uniformly large, but rather predominantly couple certain sets of transverse modes such that few-mode approximations are appropriate, and we explicate the corresponding coupled mode equations and effective potential terms arising there. 
Finally, it has been shown that the molecular multi-mode Born-Oppenheimer problem exhibits a local $\mbox{U}(N)$ gauge symmetry \cite{Pacher1989,Pacher1993}, and we investigate how this generalizes in the presence of curvature and torsion. 
This analogy immediately suggests the usefulness of the concept of a diabatic basis \cite{VanVoorhis2010} in the waveguide problem, and we give an explicit construction of the adiabatic-to-diabatic basis transformation matrix.
Besides providing a full transparent picture of nonadiabatic effects in quantum waveguides, the coupled mode equations given here could also form the starting point for numerical computations, since even today simulations of the full three-dimensional Schrödinger equation including the waveguide potential are challenging \cite{Morgan2013}.

The paper is aimed at being self-contained and is therefore structured as follows. Section \ref{sec:frame} introduces the waveguide setup and the construction of the adapted coordinate frame. Section \ref{sec:projection} evaluates the Schrödinger equation (or, equivalently, the Gross-Pitaevskii equation for noninteracting condensed bosons \cite{Pethick2008}) in this frame, and subsequently the transverse mode decomposition is performed. In Section \ref{sec:matrix}, the obtained multi-mode matrix Hamiltonian is recast into a form that is manifestly Hermitian and generalizes the molecular Born-Oppenheimer matrix Hamiltonian, 
such that in Section \ref{sec:few-mode} established few-mode approximation schemes can be applied to the quantum waveguide and are worked out for simple examples.
Section \ref{sec:diabatic} gives a detailed study of the local $\mbox{U}(N)$ gauge structure of the problem and introduces the diabatic basis.
Some calculations omitted from the main text are given in appendices \ref{app:kinetic} and \ref{app:gauge}.

\section{Parametrization of curves and adapted coordinate frame}
\label{sec:frame}
In this section we provide key features of space curves needed in this work, and of the adapted coordinate frame that in the following is used to conveniently parametrize the tubular region of space containing the quantum waveguide. The shape of the waveguide will generally be parametrized by a smooth reference curve $\vec a: \mathbb R \rightarrow \mathbb{R}^3$, combined with a potential $V_\perp$ that varies and eventually steeply ascends when moving away from this curve.
We take the space curve $\vec a$ to be parametrized by its arc length $u_1$ and assume that a comoving orthonormal tripod of vectors $\vec t = \partial \vec a/\partial u_1$ (tangential), $\vec n$ (normal), $\vec b$ (binormal) adapted to the curve exists, whose propagation along $u_1$ is determined by Frenet-Serret-type equations of motion (the dot denotes the derivative with respect to $u_1$ throughout this work)
\begin{equation}
\begin{pmatrix} \dot{\vec t}(u_1) \\ \dot{\vec n}(u_1) \\ \dot{\vec b}(u_1) \end{pmatrix} = \begin{pmatrix} 0 & \kappa(u_1) & 0 \\ -\kappa(u_1) & 0 & \tau(u_1) \\ 0 & -\tau(u_1) & 0  \end{pmatrix}  \begin{pmatrix} \vec t(u_1) \\ \vec n(u_1) \\ \vec b(u_1) \end{pmatrix}
\end{equation}
with arbitrary, but smooth, curvature $\kappa(u_1)$ and torsion $\tau(u_1)$. 
This generically holds for regular curves, with a globally nonvanishing curvature, but we can also allow for $\kappa$ to have zeros and change sign along $u_1$.
The vectors $\vec n$, $\vec b$ and any rotation of them $\vec e_2 = \cos \theta \vec n + \sin \theta \vec b$, $\vec e_3 = -\sin \theta \vec n + \cos \theta \vec b$ with arbitrary $\theta = \theta (u_1)$ span the normal plane perpendicular to the tangential vector $\vec t$ at each position along the curve.
Then, one can parametrize a region of space in the vicinity of the curve by coordinates $\vec u= (u_1, u_2, u_3)$ via 
$ \vec r(\vec u) = \vec a(u_1) + u_2 \vec e_2(u_1) + u_3 \vec e_3(u_1)$
where the ranges of $u_2$ and $u_3$ have to be restricted sufficiently to make the coordinate mapping one-to-one. 
The local natural basis of this curvilinear coordinate system is found to be $(\vec e_1, \vec e_2, \vec e_3)$ with $\vec e_2, \vec e_3$ as defined above and the basis vector $\vec e_1$ (prior to normalization) given by
\begin{equation}
\vec e_1 =  \left(1 - \kappa u_2 \cos \theta + \kappa u_3 \sin \theta \right) \vec t  +  \left(\dot \theta + \tau \right) \left(-u_2 \sin \theta - u_3 \cos \theta \right)\vec n + \left(\dot \theta + \tau \right) \left(u_2 \cos \theta - u_3 \sin \theta \right)\vec b.
\end{equation}
From this, one immediately sees that the natural basis vectors of the new coordinate system are orthogonal for nonzero values of $u_2$, $u_3$ (i.e. away from the curve) if and only if $ \dot \theta (u_1) + \tau(u_1) = 0$
is chosen. This defines the so-called Tang frame \cite{Tang1970} employed throughout this work.
Let us next introduce polar coordinates $(\rho, \vartheta)$ in the normal plane by setting $u_2 = \rho \cos \vartheta, u_3 = \rho \sin \vartheta$.
We focus here on the case of an open curve, rendering the mapping of $u_1$ onto the curve unique. However, many of the results below are independent of this assumption.

In the Tang frame global orthogonality of the basis vectors ensures that the metric tensor $g$ having entries $g_{ij} = (\partial \vec r/\partial u_i) \cdot (\partial \vec r/\partial u_j)$ is diagonal, with $g_{11} = |g| = (1- \kappa u_2 \cos \theta + \kappa u_3 \sin \theta )^2$, $g_{22}=g_{33}=1$. Thus the inverse of $g$ is also diagonal with entries $g^{11} = |g|^{-1}$, $g^{22}=g^{33}=1$.
It is obvious that for an arbitrary reference curve $\vec a$ the coordinates $u_1, u_2, u_3$ will not form a good parametrization of the full $\mathbb{R}^3$, but only of a tubular region around the curve. The local necessary condition for the coordinate transformation to be injective reads $ | g | \neq 0$.
Since we want points with $u_2 = u_3 = 0$, i.e. lying on the reference curve $\vec a$ itself, to be included in the region of space where our parametrization is well-defined, this translates to 
$1 - \kappa \rho \cos (\vartheta + \theta) > 0$ in polar coordinates.
A simple sufficient condition for this to hold is given by $\rho < 1/|\kappa|$.
This condition is not necessary, though.
In particular, at each position $u_1$, there is one direction $\vartheta$ in the normal plane for which $\cos (\vartheta + \theta)$ vanishes and $\rho$ can be chosen arbitrarily large without conflicting with the (local) injectivity of the coordinate mapping. This free direction is the direction of the binormal vector $\vec b$.
If the whole curve lies in a plane, the coordinate system can be extended infinitely far in the direction normal to this plane.
On the other hand, generally the global shape of the curve can lead to stronger limitations on the allowed range of $u_2, u_3$ than the local condition: If the curve comes close to self-intersecting, this can cause overlap of the tubular regions around the curve, inflicting injectivity of the coordinate transformation. 

\section{Transverse mode decomposition}
\label{sec:projection}
In this section we project the quantum waveguide problem onto a system of coupled longitudinal equations by means of the transverse mode decomposition, where in contrast to earlier works we allow for both nonvanishing curvature and torsion and a transverse profile that may change along the waveguide. Our starting point is the 3D Schr\"odinger equation for the single particle wavefunction $\Psi(\vec r, t)$ in the presence of an external potential $V$
\begin{equation}
 i \hbar \partial_t \Psi(\vec r, t) = H_0 \Psi(\vec r, t) = \left[ -\frac{\hbar^2}{2M} \Delta_{\mathbb{R}^3} + V(\vec r) \right] \Psi(\vec r, t)
\end{equation}
and we transform it to the Tang frame coordinates. 
Restricting the support of the wavefunction to the corresponding tubular region around the curve is achieved by applying a strongly confining potential $V$, and ultimately Dirichlet boundary conditions.
As a consequence, the particle cannot explore regions in space to which the Tang frame cannot be extended. Tunneling between different segments of the curve through the ambient space is therefore prohibited in our model. 

The Jacobian of the above coordinate transformation is given by $\sqrt{|g|}$, thus the volume element reads $\text{d}V = \sqrt{|g|} \text{d}u_1 \text{d}u_2 \text{d}u_3$.
It is well-known that the Tang frame Schr\"odinger equation greatly simplifies if the square root of the coordinate transformation's Jacobian is absorbed into the full wave function, i.e. if one works with $\chi (\vec u,t):= |g|^{1/4} \Psi$ instead of $\Psi$, such that $\int \text{d}V |\Psi|^2 = \int \text{d}u_1 \text{d}u_2 \text{d}u_3 | \chi|^2$.
Importantly, the Dirichlet boundary conditions for $\Psi$ immediately carry over to Dirichlet boundary conditions for $\chi$. 
The effective Hamiltonian $H$ for the transformed wavefunction $\chi$ is now defined via $H = |g|^{1/4}  H_0 |g|^{-1/4}$, such that $ i \hbar \partial_t \chi = H \chi$.
Only the kinetic part will be affected by this transformation. To evaluate it, we need to express the Laplace-Beltrami operator in the adapted coordinate frame, $\Delta_{\mathbb{R}^3} = \frac{g^{ij}}{2|g|}(\partial_i |g|)\partial_j + (\partial_i g^{ij})\partial_j +  g^{ij} \partial_i \partial_j$, where $\partial_i$ denotes the derivative with respect to $u_i$.
Then it is straightforward to calculate all the derivatives and we obtain the explicit form of the Schr\"odinger equation for $\chi$:
\begin{eqnarray}
 &&i \hbar \partial_t \chi
=-\frac{\hbar^2}{2M(1-\kappa u_2 \cos \theta + \kappa u_3 \sin \theta)^2} \left\{ \partial_1^2 + (1-\kappa u_2 \cos \theta + \kappa u_3 \sin \theta)^2 (\partial_2^2 + \partial_3^2) \phantom{\frac{5}{4} } + \frac{\kappa^2}{4} \right. \nonumber \\
&\quad& + \left. 2 \frac{\kappa \tau(u_2 \sin \theta + u_3 \cos \theta ) + \dot \kappa (u_2 \cos \theta - u_3 \sin \theta )}{1-\kappa u_2 \cos \theta + \kappa u_3 \sin \theta} \partial_1 +
\frac{5}{4} \frac{\left[ \kappa \tau (u_2 \sin \theta + u_3 \cos\theta ) + \dot \kappa (u_2 \cos \theta - u_3 \sin \theta ) \right]^2}{(1-\kappa u_2 \cos \theta + \kappa u_3 \sin \theta)^2} \right. \nonumber \\
&\quad& + \left.
\frac{1}{2} \frac{(2 \dot \kappa \tau + \kappa \dot \tau) (u_2 \sin \theta + u_3 \cos \theta ) +(\ddot \kappa - \kappa \tau^2 ) (u_2 \cos \theta - u_3 \sin \theta)}{1-\kappa u_2 \cos \theta + \kappa u_3 \sin \theta}  \right\} \chi + V \chi.
\label{eq:schreqchi3d}
\end{eqnarray}
The terms in Eq. (\ref{eq:schreqchi3d}) have been discussed in \cite{Clark1996}.
For a curve with vanishing torsion, $\tau \equiv 0$, one can recover the 2D result as expected. Given $\tau = 0$, then $\dot \theta = 0$, so $\theta$ is constant along the curve. Choosing it to be $\theta \equiv 0$ (in which case the Frenet tripod coincides with the basis vectors of the Tang frame along the curve, $\vec n = \vec e_2$, $\vec b = \vec e_3$), one recovers the well-known result of \cite{Exner1989}.

We now proceed to the transverse mode decomposition, paralleling the adiabatic separation of slow and fast degrees of freedom as it is e.g. applied in the Born-Oppenheimer method in molecular physics. Let us assume that the external potential separates in the form $V = V_1(u_1) + V_\perp(\vec u_\perp;u_1)$, where $V_\perp$ contains the waveguide potential that confines the particle to essentially follow the curve and limits the wavefunction support to the region of well-defined Tang frame coordinates.
This potential term depends on the transverse coordinates $\vec u_\perp = (u_2, u_3)$, and, in general, also parametrically on the arc length parameter $u_1$.
This $u_1$-dependence models arbitrary smooth deformations of the waveguide along the curve, ranging from modest variations in the cross-section to more extreme changes in the structure of $V_\perp$, for example fades between different single- and multi-well structures as are relevant e.g. for beam-splitting applications \cite{Stickney2003,Jaaskelainen2003,Bortolotti2004}.
On top, there can be an extra potential $V_1$ that only depends on the arc length coordinate $u_1$.
The crucial observation is now that the full Hamiltonian $H$ exactly contains the canonical transverse Hamiltonian $ H_\perp(u_1) = -\hbar^2/(2M) \nabla_\perp^2 + V_\perp(\vec u_\perp;u_1)$ for the fast degrees of freedom,
which parametrically depends on $u_1$ only through $V_\perp$. Here, $\nabla_\perp^2 =  \partial_2^2 + \partial_3^2 $ has been introduced.
Keeping $u_1$ fixed and diagonalizing $H_\perp(u_1)$ is the direct analogue of diagonalizing the molecular electronic Hamiltonian with the nuclei clamped at their positions.
We assume now that for each $u_1$ the transverse Hamiltonian has been diagonalized yielding eigenvalues $E_n(u_1)$ and orthonormal eigenfunctions $\phi_n(\vec u_\perp;u_1)$ according to
\begin{equation}
 H_\perp (u_1) \phi_n (\vec u_\perp;u_1) = E_n(u_1) \phi_n(\vec u_\perp;u_1),
\label{eq:schreq2d}
\end{equation}
satisfying the orthonormality relation $\int \text{d}u_2 \text{d}u_3 \phi_m^* \phi_n =: \langle \phi_m | \phi_n \rangle = \delta_{mn}$ at each $u_1$. 
Throughout this work the notation $\langle \cdots | \cdots \rangle$ is reserved for the scalar product in the normal plane.
We assume a transverse potential $V_\perp (\vec u_\perp; u_1)$ such that both the transverse eigenvalues and eigenfunctions smoothly depend on $u_1$, and their derivatives with respect to the arc length parameter are well-defined.
Any stationary wavefunction can now be expanded as $\chi(\vec u) = \sum_n \psi_n(u_1) \phi_n(\vec u_\perp;u_1)$, where the sum runs over the complete set of transverse modes. 
From now on we focus on the stationary Schr\"odinger equation $H \chi = \mathcal E \chi$, with $\mathcal E$ denoting the total energy eigenvalue.
This is reduced to a set of coupled ordinary differential equations for the longitudinal modes by projecting with $\langle \phi_m |$, yielding the exact result
\begin{eqnarray}
 \mathcal E \psi_m(u_1) &=& \left[ V_1(u_1) + E_m(u_1) \right] \psi_m(u_1) \nonumber \\
&&- \frac{\hbar^2}{2M} \sum_n \left\{  \langle \phi_m | \frac{1}{(1-\kappa \hat n)^2}|  \phi_n \rangle \partial_1^2  
 + 2 \langle \phi_m | \frac{1}{(1-\kappa \hat n)^2}| \partial_1 \phi_n \rangle \partial_1 \nonumber 
 +  \langle \phi_m | \frac{1}{(1-\kappa \hat n)^2}| \partial_1^2 \phi_n \rangle \nonumber \right.\\
&\quad& \left. +   \langle \phi_m | \frac{  2 \dot{\kappa} \hat n+   2 \kappa \tau \hat b}{(1-\kappa \hat n )^3} | \phi_n \rangle \partial_1  
 +   \langle \phi_m | \frac{2 \dot{\kappa} \hat n + 2 \kappa \tau \hat b}{(1-\kappa \hat n )^3} | \partial_1 \phi_n \rangle   
+  \frac{\kappa^2}{4}\langle \phi_m| \frac{1} {(1-\kappa \hat n)^2} | \phi_n \rangle  \nonumber \right.\\
&\quad& \left. +  \frac{1}{2}\langle \phi_m | \frac{ (\ddot{\kappa} - \kappa \tau^2) \hat n +   (2\dot{\kappa} \tau + \kappa \dot{\tau}) \hat b}{(1-\kappa \hat n)^3} | \phi_n \rangle 
 +  \frac{5}{4} \langle \phi_m | \frac{ (\dot{\kappa} \hat n + \kappa \tau \hat b)^2 }{(1-\kappa \hat n)^4} |\phi_n \rangle 
\right\} \psi_n.
\label{eq:projected3d}
\end{eqnarray}
Here we have introduced $\hat n$ and $\hat b$ which denote the projection of a vector $\vec u_\perp = u_2 \vec e_2 + u_3 \vec e_3$ in the normal plane onto the local Frenet normal vector $\vec n(u_1)$ and binormal vector $\vec b(u_1)$, respectively, i.e.
\begin{equation}
 \hat n := \vec n \cdot \vec u_\perp = u_2 \cos \theta - u_3 \sin \theta = \rho \cos( \theta + \vartheta), \quad \hat b := \vec b \cdot \vec u_\perp = u_2 \sin \theta +u_3 \cos \theta = \rho \sin( \theta + \vartheta),
\end{equation}
where for convenience also the expressions in polar coordinates have been given.
It is interesting to note that while in the construction of orthogonal adapted coordinates we had to work in terms of the Tang frame vectors $\vec e_2$, $\vec e_3$ spanning the normal plane, the normal and binormal of the Frenet tripod reenter the formalism, and we will see below that matrix elements of powers of $\hat n$ and $\hat b$ are of crucial importance when estimating which terms in the coupled ODE system (\ref{eq:projected3d}) give the dominant contributions.

\section{Hermitian matrix formulation}
\label{sec:matrix}
In this section, we study the system of equations (\ref{eq:projected3d}) in more detail. We show that it can be cast into a more compact infinite-dimensional matrix form, where also the Hermitian nature of the Hamiltonian is manifest. The nonadiabatic coupling matrix elements are analyzed, and series expansions are given that allow for their simple approximate evaluation in the limit of a thin waveguide.
In the following we employ the shorthand
\begin{equation}
 D := (1-\kappa u_2 \cos \theta + \kappa u_3 \sin \theta)^{-2} = \left(1 - \kappa \hat n \right)^{-2} \Rightarrow \dot D  = 2 \frac{\dot \kappa \hat n + \kappa \tau \hat b}{\left(1 - \kappa \hat n \right)^{3}},
\end{equation}
where the Tang frame condition $\dot \theta = - \tau$ was used. 
Then the kinetic terms involving derivatives with respect to $u_1$ in the right hand side of Eq. (\ref{eq:projected3d}) can be written in the alternative form (see appendix \ref{app:kinetic})
\begin{eqnarray}
&&\langle \phi_m | D | \phi_n \rangle \partial_1^2 + 2 \langle \phi_m | D |  \partial_1 \phi_n  \rangle \partial_1 + \langle \phi_m | \dot D | \phi_n \rangle \partial_1+ \langle \phi_m | D | \partial_1^2 \phi_n \rangle  + \langle \phi_m |  \dot D | \partial_1 \phi_n \rangle \nonumber\\
&&\qquad = \partial_1 \langle \phi_m | D | \phi_n \rangle \partial_1 - \delta_{mn} \partial_1^2 + \left[(  \partial_1 \mathbf{1}+ \mathbf F )^2\right]_{mn} + G_{mn},
\label{eq:kinetic}
\end{eqnarray}
where $\mathbf 1$ denotes the identity matrix and the matrices $\mathbf F$ and $\mathbf G$ are defined via
\begin{eqnarray}
 F_{mn} &:=& \frac{1}{2} \left( \langle \phi_m | D | \partial_1 \phi_n \rangle - \langle \partial_1 \phi_m | D | \phi_n \rangle \right),\\
 G_{mn} &:=& \frac{1}{2} \left( \langle \partial_1 \phi_m | \dot D |  \phi_n \rangle + \langle \phi_m | \dot D | \partial_1 \phi_n \rangle + \langle \partial_1 ^2\phi_m | D | \phi_n \rangle + \langle \phi_m | D | \partial_1 ^2 \phi_n \rangle\right) -
 \sum_k F_{mk}F_{kn}.
\label{eq:defFdefG}
\end{eqnarray}
Here $\sum_k$ is summed over the complete set of transverse eigenstates. These definitions immediately imply $F_{mn}^* = -F_{nm}$, $G_{mn}^* = G_{nm}$, such that $\mathbf G$ and $i \mathbf F$ are Hermitian matrices. 
Furthermore we introduce the Hermitian matrices $\mathbf V$, $\mathbf D$ and $\mathbf C$, defined by
\begin{eqnarray}
 V_{mn} &:=& \delta_{mn}\left[ V_1(u_1) +  E_m (u_1) \right], \qquad D_{mn} := \langle \phi_m | D | \phi_n \rangle = \langle \phi_m | \frac{1}{(1-\kappa \hat n)^2} | \phi_n \rangle, \\
 C_{mn} &:=&  \frac{\kappa^2}{4} \langle \phi_m | \frac{1}{(1-\kappa \hat n)^2} | \phi_n \rangle + \frac{1}{2} \langle \phi_m | \frac{(\ddot{\kappa} - \kappa \tau^2) \hat n + (2\dot{\kappa} \tau + \kappa \dot{\tau}) \hat b }{(1-\kappa \hat n)^3} | \phi_n\rangle   + \frac{5}{4} \langle \phi_m | \frac{( \dot \kappa \hat n + \kappa \tau \hat b )^2 }{(1-\kappa \hat n)^4} | \phi_n \rangle .
\end{eqnarray}
By construction we know that the multiplication operator $D > 0$ globally, which implies that the matrix $\mathbf D$ is not only Hermitian, but also positive-definite, which will become important below.

Introducing the momentum operator $ p_1 = - i \hbar \partial_1$ and the (infinite-dimensional) column vector of longitudinal wavefunctions $\vec \psi(u_1):=(\psi_1, \psi_2, \dots)^T$ we can now obtain the compact final form of the stationary Schrödinger equation in matrix representation, equivalent to the ODE sytem of Eq. (\ref{eq:projected3d}):
\begin{eqnarray}
 \mathcal E \vec \psi = \mathbf H \vec \psi := \left( \mathbf V + \frac{1}{2M} \left[  p_1 ( \mathbf D - \mathbf 1 )  p_1 + (  p_1 \mathbf{1} - i \hbar \mathbf F )^2  - \hbar^2 \mathbf G - \hbar^2 \mathbf C \right] \right) \vec \psi.
\label{eq:matrixform}
\end{eqnarray}
As desired, the matrix Hamiltonian $\mathbf H$ is immediately seen to be Hermitian with respect to the scalar product in the projected space $( \vec \psi | \vec \psi' ) := \int \text d u_1 \vec \psi^\dagger(u_1) \vec \psi' (u_1)$, as $p_1$ and all matrix operators in $\mathbf H$ are Hermitian with respect to the scalar product $(  \cdots  | \cdots )$. 
We should remark that, rather unusually, the Hamiltonian of Eq. (\ref{eq:matrixform}) features two kinetic terms. The first one, $p_1 ( \mathbf D - \mathbf 1 ) p_1$, vanishes in the limit of a straight waveguide with $\kappa \equiv 0$, implying $D\equiv1$ globally, and has no counterpart in the molecular Born-Oppenheimer problem. On the other hand, if the transverse potential is $u_1$-independent, we can choose $\mathbf F \equiv \mathbf 0$, and both terms can be merged to $ p_1 \mathbf D  p_1$. 
In fact, it is always possible to merge the kinetic terms into one, at the price of also changing the scalar part of the matrix Hamiltonian in a suitable way. This is explored in section \ref{sec:few-mode} and appendix \ref{app:kinetic}.

So far, we have not made any specific assumptions on the shape of the waveguide. Generally, the exact matrix Hamiltonian $\mathbf H$ will have both diagonal and off-diagonal elements, such that the various transverse modes are intimately coupled. As a generic feature of a quantum waveguide, we can expect, however, that there is a separation in length scales, such that the potential $V_\perp$ stronly localizes the wavefunction in the normal plane and forces it to stay close to the reference curve. In the limit of the waveguide becoming ultrathin the transverse modes decouple. 
We are now in a position to systematically study the nonadiabatic couplings elements as well as the adiabatic decoupling limit of the exact, general multi-mode equation. To obtain further insight into the coupling matrix elements, we need the following expansions which are easily derived from the geometric series:
\begin{equation}
 \frac{1}{(1-\xi)^2} = \sum_{l=0}^\infty (l+1) \xi^l, \quad 
 \frac{1}{(1-\xi)^3} = \sum_{l=0}^\infty \frac{(l+1)(l+2)}{2} \xi^l, \quad
 \frac{1}{(1-\xi)^4} = \sum_{l=0}^\infty \frac{(l+1)(l+2)(l+3)}{6} \xi^l, \qquad \xi \in (-1,1).
\end{equation}
The strategy is to use these to expand the coupling matrix elements in the multi-mode Hamiltonian into powers of suitable small parameters, such that approximations can be obtained by simply truncating the series or using more refined perturbative schemes.
In particular, the problem of finding the coupling matrix elements will be reduced to calculating matrix elements of products and powers of the projected position operators in the normal plane, $\hat n$ and $\hat b$.
For the special case of a Dirichlet waveguide of constant cross-section, such expansions have been given in \cite{Duclos1995}. Let us start with the matrix $\mathbf D$ from the kinetic part of the Hamiltonian.
Assuming that the limit of the sum and the integration with respect to $u_2, u_3$ can be interchanged, we can write for its matrix elements
\begin{equation}
 D_{mn} = \langle \phi_m | \frac{1}{(1-\kappa \hat n )^2} | \phi_n \rangle = \delta_{mn} + \sum_{l=1}^\infty (l+1) \kappa^l \langle \phi_m | \hat n^l | \phi_n \rangle.
\label{eq:Dexp}
\end{equation}
The first term $\delta_{mn}$ simply reflects the orthonormality of the transverse modes. The terms in the remaining sum are proportional to $\kappa^l$ multiplied by a matrix element of the $l$-th power of the transverse position operator projected onto the Frenet normal, $\hat n^l$. If the confining potential in the normal direction is very tight and strong, such that it localizes the transverse modes on a length scale much smaller than the radius of curvature $1/\kappa$, then $\kappa^l \langle \phi_m | \hat n^l | \phi_n \rangle$ will be suppressed with increasing $l$, and in the lowest order approximation $D_{mn} \approx \delta_{mn}$ becomes diagonal, such that the transverse modes decouple. 
While quite different arguments for this decoupling limit have been given, often on the basis of Eq. (\ref{eq:schreqchi3d}) only and arguing $u_2$, $u_3$ to be negligible by themselves, a crucial advantage of the above approach is that it transparently highlights the role of the transverse modes: Which of the terms $\kappa^l \langle \phi_m | \hat n^l | \phi_n \rangle$ can be neglected at which level of precision is, in general, not only a matter of a simple length scale comparison, but also depends on the shape of the wavefunctions $\phi_m$, $\phi_n$. Given the potential $V_\perp$, these can be calculated explicitly, and one can precisely monitor which terms in the expansion of each matrix element $D_{mn}$ should be kept and which can be safely ignored. In particular, by symmetry reasons certain matrix elements of $\hat n^l$ may vanish exactly, which cannot be captured by a simple scaling analysis.

In a similar fashion we proceed for $\mathbf C$. Expanding its matrix elements we find 
\begin{eqnarray}
 C_{mn} 
&=&  \frac{\kappa^2}{4} \left[ \delta_{mn} + \sum_{l=1}^\infty (l+1) \kappa^l \langle \phi_m | \hat n^l | \phi_n \rangle \right]  
+ \frac{1}{4} \sum_{l=0}^\infty (l+1) (l+2) \kappa^l \langle \phi_m | \left[ (\ddot{\kappa} - \kappa \tau^2) \hat n + (2\dot{\kappa} \tau + \kappa \dot{\tau} )\hat b \right] \hat n^l | \phi_n \rangle  \nonumber \\
&\quad& + \frac{5}{24 } \sum_{l=0}^\infty (l+1) (l+2)(l+3) \kappa^{l} \langle  \phi_m |(\dot \kappa \hat n + \kappa \tau \hat b )^2 \hat n^{l} | \phi_n \rangle.
\end{eqnarray}
If, as discussed above, strong and tight confinement in the normal direction is assumed and all products of $\kappa^l$ and a matrix element containing $\hat n^l$ are neglected for $l\geq 1$, this expansion breaks down to
\begin{eqnarray}
  C_{mn} &\approx&  \frac{\kappa^2}{4} \delta_{mn}  + \frac{\ddot{\kappa} }{2}   \langle \phi_m |  \hat n | \phi_n \rangle  + \dot{\kappa} \tau \langle \phi_m | \hat b | \phi_n \rangle 
+ \frac{ \kappa }{2} \dot{\tau} \langle \phi_m | \hat b | \phi_n \rangle  + \frac{5}{4} \dot \kappa ^2  \langle  \phi_m | \hat n^2 | \phi_n \rangle + \frac{5}{4} \kappa^2 \tau ^2   \langle  \phi_m | \hat b^2 | \phi_n \rangle,
\label{eq:Capprox}
\end{eqnarray}
indicating that even if the extension of the waveguide in the normal direction is small compared to the radius of curvature $1/\kappa$, $\mathbf C$ may contain relevant off-diagonal couplings. If, as can be explicitly checked once the transverse potential is specified, the last five terms on the right hand side of Eq. (\ref{eq:Capprox}) which contain matrix elements of powers of $\hat n$ and $\hat b$ and are therefore suppressed in the ultrathin waveguide limit can also be neglected, only the diagonal contribution $\kappa^2/4 $ remains. This leading order term is independent of the transverse modes and thus the details of the confinement $V_\perp$. This attractive, purely geometric potential term in the decoupled single-mode longitudinal equations was first identified in \cite{Costa1981}, 
under the assumption of the extension of the wavefunction away from the curve along any direction in space being small compared to any other length scale in the system. 
Keeping the full matrix elements as done above has the advantage of preserving limiting cases that go lost in the standard procedure where a single confinement length scale in the normal plane is assumed. In particular, it allows to see that in the special case of vanishing (or very small) torsion $\tau$ and change of torsion $\dot \tau$, the binormal direction is (essentially) unrestricted, i.e. matrix elements of $\hat b$ and $\hat b^2$ do not have to be small to obtain adiabatic decoupling.

If the potential $V_\perp$ is independent of $u_1$, and the transverse mode are correspondingly also chosen to be $u_1$-independent, then $\mathbf D$ and $\mathbf C$ are the only terms in the Hamiltonian that can induce nonadiabatic couplings. 
Introducing a single formal small parameter that comes with each power of $\hat n$ or $\hat b$, then the above expansions naturally lead to a perturbation theory beyond the lowest decoupled order, as detailed in \cite{Duclos1995,Grushin2008,Grushin2009} and similarly in \cite{Willatzen2010}.
This approach has its limitations, though, since generally $\kappa$, $\tau$, their derivatives and the various matrix elements of $\hat n$, $\hat b$ may induce very different length scales.

We now turn to the matrix elements in the Hamiltonian that contain derivatives of the transverse modes. These enter the multi-mode equations if the transverse potential (as seen from the Tang frame) changes when moving along the curve, i.e. when changing $u_1$. For $F_{mn}$ we immediately find
\begin{equation}
F_{mn} = \langle\phi_m |\partial_1 \phi_n \rangle + \frac{1}{2} \sum_{l=1}^\infty (l+1)\kappa^l \left( \langle \phi_m | \hat n^l | \partial_1 \phi_n \rangle - \langle \partial_1 \phi_m | \hat n^l | \phi_n \rangle \right),
\label{eq:F}
\end{equation}
where $\partial_1 \langle \phi_m | \phi_n \rangle = 0$ has been used.
Assuming again that the normal confinement is tight such that for any $l\geq 1$ products of $\kappa^l$ and matrix elements $\langle \phi_m | \hat n^l | \partial_1 \phi_n \rangle$ (now between one transverse mode and one mode's first derivative) can be neglected, this reduces to $F_{mn} \approx \langle \phi_m | \partial_1 \phi_n \rangle$, which is the familiar, plain Born-Oppenheimer-type result in the absence of any curvature. Even in the limit of a straight waveguide, there will be off-diagonal derivative couplings between the transverse channels if the shape of the waveguide changes along $u_1$. Only if this modulation is ''spatially slow'' and the modes are energetically well separated, the couplings can be neglected and $F_{mn} \approx 0$, which is the essence of the adiabatic approximation, see also Eq. (\ref{eq:hellfeyn}) and the discussion below. Beyond the ultrathin limit, the derivative couplings known from the molecular Born-Oppenheimer framework are modified by the curvature, in particular allowing for nonvanishing derivative couplings between modes for which $\langle \phi_m | \partial_1 \phi_n \rangle$ itself vanishes due to symmetry (e.g. parity) reasons.

Finally, we consider the matrix $\mathbf G$. In a straightforward calculation, we find for its matrix elements
\begin{eqnarray}
G_{mn} 
&=& \frac{\dot \kappa}{2} \sum_{l=1}^\infty l(l+1)\kappa^{l-1}  \left( \langle \partial_1 \phi_m | \hat n^l |  \phi_n \rangle + \langle  \phi_m | \hat n^l | \partial_1 \phi_n \rangle \right) \nonumber 
  + \frac{\tau}{2} \sum_{l=1}^\infty l(l+1)\kappa^{l}  \left( \langle \partial_1 \phi_m | \hat n^{l-1} \hat b|  \phi_n \rangle + \langle  \phi_m | \hat n^{l-1} \hat b | \partial_1 \phi_n \rangle \right) \nonumber\\
&\quad& +\frac{1}{2} \sum_{l=1}^\infty (l+1)\kappa^{l}  \left( \langle \partial_1^2 \phi_m | \hat n^l|  \phi_n \rangle + \langle  \phi_m | \hat n^l | \partial_1^2 \phi_n \rangle \right)\nonumber\\
&\quad& + \frac{1}{4} \sum_k \sideset{}{'}\sum_{l,l'=0}^\infty (l+1)(l'+1)\kappa^{l+l'} \left( \langle  \phi_m | \hat n^l | \partial_1 \phi_k \rangle \langle  \partial_1 \phi_k  |\hat n^{l'} | \phi_n \rangle + \langle \partial_1  \phi_m | \hat n^l | \phi_k \rangle \langle  \phi_k  | \hat n^{l'} |  \partial_1\phi_n \rangle \right. \nonumber\\
&\qquad& \qquad\qquad \qquad\qquad \left. -\langle  \phi_m | \hat n^l | \partial_1 \phi_k \rangle \langle \phi_k  |\hat n^{l'} | \partial_1\phi_n \rangle - \langle \partial_1  \phi_m | \hat n^l | \phi_k \rangle \langle  \partial_1 \phi_k  | \hat n^{l'} | \phi_n \rangle \right), 
\end{eqnarray} 
where the primed sum means that $l=l'=0$ is excluded. Here, the identities $\partial_1 \langle \phi_m | \phi_n \rangle = 0$, $\partial_1^2 \langle \phi_m | \phi_n \rangle = 0$ and $\sum_k |\phi_k \rangle \langle \phi_k | = 1$ have been employed.
Neglecting, as before, any products of $\kappa^l$ and matrix elements of $\hat n^l$ for $l \geq 1$, and in a second step also matrix elements of $\hat n$ multiplied by $\dot \kappa$ and matrix elements of $\hat b$ multiplied by $\kappa \tau$, one finds that the lowest order of $G_{mn}$ vanishes. This coincides with the corresponding result from the molecular Born-Oppenheimer framework: In the ultrathin waveguide limit, the only coupling is through nonzero $F_{mn}$. Beyond this limit, $\mathbf G$ includes both diagonal corrections and off-diagonal couplings which depend on the curvature, its first derivative, and the torsion.

If couplings are present, a natural question to ask is which modes are dominantly coupled to which. A general question to this answer cannot be expected as long as $V_\perp$ is not specified. The coupling elements in $\mathbf D$ and $\mathbf C$ emerge from matrix elements such as $\langle \phi_m | \hat n^l | \phi_n \rangle$. For important model classes of $V_\perp$, such as box or harmonic oscillator confinement \footnote{We remark that strictly speaking a pure harmonic oscillator confinement is not consistent with the need to restrict the wavefunction to the region of well-defined Tang frame coordinates. However, adding Dirichlet boundary conditions at a distance much larger than the oscillator length only weakly deforms energetically low-lying harmonic oscillator modes, since it mainly affects their exponentially suppressed tail beyond the classical turning point. The important expectation values of $\hat n^l$, $\hat b^l$ etc. for small $l$ are almost insensitive and can be evaluated neglecting the Dirichlet boundary conditions.}, one can explicitly check that there is a tendency for the matrix elements to be most significant for values of $m$, $n$, for which $E_m$ and $E_n$ are not too far apart, i.e. modes that are close in energy are predominantly coupled, as is also expected from perturbation theory. On the other hand, symmetry selection rules can suppress certain couplings, even if these were expected from the simple energetic argument.
In a more general way, the leading order derivative coupling term can be argued to be centered around the diagonal (in terms of neighbouring energies) using the off-diagonal Hellmann-Feynman argument \cite{Worth2004}
\begin{equation}
\langle \phi_m | \partial_1 \phi_n \rangle = \frac{\langle \phi_m | \dot H_\perp | \phi_n \rangle}{E_n - E_m}, \qquad E_m \neq E_n 
\label{eq:hellfeyn}
\end{equation}
indicating that as long as $H_\perp$ is slowly varying with $u_1$, to leading order the derivative coupling $F_{mn}$ between two states is suppressed by their energy difference.
We remark that in writing Eq. (\ref{eq:hellfeyn}) we assume that the $u_1$-derivative of $H_\perp$ is well-defined, which is stronger than assuming that its eigenmodes and eigenvalues are differentiable with respect to $u_1$. For instance, if $V_\perp$ contains a hard-wall potential of spatially varying shape, imposing different Dirichlet boundary conditions as a function of $u_1$, $\dot H_\perp$ may not be a meaningful notion, but all the $E_m$ and $\phi_m$ may well be differentiable with respect to $u_1$ such that the overall formalism applies. 

\section{Few-mode approximations: Beyond the adiabatic approach}
\label{sec:few-mode}
In this section we focus on cases in which nonadiabatic corrections within a certain subset of transverse modes need to be taken into account, while the coupling to modes outside of this set is negligible. In this common situation, one may approximately resort to a restriction of the matrix Hamiltonian obtained above to the set of coupled modes. The coupled equations resulting from such a few-mode approximation are worked out in the following, leading to the emergence of new effective potential terms. These are explicated for  the simple case of a twisting waveguide, and a shifting-based waveguide design to be introduced below.

The expansions presented in the previous sections indicate that in the limit of the waveguide confinement being sufficiently tight and strong in all spatial directions and only slowly varying along $u_1$ (with no pair of transverse modes coming close to degeneracy) the following lowest order approximations hold:
\begin{equation}
 \mathbf F \approx \mathbf 0, \quad \mathbf G \approx \mathbf 0, \quad \mathbf D \approx \mathbf 1, \quad \mathbf C \approx \frac{\kappa^2}{4} \mathbf 1, \quad V_{mn} = \delta_{mn} \left( V_1 + E_m \right).
\label{eq:BOapproxs}
\end{equation}
This implies that all matrices entering the Hamiltonian $\mathbf H$ are approximately diagonal, such that the various modes are fully decoupled, and each of them approximately is governed by
\begin{equation}
 \mathcal E \psi_m = \left[  - \frac{\hbar^2}{2M} \partial_1^2 + V_{1}(u_1) + E_m(u_1) - \frac{\hbar^2 \kappa^2}{8M} \right] \psi_m.
\label{eq:BO}
\end{equation}
The different $\psi_m$ are uncoupled and the mode's transverse eigenvalue $E_m$ enters the effective Schrödinger equation in the form of a potential energy term. Additionally, the equation contains the attractive geometric potential proportional to $\kappa^2$. 
The analogue of this approximation in the framework of molecular physics is the plain, lowest order Born-Oppenheimer approximation: There is no vibrational coupling between the different electronic configurations, and the nuclei experience the electronic eigenvalue as an adiabatic potential energy surface. The geometric potential, on the other hand, originates from the curved geometry of the waveguide and is absent in the molecular problem.
Let us remark that in fact not only the term $\mathbf D - \mathbf 1$ in the Hamiltonian is neglected in the Born-Oppenheimer-type approximation leading to Eq. (\ref{eq:BO}), but rather the combination $ p_1 \left( \mathbf D - \mathbf 1 \right) p_1$, involving the longitudinal momentum also. Thus, the validity of the lowest order adiabatic approximation is not only limited by the smallness of the transverse length scales with respect to curvature, torsion and their derivatives, but also by the longitudinal momentum: If the latter becomes too large, this may tend to lead to mode-coupling even for only weakly curved waveguides. The same is true for the derivative couplings in $\mathbf F$, which also come in combination with the longitudinal momentum $ p_1$.

When going beyond the lowest order terms in the ultrathin, slowly varying waveguide limit, the matrices in $\mathbf H$ are no longer diagonal. Accordingly, different modes $\psi_m$ are coupled and one needs to go back to the infinite-dimensional matrix Schrödinger equation (\ref{eq:matrixform}). Of course, one cannot hope to work with an infinite number of transverse modes $\phi_m$ in practice. Fortunately, in many cases one is allowed to restrict to a not too large finite subset of coupled transverse modes, since typically the non-negligible off-diagonal couplings will not be distributed uniformly: There will often be dominant couplings among certain subsets of transverse states, while each subset as a whole is essentially decoupled from the remaining modes. In the case of the derivative couplings $F_{mn}$, we have seen in Eq. (\ref{eq:hellfeyn}) that to leading order they are suppressed by the energy difference between the modes $\phi_m$ and $\phi_n$, so if a subset of modes is energetically well separated from the rest, the couplings leading out of this subset are small. Similarly, the largest matrix elements of the position operator projections $\langle \phi_m | \hat n^l | \phi_n \rangle$, $\langle \phi_m | \hat b^l | \phi_n \rangle$ etc. for small $l$, i.e. for low orders in the thin waveguide expansions of section \ref{sec:matrix}, typically also tend to cluster around the diagonal in terms of energy, where however this assumption depends on the shape of the transverse eigenmodes determined by $V_\perp$ and when in doubt can be checked explicitly once this is specified. 

These considerations give rise to an approximation scheme which drops the strict assumptions of Eqs. (\ref{eq:BOapproxs}), but still asserts that there is a tractable number of modes that are allowed to be coupled to each other, but decoupled from the rest. Let $\mathcal S$ denote the subset of mode indices labelling the modes that are taken to be decoupled from the rest. Then approximately it holds for all $m \in \mathcal S$ 
\begin{equation}
 \mathcal E \psi_m = \sum_{n \in \mathcal S} H_{mn} \psi_n = \sum_{n \in \mathcal S}\left[ V_{mn} -\frac{\hbar^2}{2M} \left\{ \partial_1 \langle \phi_m | D -1 | \phi_n \rangle \partial_1 + \left[(  \partial_1 \mathbf{1}+ \mathbf F )^2\right]_{mn}+ G_{mn}+C_{mn} \right\} \right] \psi_n.
\label{eq:BH}
\end{equation}
In molecular physics, the analogue of this scheme is commonly referred to as Born-Huang or group-Born–Oppenheimer approximation \cite{Domcke2004}. 
Now both the matrix square $(  \partial_1 \mathbf{1}+ \mathbf F )^2$ and $G_{mn}$ implicitly contain summations $\sum_k$ over all transverse modes, also the ones not within the subset $\mathcal S$. It is desirable to recast these terms into a form in which modes outside of $\mathcal S$ no longer appear and the full matrix $\mathbf F$ is replaced by its restriction to the modes in $\mathcal S$. We denote this smaller matrix by $\mathbf F^{(\mathcal S)}$, and correspondingly $\mathbf 1^{(\mathcal S)}$ denotes the unit matrix whose dimension equals the number of modes in $\mathcal S$.
Then one can check for $m, n \in \mathcal S$:
\begin{eqnarray}
 &&\left[(  \partial_1 \mathbf{1}+ \mathbf F )^2\right]_{mn}+ G_{mn} \nonumber\\
= &&\left[(  \partial_1 \mathbf{1}^{(\mathcal S)}+ \mathbf F^{(\mathcal S)} )^2\right]_{mn} - \sum_{k \in \mathcal S}F_{mk}F_{kn} + \frac{1}{2} \left( \langle \partial_1 \phi_m | \dot D |  \phi_n \rangle + \langle \phi_m | \dot D | \partial_1 \phi_n \rangle + \langle \partial_1 ^2\phi_m | D | \phi_n \rangle + \langle \phi_m | D | \partial_1 ^2 \phi_n \rangle\right),
\end{eqnarray}
which, as desired, no longer depends on modes outside of $\mathcal S$. The scalar potential terms emerging here henceforth will be called the Born-Huang potential:
\begin{equation}
 V_{mn}^\text{BH} := -\frac{\hbar^2}{2M} \left[ - \sum_{k \in \mathcal S}F_{mk}F_{kn} + \frac{1}{2} \left( \langle \partial_1 \phi_m | \dot D |  \phi_n \rangle + \langle \phi_m | \dot D | \partial_1 \phi_n \rangle + \langle \partial_1 ^2\phi_m | D | \phi_n \rangle + \langle \phi_m | D | \partial_1 ^2 \phi_n \rangle\right) \right],
\end{equation}
such that the effective coupled Schrödinger equation system in the subset $\mathcal S$, Eq. (\ref{eq:BH}), now reads
\begin{equation}
 \mathcal E \psi_m = \sum_{n \in \mathcal S} H_{mn} \psi_n = \sum_{n \in \mathcal S}\left[ V_{mn}+ V_{mn}^\text{BH} -\frac{\hbar^2}{2M} \left\{ \partial_1 \langle \phi_m | D -1 | \phi_n \rangle \partial_1 + \left[(  \partial_1 \mathbf{1}^{(\mathcal S)}+ \mathbf F^{(\mathcal S)} )^2\right]_{mn} +C_{mn}  \right\} \right]  \psi_n.
\label{eq:BH2}
\end{equation}
One can immediately verify that in the limit of $\kappa=0$, 
the Born-Huang potential matrix reduces to $V_{mn}^\text{BH} = \hbar^2/(2M) \langle \partial_1 \phi_m | \left( 1 - \sum_{k \in S}| \phi_k \rangle \langle \phi_k |\right) | \partial_1 \phi_n \rangle$,
reproducing the corresponding result from the molecular framework, see e.g. \cite{Domcke2004}.

When having restricted to a subset $\mathcal S$, one can additionally use truncations of the ultrathin waveguide expansions introduced in section \ref{sec:matrix} to evaluate the matrix elements in Eq. (\ref{eq:BH2}), such that only a finite number of matrix elements of $\hat n$, $\hat b$ and their powers is required. Then, the accuracy of the combined approximation can be systematically checked by ensuring that neither enlarging $\mathcal S$ nor taking into account higher order terms in the expansions alters the result.
We now return to the task of merging the two kinetic terms into one.
For notational simplicity, we suppress the superscripts $(\mathcal S)$ in the following. In appendix \ref{app:kinetic} it is shown that matrices $\mathbf F'$ and $\mathbf V^{\prime \text{BH}}$ exist such that
\begin{equation}
 -\frac{\hbar^2}{2M} \left[ \partial_1 (\mathbf{D-1})\partial_1 + (\partial_1 \mathbf 1  + \mathbf F)^2  \right] + \mathbf{V}^\text{BH} = -\frac{\hbar^2}{2M} (\partial_1\mathbf 1  + \mathbf F') \mathbf D (\partial_1\mathbf 1  + \mathbf F')+ \mathbf V^{\prime \text{BH}},
\label{eq:kineticrewritten}
\end{equation}
where the skew-Hermitian matrix $\mathbf F'$ is implicitly fixed by $ 2 \mathbf F =  \mathbf F' \mathbf D + \mathbf D \mathbf F'$ and then the Hermitian matrix $\mathbf V^{\prime \text{BH}}$ is obtained as
\begin{equation}
\mathbf V^{\prime \text{BH}} = \mathbf V^{\text{BH}} -\frac{\hbar^2}{2M} \left\{ \mathbf F^2 +\frac{1}{2} \left( \partial_1 \left[{\mathbf F'}, \mathbf D \right] \right) - \mathbf F' \mathbf D \mathbf F' \right\}.
\label{eq:BH_absorbed}
\end{equation}
The matrix equation $2 \mathbf F =  \mathbf F' \mathbf D + \mathbf D \mathbf F'$ will be encountered again below in Section \ref{sec:diabatic} where we will show that it has a unique solution $\mathbf F'$ 
and also give a way of constructing this solution from $\mathbf F$ and $\mathbf D$. Even if the right hand side of Eq. (\ref{eq:kineticrewritten}) is appealing due to its compact symmetric form, it has the drawback that the terms $\mathbf F$ and $\mathbf V^\text{BH}$ on the left hand side are in general much simpler to obtain than their primed counterparts. When discussing transformations of the matrix Hamiltonian induced by unitary transformations in the space of transverse wavefunctions in Section \ref{sec:diabatic}, we will however see that the primed matrices transform in a more natural way. Depending on the context, it may be advantageous to work in either representation, both being, of course, fully equivalent.

Let us now apply Eq. (\ref{eq:BH2}) to the special case of only one single mode decoupled from the rest, $\mathcal S = \{m\}$. If all transverse modes are taken to be $u_1$-independent, this reproduces the ``adiabatic operator`` introduced in \cite{Duclos1995}.
If we allow for the transverse mode to change along $u_1$, we can invoke the ultrathin waveguide limit again and keep only the lowest order terms of the expansions introduced in the previous section, leading to $D_{mm} \approx 1$, $C_{mm} \approx \kappa^2/4$, $F_{mm} \approx \langle \phi_m | \partial_1 \phi_m \rangle$, while $ V_{mm}^\text{BH} \approx \hbar^2/(2M) \left[\langle \partial_1 \phi_m | \partial_1 \phi_m \rangle + \langle \phi_m |\partial_1 \phi_m \rangle^2  \right]$.
The lowest order single-mode effective Schrödinger equation then reads
\begin{equation}
 \mathcal E \psi_m = \left[ V_{1}(u_1) + E_m(u_1) - \frac{\hbar^2}{2M} \left( \partial_1 +  \langle \phi_m | \partial_1 \phi_m \rangle \right)^2 - \frac{\hbar^2 \kappa^2}{8M}  + \frac{\hbar^2}{2M} \left( \langle \partial_1 \phi_m | \partial_1 \phi_m \rangle + \langle \phi_m |  \partial_1\phi_m\rangle^2 \right) \right] \psi_m.
\label{eq:singlemode}
\end{equation}
Assuming that the transverse wavefunction $\phi_m$ is chosen real at each $u_1$, the term $\langle \phi_m | \partial_1 \phi_m \rangle$ vanishes, and there are three contributions due to the quantum waveguide: (i) The attractive geometric potential, proportional to $\kappa^2$, (ii) the transverse eigenenergy $E_m(u_1)$, acting as a potential energy surface for the longitudinal motion, (iii) a repulsive contribution due to the change of the transverse wavefunction, proportional to the norm $ \langle \partial_1 \phi_m | \partial_1 \phi_m \rangle = \| \partial_1 \phi_m \|^2$. 
The first term is responsible for the emergence of bound states in regions of large curvature, while the second term can support bound states at minima of the transverse energy $E_m(u_1)$, for instance at ''bulges`` of Dirichlet waveguides. 
The last term in Eq. (\ref{eq:singlemode}) is absent in the Born-Oppenheimer-type Eq. (\ref{eq:BO}) and arises from the Born-Huang potential. It  has been identified before for the special setting of a waveguide whose cross-section preserves its shape along $u_1$ but twists with respect to the Tang frame \cite{Mitchell2001,Bouchitte2007,Krejcirik2012} (as will be explored in more detail below), and also in its general lowest order form in \cite{Wachsmuth2010}. 
Again we can make the connection to the Born-Huang approximation of molecular physics here: If $\phi_m$ is chosen real throughout, $F_{mm} \equiv 0$, $\dot{F}_{mm} \equiv 0$,
then the effective potential term in the Hamiltonian reads $+\hbar^2/(2M)\langle \partial_1 \phi_m | \partial_1 \phi_m \rangle =-\hbar^2/(2M)\langle \phi_m | \partial_1^2 \phi_m \rangle$, reproducing the diagonal contribution of the longitudinal kinetic energy operator as expected \cite{Domcke2004}. 

\subsection*{Applications of the lowest order single-mode Born-Huang approximation: Twisting and shifting}
In this subsection we specify the single-mode Born-Huang approximation, Eq. (\ref{eq:singlemode}), for two scenarios where the confining potential $V_\perp$ depends on $u_1$ in particular ways.
First we assume that $V_\perp$ preserves its shape along the curve, but rotates with respect to the Tang frame. This has been termed ``twisting``, and we briefly illustrate here how the more general Eq. (\ref{eq:singlemode}) reproduces the known lowest order result. 
The relative angle between some fixed axis of the rotating potential and the Tang frame will be denoted by the smooth function $\alpha(u_1)$. Of course, $\alpha$ is only well defined up to a constant, i.e. one is free to choose $\alpha (u_{1,0})=0$ at some position $u_{1,0}$. Introducing the rotation operator in the normal plane $\mathcal R_\perp\left[\alpha\right] = \exp \left( -(i/\hbar) \alpha \mathcal J_\perp \right)$, with $\mathcal J_\perp$ denoting the corresponding angular momentum operator, the twisting assumption means that the transverse Hamiltonian at a given position $u_1$ reads $H_\perp(\vec u_\perp; u_1) = \mathcal R_\perp \left[\alpha(u_1)\right]  H_\perp(\vec u_\perp; u_{1,0})  \mathcal R_\perp^{-1} \left[\alpha(u_1)\right]$.
Thus, the transverse eigenmodes $\phi_m (\vec u_\perp; u_{1,0} )$ at the reference position $u_{1,0}$ can be continued to eigenmodes for arbitrary $u_1$ by setting 
$\phi_m (\vec u_\perp ;u_1 ) := \mathcal R_\perp \left[\alpha(u_1)\right] \phi_m (\vec u_\perp; u_{1,0} ) $.
The transverse eigenvalues $E_m$ will be independent of $u_1$ then.
Starting from a real wavefunction at $u_{1,0}$, this choice will ensure that the mode remains real along $u_1$.
From this we find $\langle \phi_m | \partial_1 \phi_m \rangle = - (i/{\hbar}) \dot{\alpha}(u_1) \langle \phi_m | \mathcal J_\perp \phi_m \rangle$, while $\langle \partial_1 \phi_m | \partial_1 \phi_m \rangle = ({\dot{\alpha}^2}/{\hbar^2})  \langle \phi_m | \mathcal J_\perp^2 \phi_m \rangle$. 
Note that the matrix elements of $\mathcal J_\perp$ and $\mathcal J_\perp^2$ do not change along $u_1$ and can therefore be evaluated at any desired position. 
In polar coordinates of the normal plane $\mathcal J_\perp = -i \hbar \partial_\vartheta$ and assuming a real transverse wavefunction $\phi_m$, we have $\langle \phi_m | \mathcal J_\perp \phi_m \rangle = 0$, 
such that the only contribution of the twisting to the effective potential in Eq. (\ref{eq:singlemode}) is
\begin{equation}
 V_\text{twist}(u_1) = \frac{\hbar ^2 \dot{\alpha}(u_1)^2}{2M} \| \partial_\vartheta \phi_m \|^2,
\label{eq:repulsivetwist}
\end{equation}
reproducing the known result from the literature \cite{Mitchell2001,Bouchitte2007,Krejcirik2012}: The longitudinal wavefunction is repelled by regions of strong twist (large absolute value of $\dot \alpha$), and the strength of the repulsion is controlled by the shape of the transverse potential through the norm of the angular derivative $\|\partial_\vartheta \phi_m\|^2$. This also implies that a local minimum of $|\dot \alpha|$ creates a minimum in the effective potential that can lead to the existence of bound states as has been seen in \cite{Exner2005}.

\begin{figure}[ht]
\centering
\includegraphics[height=0.3\textheight]{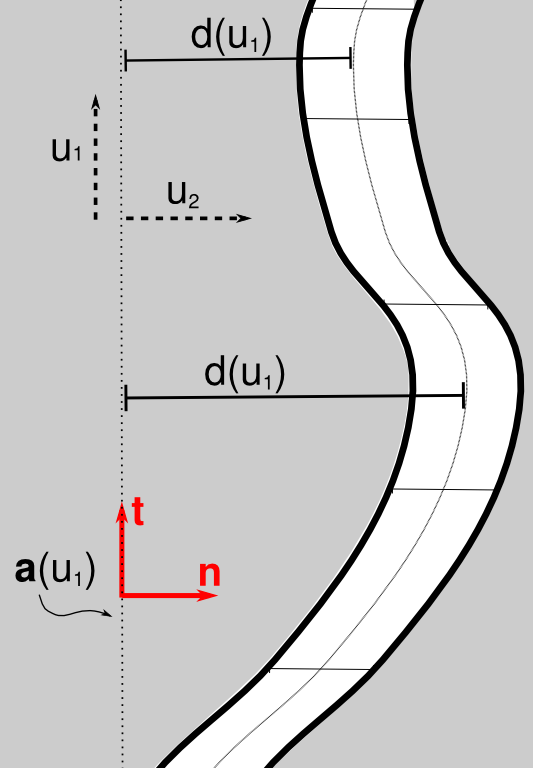}
\hspace{5mm}
\includegraphics[height=0.3\textheight]{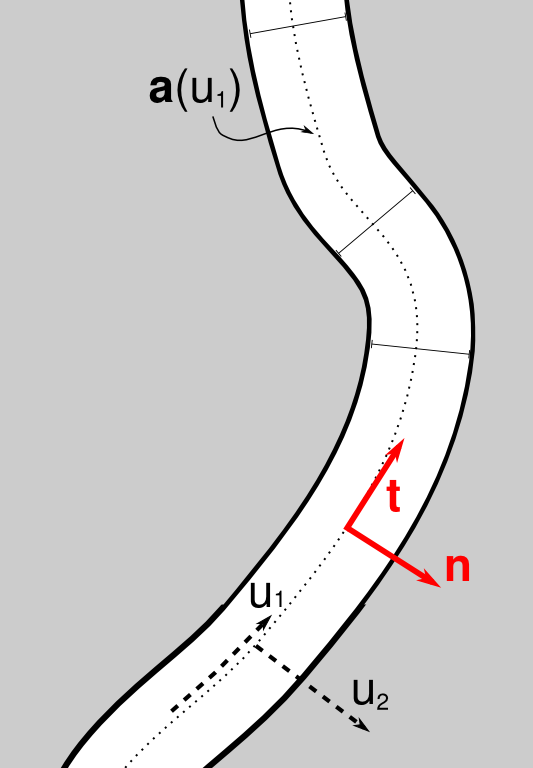}
\caption{Left: An alternative type of quantum waveguide, obtained using a straight planar reference curve combined with a \mbox{$u_1$-dependent} displacement of the confining potential. This results in a repulsive effective potential proportional to $\dot d(u_1)^2$. Right: A corresponding waveguide with the reference curve in its center and a $u_1$-independent confining potential perpendicular to it, resulting in an attractive effective potential proportional to $\kappa^2$ and independent of the transverse mode under consideration.}
\label{fig:displ}
\end{figure}

The second case we discuss here addresses a potential $V_\perp$ that preserves its shape along $u_1$, but is shifted in the normal plane by a smooth displacement function $\vec d(u_1)$ without changing its orientation relative to the Tang frame basis vectors. This gives rise to a waveguide as sketched for a simple two-dimensional case in the left panel of Fig. \ref{fig:displ}. Formally, it means that with respect to some reference position $u_{1,0}$ where $\vec d (u_{1,0})$ vanishes, the transverse Hamiltonian reads $ H_\perp(\vec u_\perp; u_1) = \mathcal T [\vec d(u_1)]  H_\perp(\vec u_\perp; u_{1,0})  \mathcal T^{-1} [\vec d(u_1)]$. Here $\mathcal T[\vec d(u_1)] = \exp [ -(i/\hbar)\vec d(u_1) \cdot {\vec p}_\perp ]$ and $ {\vec p_\perp}$ denote the displacement and momentum operators in the normal plane, respectively.
Again, the transverse eigenmodes $\phi_m (\vec u_\perp; u_{1,0} )$ can be continued to eigenmodes for arbitrary $u_1$ by setting
 $\phi_m (\vec u_\perp ;u_1 ) = \mathcal T [\vec d(u_1)] \phi_m (\vec u_\perp; u_{1,0} ) $ 
and the transverse eigenvalues $E_m$ will be independent of $u_1$.
One thus obtains
\begin{equation}
\langle \phi_m | \partial_1 \phi_m \rangle = - \frac{i}{\hbar} \dot{\vec d}(u_1) \cdot \langle \phi_m | {\vec p}_\perp |\phi_m \rangle, \qquad
 \langle \partial_1 \phi_m | \partial_1 \phi_m \rangle =  \hbar^{-2} \left|\dot{\vec d}(u_1)\right| ^2 \langle \phi_m | \left[\vec e(u_1) \cdot {\vec p}_\perp \right]^2|\phi_m \rangle,
\end{equation}
where $\vec e(u_1)$ is a unit vector parallel to $\dot{\vec d}(u_1)$. 
Since $[\mathcal T, {\vec p}_\perp] = 0$, the matrix elements are again independent of $u_1$ and need to be evaluated at one reference point only.
As before, the matrix element $\langle \phi_m |  {\vec p}_\perp | \phi_m \rangle$ vanishes if $\phi_m$ is chosen real. In this case the shift-induced Born-Huang potential in Eq. (\ref{eq:singlemode}) reads
\begin{equation}
 V_\text{shift}(u_1) = \frac{\left|\dot{\vec d}(u_1)\right| ^2}{2M} \langle \phi_m | \left[ \vec e(u_1) \cdot {\vec p}_\perp \right]^2 | \phi_m \rangle = \frac{\hbar^2 \left|\dot{\vec d}(u_1)\right| ^2}{2M} \| \left[ \vec e(u_1) \cdot \nabla_\perp \right]\phi_m \|^2.
\end{equation}
In analogy to the twist-induced potential discussed before, this contribution is always repulsive, and it is so most strongly in regions where the displacement of the transverse potential from the reference line changes most quickly, i.e. where the norm of the derivative of ${\vec d}$ with respect to $u_1$ is largest. On the other hand, the shape of the transverse mode may also be important if the unit vector $\vec e(u_1)$ changes its direction relative to the Tang frame axes when moving along $u_1$: Then, different projections of the transverse momentum squared enter in the matrix element for different $u_1$, and for anisotropic transverse wavefunctions this will generally make a difference. Local minima of $V_\text{shift}(u_1)$, for instance due to a minimum of the norm of $\dot{\vec d}$ at fixed $\vec e$, can again be expected to support bound states.

Note that this $u_1$-dependent displacement can lead to interesting consequences even in the most simple case of a straight reference line, i.e. $\kappa \equiv 0$. Consider for simplicity the planar case here, where the $u_3$ direction separates, and $V_1 = 0$.
Within the lowest order single-mode Born-Huang approximation single-mode approximation, the effective Hamiltonian for the longitudinal motion reads
\begin{equation}
 H_{mm} = \frac{ p_1^2}{2M} + \frac{\dot{d}(u_1)^2}{2M} \langle \phi_m | p_2^2 | \phi_m \rangle,
\end{equation}
where $p_2 = -i \hbar \partial_2$ is the momentum operator of the normal coordinate $u_2$.
Although this Hamiltonian also describes a curved waveguide, it is crucially different from the waveguides usually studied: Here, slices through the potential that confines the particle to stay inside the waveguide look the same when taken perpendicular to a straight reference curve outside the waveguide itself, see Fig. \ref{fig:displ}. This is, of course, fundamentally different from a situation where slices through the confining potential look the same when taken perpendicular to a reference curve of nonzero $\kappa$ inside the waveguide, as is usually assumed and captured by a $u_1$-independent $V_\perp$ in the formalism. Both ways of designing the waveguide produce effective potential terms in the single-mode approximation Hamiltonian, and -- beyond this adiabatic limit -- also different nonadiabatic coupling mechanisms. In the former, shifting-based design the effective potential depends on the expectation value of $p_2^2$ in the transverse mode under consideration, thus different transverse components of the full wavefunction will experience different shift-induced repulsive potentials. In the latter, ribbon-type setting where the confining potential is carried along perpendicularly to the central curve, the effective potential depends on the curvature only, i.e. on a geometric property of the curve traced out by the center of the waveguide, and independent of the transverse modes. While this is a desirable feature from a conceptual point of view, it is conceivable that some of the waveguides encountered in experiments are more of the shift-induced than of the idealized ribbon-type design, leading to the necessity of a different theoretical description. 

\section{Transverse basis transformations and gauge theoretical structure}
\label{sec:diabatic}
This section is devoted to an analysis of basis transformations in the space of transverse modes which, as will be seen, can be understood as local gauge transformations. Indications of gauge theory being relevant for constrained quantum systems have been presented in \cite{Mitchell2001,Maraner1996,Fujii1997,Ulreich1998,Schuster2003}. We show that the quantum waveguide Hamiltonian exhibits a local $\mbox{U}(N)$ gauge structure, naturally extending the one that has been identified in the molecular Born-Oppenheimer Hamiltonian \cite{Pacher1989,Pacher1993}. We work out the transformation leading to the so-called diabatic basis, in which certain nonadiabatic couplings are eliminated from the formalism.

To start, we recall that in the derivation of the multi-mode Hamiltonian of Eq. (\ref{eq:matrixform}) we have made use of the expansion $\chi = \sum_m \psi_m \phi_m$, where $\phi_m$ were taken to be the eigenfunctions of the transverse Hamiltonian $H_\perp$. This adiabatic basis is a convenient choice because it leads to a diagonal matrix $\mathbf V$, comprising the potential energy surface terms $E_m (u_1)$. However, this choice is not unique. Any unitary transformation of the transverse modes $\phi_m \rightarrow \tilde \phi_m$ preserves the orthonormality $\langle \tilde \phi_m | \tilde \phi_n \rangle = \delta_{mn}$, and one could equally well expand $\chi = \sum_m \tilde \psi_m \tilde \phi_m$, leading to a coupled Schrödinger equation for the modes $\tilde \psi_m$, governed by a matrix Hamiltonian $\tilde{\mathbf H}$. In the molecular Born-Oppenheimer framework whose lowest-dimensional version is included in our formalism as a limiting case, it is well-known that such unitary basis transformations can be understood as gauge transformations \cite{Pacher1989,Pacher1993}. In particular, there is one distinct choice of basis that minimizes certain undesired couplings and is of central importance both conceptually and in numerical applications: the so-called diabatic basis, see e.g. \cite{VanVoorhis2010} and references therein. We will in the following develop a generalized theory and corresponding concepts to take into account the effects induced by curvature and torsion. 

Our starting point is Eq. (\ref{eq:BH2}). We assume that a subset $\mathcal S =\{1, \dots N \}$ of relevant adiabatic modes has been singled out that can be taken to be decoupled from the rest.
Then, within the space of these functions at each $u_1$ we can perform a unitary transformation to a different set of orthogonal basis functions $\tilde \phi_n$. We introduce the local unitary $N\times N$ transformation matrix $\mathbf A(u_1)$ via
\begin{equation}
 \tilde \phi_n(\vec u_\perp; u_1) = \sum_{j=1}^N A_{nj}^*(u_1) \phi_j(\vec u_\perp; u_1), \quad n = 1,\dots,N,
\label{eq:diabtrafo}
\end{equation}
The choice of the complex conjugate matrix element $A_{nj}^* = A_{jn}^\dagger$ in Eq. (\ref{eq:diabtrafo}) is purely conventional and will simplify some expressions below. 
Of course, the $\tilde \phi_n$ will in general no longer be eigenstates of $H_\perp$, leading to a more complicated form of the matrix elements $\tilde V_{mn} =  V_1 \delta_{mn}+ \langle \tilde \phi_m | H_\perp | \tilde \phi_n \rangle$ than in the adiabatic basis. 
Still, it can be advantageous to work in such a rotated basis to get rid of (potentially singular) nonadiabatic coupling terms that can be handled much more easily in the new frame. 
To see this, let us look more closely at the derivative couplings in the matrix $\mathbf F$. 
Extending the usual off-diagonal Hellmann-Feynman argument, we take the derivative of the transverse Schrödinger equation, Eq. (\ref{eq:schreq2d}), with respect to $u_1$ to find after some manipulation that 
\begin{eqnarray} 
F_{mn} &=& \frac{1}{2} \frac{\langle \phi_m | \{ D , \dot H_\perp \} | \phi_n \rangle + \langle \phi_m | [ D , H_\perp ] | \partial_1 \phi_n \rangle - \langle \partial_1 \phi_m | [ D , H_\perp ] |  \phi_n \rangle - (\dot E_m + \dot E_n) \langle \phi_m | D | \phi_n \rangle}{E_n - E_m}
\label{eq:hellfeynD}
\end{eqnarray}
for $E_m \neq E_n$, where $\{ a, b\}:=a b + ba$ denotes the anti-commutator. This equation holds in general, while for our class of Hamiltonians $H_\perp$ one furthermore has the simplifications $\{D, \dot H_\perp \} = 2 D \dot V_\perp$ and $[ D , H_\perp ] = -\hbar^2/(2M) [ D , \nabla_\perp^2]$.
In the limit of a straight waveguide, $D=1$, the expression for $F_{mn}$ reduces to Eq. (\ref{eq:hellfeyn}).
Eq. (\ref{eq:hellfeynD}) suggests that the $F_{mn}$ nonadiabatic coupling between different transverse modes can become arbitrarily large if at some position of the curve two transverse eigenvalues come arbitrarily close (or cross). Generically, tuning $u_1$ as the single parameter in the transverse Hamiltonian $H_\perp$ will not lead to degeneracies in the spectrum, as stated by the classical no-crossing theorem \cite{Neumann1929}, but it is more generic to encounter avoided crossings where two eigenvalue bands almost touch. 
In the vicinity of an (avoided) crossing, one will encounter very large (and potentially, as a function of $u_1$, also quickly varying) couplings $F_{mn}$ which indicate that the adiabatic basis is not well suited for a description of the problem. 
After a suitable basis transformation $\mathbf A(u_1)$, the transformed couplings $\tilde F_{mn}$ can vanish (or become very small) and will not show the singular behaviour any more. The transformed basis with this property is a diabatic (or quasi-diabatic) one, and its functions $\tilde \phi_m$ may offer -- beyond numerical advantages -- also a more insightful understanding of the relevant modes in the vicinity of the avoided crossing \cite{VanVoorhis2010}.

In the following it is shown how the components of the matrix Hamiltonian $\mathbf H$ transform under an arbitrary local basis transformation $\mathbf A(u_1)$, and the transformation that leads to a vanishing $\tilde{\mathbf F}^{(\mathcal S)}$ is explicitly provided. From here on, only $N \times N$ matrices restricted to $\mathcal S$ will be considered, and the superscript $(\mathcal S)$ will be omitted.
Having introduced the transformed basis, for the full rescaled wavefunction $\chi$ there are now two equivalent expansions $\chi = \sum_{n=1}^N \psi_n\phi_n  = \sum_{n=1}^N \tilde \psi_n  \tilde \phi_n$, which implies $\tilde{\vec \psi} = \mathbf A \vec \psi$. 
It is readily observed that the multi-mode Schrödinger equation is form-invariant under the basis transformation, i.e. $\mathcal E \tilde{\vec \psi} = \tilde{\mathbf H} \tilde{\vec \psi} $, if the transformed Hamiltonian $\tilde{\mathbf H}=\mathbf{ AH A^\dagger}$. 
In this section, along with constructing the diabatic basis, we will show that this form-invariance under local $ \mbox{U}(N)$ transformations holds for the full Hamiltonian.
First we note that the matrix representation in the transformed basis of any operator $O$ that does not act on $\mathbf A$ can simply be obtained from the adiabatic one by conjugation with $\mathbf A$:
\begin{equation}
 \tilde O_{mn}:=\langle \tilde \phi_m | O | \tilde \phi_n \rangle = \left( \mathbf A \mathbf  O \mathbf A^\dagger \right)_{mn},
\end{equation}
where $\mathbf O$ denotes the matrix representation of $O$ in the adiabatic basis. This fixes, in particular, the transformation property of $\mathbf V$, $\mathbf D$ and $\mathbf C$.
Next we turn to the derivative coupling matrix $\mathbf F$. In the transformed basis, we find
\begin{eqnarray}
\tilde{F}_{mn} = \frac{1}{2} \left( \langle \tilde \phi_m | D | \partial_1 \tilde \phi_n\rangle - \langle  \partial_1\tilde \phi_m | D | \tilde \phi_n\rangle \right) 
&=&  \sum_{j,k=1}^N \left[ A_{mj} F_{jk} A^\dagger_{kn}  + \frac{1}{2} \left( A_{mj} D_{jk} \dot A^\dagger_{kn}  - \dot A_{mj} D_{jk} A^\dagger_{kn}  \right) \right],
\end{eqnarray}
which in matrix form reads $\tilde{\mathbf{F}} =  \mathbf {A F A^\dagger }+ \frac{1}{2} ( \mathbf{ A D} \dot{\mathbf A}^\dagger-\dot{\mathbf A} \mathbf{ D A^\dagger} )$. The strictly diabatic basis is characterized by $\tilde{\mathbf {F}} = \mathbf 0$.                                          
In marked contrast to the situation in polyatomic molecular systems, since we consider one-dimensional open curves a strictly diabatic basis always exists. We will demonstrate this in the following by constructing the corresponding unitary transformation matrix $\mathbf A$ explicitly. Without loss of generality, the derivative of $\mathbf {A}$ can be written as $\dot{ \mathbf{ A} }(u_1)= \mathbf{A}(u_1) \mathbf{S}(u_1)$, where $\mathbf{S}(u_1)$ is to be determined. 
Preservation of the unitarity of $\mathbf{A}$ along $u_1$ is ensured if
$\mathbf S$ is skew-Hermitian, i.e. $\mathbf{S^\dagger = - S}$,
which implies $\dot{\mathbf{A}}^\dagger = \mathbf{-S A^\dagger}$. Inserting this yields
\begin{equation}
\tilde{ \mathbf { F}} = \mathbf {A F A^\dagger }- \frac{1}{2} \left( \mathbf{ A D S A^\dagger} + \mathbf{ A S D A^\dagger} \right) = \mathbf{0} \qquad 
\Leftrightarrow \qquad 2 \mathbf {F } =\mathbf{ D S}+\mathbf{ S D} = \{ \mathbf{D, S} \}.
\label{eq:sylv}
\end{equation}
The latter is an instance of the Sylvester matrix equation, and more specifically a Lyapunov matrix equation \footnote{After multiplication with $i$, Eq. (\ref{eq:sylv}) assumes the standard form of a Lyapunov matrix equation $ \mathbf Q + \mathbf {DX + X D^\dagger} = \mathbf 0$, where $\mathbf Q := -2i \mathbf F$ is Hermitian and $\mathbf X := i \mathbf S$ is to be determined.}. It can be uniquely solved for $\mathbf S$ without any further approximation in the following explicit way \cite{Jameson1968}. Note that $\mathbf D$ is Hermitian in our problem, so it can be diagonalized, i.e. there is a unitary matrix $\mathbf U_d$, such that
 $\mathbf U_d \mathbf D \mathbf U_d^{-1} = \text{diag} (d_1, \dots, d_N)$. Since $D > 0$ globally, the matrix $\mathbf D$ is positive definite and all its eigenvalues $d_i$ are strictly positive.
Then, setting $\mathbf{S}' = \mathbf{U_d S U_d^{-1}}$, we have the following equivalences
\begin{eqnarray}
 2 \mathbf {F } &=&\mathbf{ D S}+\mathbf{ S D} \Leftrightarrow  
 2 \mathbf U_d \mathbf F \mathbf U_d^{-1} = \text{diag} (d_1, \dots, d_N) \mathbf{S}'+ \mathbf{S}'\text{diag} (d_1, \dots, d_N) 
\Leftrightarrow \forall i,j: 2 \left[ \mathbf{U}_d  \mathbf{F} \mathbf{U}_d^{-1} \right]_{ij} = (d_i + d_j) S_{ij}' \nonumber.
\end{eqnarray}
Since $d_i + d_j > 0$ for all $i, j$, this yields the solution $\mathbf S$ as a function of $\mathbf D, \mathbf F$ via
\begin{equation}
 S_{ij}' = \frac{2}{d_i + d_j} \left[ \mathbf{U}_d  \mathbf{F} \mathbf{U}_d^{-1} \right]_{ij}, \quad \mathbf S = \mathbf{U}_d^{-1} \mathbf{S}' \mathbf{U}_d,
\label{eq:diabS}
\end{equation}
and this solution is unique and independent of the choice of the unitary matrix $\mathbf U_d$. 
It should be noted that standard numerical solvers for Sylvester matrix equations exist which circumvent matrix diagonalization, see e.g. \cite{Bartels1972}.
We also remark that if no few-mode restriction has been performed, Eq.~(\ref{eq:sylv}) immediately has the simple solution $S_{mn} = \langle \phi_m | \partial_1 \phi_n \rangle$, since using completeness $\sum_k | \phi_k \rangle \langle \phi_k | = 1$ one finds $\sum_{k} \left( \langle \phi_m | \partial_1 \phi_k \rangle D_{kn} + D_{mk} \langle \phi_k | \partial_1 \phi_n \rangle \right) = 2 F_{mn}$.
If, however, one has restricted to a subset of modes $\mathcal S$ first, the same only holds when assuming $\sum_{k \in \mathcal S} | \phi_k \rangle \langle \phi_k | \approx 1$ in the matrix product, which introduces an additional approximation.

Let us point out that indeed it is ensured that the matrix $\mathbf S$ obtained in Eq. (\ref{eq:diabS}) is skew-Hermitian, since this is true for $\mathbf F$ and the prefactor $2/(d_i + d_k)$ is real and symmetric.
Once $\mathbf S$ is found, given some initial condition $\mathbf A^\dagger (u_{1,0})$ the equation of motion along $u_1$ for the diabatic basis transformation $\mathbf A^\dagger$ can immediately be integrated to the path-ordered exponential (or, numerically, using a standard ODE solver)
\begin{equation}
 \mathbf{A^\dagger}(u_1) = \mathcal P \exp \left(- \int_{u_{1,0}}^{u_1} \text{d}u_1' \mathbf {S}(u_1') \right) \mathbf{A^\dagger}(u_{1,0}),
\end{equation}
where $\mathbf{A^\dagger}(u_{1,0})$ fixes the diabatic-to-adiabatic transformation at a reference point $u_{1,0}$, which is free to be chosen. An immediate choice is $\mathbf{A^\dagger}(u_{1,0}) = \mathbf 1$, such that the adiabatic and diabatic bases coincide at $u_{1,0}$. Note that in the limiting case of $\kappa \equiv 0$ everywhere, the matrix $\mathbf D$ simplifies to the unit matrix. 
Thus $\mathbf S = \mathbf F$ immediately solves the Lyapunov equation (\ref{eq:sylv}) and $\mathbf{A^\dagger}(u_1) = \mathcal P \exp [- \int_{u_{1,0}}^{u_1} \text{d}u_1' \mathbf {F}(u_1') ] \mathbf{A}^\dagger(u_{1,0})$ is recovered, as in the framework of molecular physics \cite{Baer2006}, where transforming to the diabatic basis via this explicit integration has been used for instance to study the photodissociation of OH molecules \cite{Dishoeck1984}.

A remark is in order here. In contrast to the traditional Born-Oppenheimer molecular setting, in our case the integration yielding $\mathbf A(u_1)$ is performed along the one-dimensional reference curve, which is also assumed not to be closed, such that for any $u_1$ there is only one integration path that connects $u_{1,0}$ and $u_1$. This uniqueness property usually does not hold in the high dimensional nuclear coordinate space where the analogous integration needs to be performed in the molecular Born-Oppenheimer problem, and one has additional compatibility conditions to be fulfilled which turn out to be very restrictive, such that usually a strictly diabatic basis cannot be constructed \cite{Domcke2004}. Similar complications occur when we allow the curve to be closed. Then there are multiple values of $u_1$ corresponding to the same point along the curve, and it must be ensured that the adiabatic-to-diabatic basis transformation is still unique at each point, cf. \cite{Wachsmuth2010}.

We next focus on the important special case of only two modes (which we take to be real here) that are coupled to each other, but decoupled from the rest. 
There is a simplified way of constructing the diabatic basis transformation in this case.
Any real skew-Hermitian $2 \times 2$ matrix, such as $\mathbf F$ and $\mathbf S$, is determined by a single real parameter: it is a multiple of the standard skew-symmetric matrix $\mathbf J = \bigl( \begin{smallmatrix} 0 & 1\\ -1 & 0 \end{smallmatrix} \bigr)$, which in turn has the property $\mathbf M \mathbf J \mathbf M^T = \det \left( \mathbf M \right) \mathbf J  $ for any $2 \times 2$ matrix $\mathbf M$. Thus, with the real orthogonal matrix $\mathbf {U}_d$ that diagonalizes $\mathbf D$, it is found using Eq. (\ref{eq:diabS})
\begin{eqnarray}
&& \mathbf F = F_{12} \mathbf J  
\quad \Rightarrow \quad \mathbf{S}' = \frac{2 F_{12}}{d_1 + d_2} \det{\left(\mathbf{U}_d \right) } \mathbf J 
 \quad \Rightarrow \quad \mathbf{S} = \frac{2 F_{12}}{d_1 + d_2} \det{ \left( \mathbf{U}_d \right) } \det{\left( \mathbf{U}_d^{-1} \right) } \mathbf J  = \frac{2 F_{12}}{D_{11} + D_{22}} \mathbf J, 
\end{eqnarray}
where in the final step it was used that $d_1 + d_2 = D_{11} + D_{22}$ is the invariant trace of $\mathbf D$. 
This immediately gives $\mathbf{S}(u_1)$ starting from $\mathbf{F}(u_1)$, $\mathbf{D}(u_1)$. 
Furthermore, due to the property $\mathbf J ^2 = - \mathbf 1 $, arbitrary powers of $\mathbf S$ at arbitrary $u_1$ collapse and commute, such that the path-ordered exponential can also be considerably simplified (taking $u_{1,0} = 0$ and $\mathbf{A^\dagger}(u_{1,0}) = \mathbf 1$ here):
\begin{eqnarray}
 \mathbf{A^\dagger}(u_1) &=& \begin{pmatrix} \cos \gamma & -\sin \gamma \\ \sin \gamma & \cos \gamma \end{pmatrix}, \qquad \gamma (u_1) := 2 \int_0^{u_1} \text{d}u_1' \frac{F_{12}}{D_{11}+D_{22}}.
\end{eqnarray}
This provides a convenient way of constructing the diabatic basis in the special case of only two real modes involved.

Let us now return to arbitrary basis transformations in a subset $\mathcal S$ containing an arbitrary number of transverse modes. The transformation properties of $\mathbf D$, $\mathbf C$, $\mathbf V$ and $\mathbf F$ have been discussed above, but we have not yet supplied the transformed Born-Huang potential term. Its transformation according to the basis change is more complicated than plain matrix conjugation with $\mathbf A$, since it involves derivatives of the transverse wavefunctions. The Born-Huang potential in the transformed basis reads (see appendix \ref{app:gauge})
\begin{equation}
  \tilde V_{mn}^\text{BH} =\left[ \mathbf{A} \mathbf{V^\text{BH}}\mathbf{A^\dagger} \right]_{mn}
 -\frac{\hbar^2}{2M}  \left[ \mathbf{A}\left(  - \frac{1}{4} \{\mathbf D, \mathbf S \}^2  +  \frac{1}{2} \{ \{\mathbf D, \mathbf S \}, \mathbf F \}+ \frac{1}{2}  \left(\partial_1 \left[\mathbf { S}, \mathbf D \right] \right)  - \{ \mathbf{S},\mathbf{F} \}
 + \frac{1}{2}\left\{\mathbf {S^2}, \mathbf D \right\}  \right) \mathbf{A^\dagger} \right]_{mn},
\label{eq:BHtrafo2}
\end{equation}
where again $\dot{\mathbf{ A}} = \mathbf { A S}$ defines the skew-Hermitian matrix $\mathbf S (u_1)$.
In the limit of a straight waveguide, $D = 1$, this collapses to
\begin{eqnarray*}
 - \frac{1}{4} \{\mathbf D, \mathbf S \}^2  +  \frac{1}{2} \{ \{\mathbf D, \mathbf S \}, \mathbf F \}+ \frac{1}{2}  \left( \partial_1\left[\mathbf { S}, \mathbf D \right] \right)  - \{ \mathbf{S},\mathbf{F} \}
 + \frac{1}{2}\left\{\mathbf {S^2}, \mathbf D \right\} 
= \mathbf 0,
\end{eqnarray*}
such that the transformed Born-Huang potential in this limit is simply obtained by conjugating with $\mathbf A$ as expected from the known result in molecular physics \cite{Domcke2004}.
Eq. (\ref{eq:BHtrafo2}) indicates that the Born-Huang potential by itself is not form-invariant under general gauge transformations $\mathbf A$, but picks up a number of extra terms that only vanish in the limit of the waveguide being straight. We show now that the same is true for the kinetic terms in the Hamiltonian, and that in total the Hamiltonian is gauge-invariant in the proper way. To see this, we note that (see appendix \ref{app:gauge})
\begin{eqnarray}
 -\frac{\hbar^2}{2M}  \left[ \left(\partial_1 + \tilde{\mathbf F} \right)^2  + \partial_1 ( \tilde{\mathbf D} - \mathbf  1) \partial_1 \right]
&&= \mathbf A \left( -\frac{\hbar^2}{2M} \left[\left( \partial_1 + \mathbf F \right)^2 + \partial_1 (\mathbf D  - \mathbf  1)  \partial_1 \right] \right) \mathbf A^\dagger \nonumber \\
&&\kern-4em + \frac{\hbar^2}{2M}  \mathbf A \left( - \frac{1}{4} \{\mathbf D, \mathbf S \}^2  +  \frac{1}{2} \{ \{\mathbf D, \mathbf S \}, \mathbf F \}+ \frac{1}{2}  \left( \partial_1\left[\mathbf { S}, \mathbf D \right] \right)  - \{ \mathbf{S},\mathbf{F} \}
 + \frac{1}{2}\left\{\mathbf {S^2}, \mathbf D \right\}  \right) \mathbf A^\dagger,
\label{eq:trafo-kinetic}
\end{eqnarray}
giving exactly the same extra terms as originating from the transformation of the Born-Huang potential, see Eq.~(\ref{eq:BHtrafo2}), but with opposite sign. Thus, these contributions cancel when the full Hamiltonian is considered, and one indeed has the expected property that the Hamiltonian is form-invariant under local $ \mbox{U}(N)$ transformations, i.e. $ \tilde{\mathbf H} = \mathbf {A H A^\dagger}$.
Note however that, as seen above, the individual terms in the Hamiltonian of Eq. (\ref{eq:BH2}) have more involved transformation properties, only if their sum is considered the form-invariance is unveiled.

Finally, we remark that from the point of view of gauge transformation properties, the alternative representation of the matrix Hamiltonian with only one kinetic term as given in Eq. (\ref{eq:kineticrewritten}) turns out to be the more natural object.
Even if $\mathbf F'$, and consequently also $\mathbf C'$, are only specified rather implicitly from the solution of the Lyapunov equation, we can study their behaviour under gauge transformations (i.e., basis transformations in the transverse states). It turns out that $\mathbf F'$ transforms in a more generic way than $\mathbf F$ encountered above, and also the Born-Huang potential $\mathbf V^{' \text{BH}}$ has the usual tensorial transformation behaviour, i.e. it is simply conjugated with $\mathbf A$. So from the point of view of gauge invariance, the compact representation with only one kinetic term may be thought of as the more generic one, since its individual terms are form-invariant by themselves, not only in combination. 

Let us now demonstrate these statements. First we have $\mathbf D' = \mathbf D$, so $\tilde{\mathbf D}' = \tilde{\mathbf D} = \mathbf {A D A^\dagger}$.
For $\mathbf F$ we know that $\tilde{\mathbf F} = \mathbf{A F A^\dagger} - \frac{1}{2} \mathbf{A} \{\mathbf D, \mathbf S \} \mathbf{A^\dagger}$. The Lyapunov equation $2 \mathbf F = \{ \mathbf D, \mathbf F'\}$ implicitly fixes $\mathbf F'$, and the corresponding equation must hold after the transformation. One can immediately check that $ 2 \tilde{\mathbf F} = \{ \tilde{\mathbf D}, \tilde{\mathbf F}' \} $ is ensured by $\tilde{\mathbf F}' = \mathbf{A ( F' -S) A^\dagger}  = \mathbf{A F' A^\dagger} + \mathbf A \dot{\mathbf{ A}}^\dagger$, reproducing the transformation behaviour of the derivative coupling matrix in the molecular framework \cite{Pacher1989}. Then one immediately sees that the covariant derivative $\partial_1 \mathbf 1 + \mathbf F'$ transforms canonically, according to
$ \partial_1 \mathbf 1 + \tilde{\mathbf F}' = \mathbf A \left(\partial_1 \mathbf 1 + \mathbf F' \right) \mathbf A^\dagger$,
ensuring that each factor in the kinetic part of the Hamiltonian $\left(\partial_1 \mathbf 1 + \mathbf F' \right) \mathbf D \left(\partial_1 \mathbf 1 + \mathbf F' \right)$ is form-invariant under local gauge transformations by itself (and, of course, so is the whole product).
This already ensures that $\mathbf V^{\prime \text{BH}}$ is form-invariant, since we know that the full Hamiltonian is, and indeed one can check that the transformation of $\mathbf C'$ as defined in appendix \ref{app:kinetic} gives
extra terms cancelling exactly those arising in the transformation of $\mathbf V^\text{BH}$, cf. Eq. (\ref{eq:BHtrafo2}). Thus, the modified Born-Huang potential $\mathbf V^{\prime \text{BH}}$ that contains $\mathbf C'$ transforms canonically, by matrix conjugation with $\mathbf A$. In total, we find that, of course, the Hamiltonian is still invariant under local gauge transformations after merging the two kinetic terms, but now also its individual components are form-invariant, not only their combination.
\section{Brief conclusions}
To summarize, we employed the transverse mode decomposition to obtain the exact multi-mode matrix Hamiltonian for a quantum waveguide of arbitrary curvature, torsion, and spatially varying transverse profile, identifying the adiabatic limit and nonadiabatic coupling matrix elements. Series expansions of the coupling matrix elements were provided that may be truncated at low orders when approaching the ultrathin waveguide limit. For the common scenario of a subset of modes being nonadiabatically coupled to each other, but decoupled from the rest, sytematic few-mode approximation schemes were given. The resulting effective potential terms were worked out in simple special cases, reproducing the known result for twisting, and revealing a similar effect in shift-induced waveguides. It was demonstrated that the quantum waveguide exhibits a natural generalization of the local $\mbox{U}(N)$ gauge structure of the molecular Born-Oppenheimer problem, and the possibility of a strictly diabatic basis was explored, resulting in an explicit construction of the adiabatic-to-diabatic basis transformation matrix. 
The theoretical framework put forward here represents a general and very natural starting point for future investigations of nonadiabatic coupling effects in quantum waveguides. Promising applications include the design of novel waveguide structures with enhanced control over the longituditnal dynamics and the transverse mode profile of the guided matter waves.

\begin{acknowledgments}
The authors thank Sven Krönke for helpful discussions. J.S. gratefully acknowledges a scholarship by the Studienstiftung des deutschen Volkes.
\end{acknowledgments}

\appendix
\section{Derivation of the generalized Born-Oppenheimer kinetic matrix operator and alternative representations}
\label{app:kinetic}
In this appendix we give the detailed calculations postponed in Sections \ref{sec:matrix} and \ref{sec:few-mode}, starting with the derivation of Eq. (\ref{eq:kinetic}). By the definitions of $\mathbf F$ and $\mathbf G$, Eq. (\ref{eq:defFdefG}), we have
\begin{eqnarray*}
&&\left[(  \partial_1 \mathbf{1}+ \mathbf F )^2\right]_{mn} + G_{mn} = \delta_{mn}\partial_1^2 + 2 F_{mn}\partial_1 + \dot F_{mn} + \sum_k F_{mk}F_{kn} + G_{mn} \\
&=& \delta_{mn}\partial_1^2 + \left( \langle \phi_m | D | \partial_1 \phi_n \rangle - \langle \partial_1 \phi_m | D | \phi_n \rangle \right) \partial_1 + \langle  \phi_m | \dot D | \partial_1\phi_n \rangle + \langle \phi_m | D | \partial_1^2 \phi_n \rangle ,
\end{eqnarray*}
implying 
\begin{eqnarray*}
 &&\partial_1 \langle\phi_m | D -1  |\phi_n \rangle \partial_1 +\left[(  \partial_1 \mathbf{1}+ \mathbf F )^2\right]_{mn} + G_{mn} \\
&=&  \langle \partial_1 \phi_m | D |\phi_n \rangle \partial_1 +  \langle\phi_m | \dot D |\phi_n \rangle \partial_1 +  \langle\phi_m | D |\partial_1 \phi_n \rangle \partial_1 +  \langle\phi_m | D |\phi_n \rangle \partial_1^2 \\
&\qquad&+ \left( \langle \phi_m | D | \partial_1 \phi_n \rangle - \langle \partial_1 \phi_m | D | \phi_n \rangle \right) \partial_1 + \langle  \phi_m | \dot D | \partial_1\phi_n \rangle + \langle \phi_m | D | \partial_1^2 \phi_n \rangle\\
&=&    \langle\phi_m | \dot D |\phi_n \rangle \partial_1 + 2 \langle\phi_m | D |\partial_1 \phi_n \rangle \partial_1 +  \langle\phi_m | D |\phi_n \rangle \partial_1^2 +\langle  \phi_m | \dot D | \partial_1\phi_n \rangle + \langle \phi_m | D | \partial_1^2 \phi_n \rangle,
\end{eqnarray*}
which proves Eq. (\ref{eq:kinetic}). Consequently, the ensuing matrix form of the Hamiltonian contains two kinetic terms that read $\partial_1 \left( \mathbf D - \mathbf 1 \right) \partial_1 + \left( \partial_1 \mathbf 1 + \mathbf F \right)^2$, and after a few-mode-restriction to the subset $\mathcal S$ becomes $\partial_1 \left( \mathbf D^\mathcal{(S)} - \mathbf 1^\mathcal{(S)} \right) \partial_1 + \left( \partial_1 \mathbf 1^\mathcal{(S)} + \mathbf F^\mathcal{(S)} \right)^2$ plus a scalar contribution which forms a part of the Born-Huang potential, as indicated in the text.
We will now show that, as claimed in Section \ref{sec:few-mode}, these two kinetic terms can be merged into one (which still has the property of being manifestly Hermitian) at the price of also modifying the scalar part of the Hamiltonian. To this end, we demand that (omitting the superscripts $(\mathcal S)$):
\begin{equation*}
  \partial_1 \left( \mathbf D - \mathbf 1 \right) \partial_1 + \left( \partial_1 \mathbf 1 + \mathbf F \right)^2 \stackrel{!}{=} \left( \partial_1 \mathbf 1 + \mathbf F' \right) \mathbf D' \left( \partial_1 \mathbf 1 + \mathbf F' \right) + \mathbf C'
\end{equation*}
with matrices $\mathbf F'$, $\mathbf D'$ and $\mathbf C'$ to be determined. 
Expanding both sides and equating order by order in the derivative $\partial_1$ yields the system
\begin{eqnarray}
 \mathbf D &=& \mathbf D' \label{eq:partial1^2}\\ \dot {\mathbf D} + 2 \mathbf F &=& \dot {\mathbf D'} + \mathbf F' \mathbf D' +  \mathbf D' \mathbf F' \label{eq:partial1^1}\\
 \mathbf F^2 +\dot {\mathbf F } &=&\mathbf F' \mathbf D' \mathbf F' +\mathbf D' \dot{\mathbf F'} +\dot{\mathbf D'} \mathbf F'+ \mathbf C' \label{eq:partial1^0}.
\end{eqnarray}
which needs to be solved for $\mathbf D'$, $\mathbf F'$, $\mathbf C'$ as functions of $\mathbf D$, $\mathbf F$.
Inserting Eq. (\ref{eq:partial1^2}) into Eq. (\ref{eq:partial1^1}) results in $2 \mathbf F = \mathbf F' \mathbf D + \mathbf D \mathbf F'$,
which is the same Lyapunov equation encountered in the adiabatic-to-diabatic basis transformation in Section \ref{sec:diabatic}, and by the same arguments can be uniquely solved for the anti-Hermitian matrix $\mathbf F'$ as desired. 
Finally, using the previous results, Eq. (\ref{eq:partial1^0}) can be recast:
\begin{eqnarray*}
 \mathbf C' &=& 
 \mathbf F^2 +\frac{1}{2} \left(\dot {\mathbf F' } \mathbf D + \mathbf F' \dot{\mathbf D} +\dot {\mathbf D } \mathbf F' + \mathbf D \dot{\mathbf F'} \right)  -\mathbf D \dot{\mathbf F'} -\dot{\mathbf D} \mathbf F'- \mathbf F' \mathbf D \mathbf F' \nonumber 
= \mathbf F^2 +\frac{1}{2} \left( \partial_1 \left[{\mathbf F'}, \mathbf D \right] \right) - \mathbf F' \mathbf D \mathbf F' .
\end{eqnarray*}
Here it can immediately be seen that $\mathbf C'$ is Hermitian as required, since $\mathbf F$, $\mathbf F'$ are anti-Hermitian, $\mathbf D$ is Hermitian, and the commutator of a Hermitian and an anti-Hermitian matrix is Hermitian itself. 
Finally, we can absorb $\mathbf C'$ into the Born-Huang potential by setting $\mathbf V^{\prime \text{BH}}:=\mathbf V^{\text{BH}} - \hbar^2/(2M) \mathbf C'$, resulting in Eq. (\ref{eq:BH_absorbed}).

Restoring the superscript, the Lyapunov equation to be solved reads $2 \mathbf F^\mathcal{(S)} = \mathbf F^\mathcal{\prime (S)} \mathbf D^\mathcal{(S)} + \mathbf D^\mathcal{(S)} \mathbf F^\mathcal{\prime (S)}$. As seen in Section \ref{sec:diabatic}, if no few-mode restriction is performed and $\mathcal S$ is a complete set of transverse modes, this immediately has the solution $F'_{mn} = \langle \phi_m | \partial_1 \phi_n \rangle$.
So without any few-mode restriction, the derivative coupling $\mathbf F'$ simply assumes the form known from the molecular Born-Oppenheimer problem. If a subset $\mathcal S$ of transverse modes is singled out first, this holds only approximately after inserting $\sum_{k \in \mathcal S} | \phi_k \rangle \langle \phi_k | \approx 1$ into the Lyapunov equation. We have seen before that, by the same reasoning, when looking for the adiabatic-to-diabatic basis transformation without performing a few-mode approximation first -- which is of course not the generic scenario -- the Lyapunov equation can also be immediately solved for $S_{mn} = \langle \phi_m | \partial_1 \phi_n \rangle$, and the basis transformation has the same form as in the absence of curvature. This is expected, since a complete diabatic basis is just \emph{any} complete basis of the normal plane where no transverse mode depends on $u_1$, i.e. which is not at all adapted to changes along the waveguide. This is true both with and without curvature and torsion. 

\section{Detailed calculation of gauge transformation properties}
\label{app:gauge}
In this appendix we comprise the derivation of the transformation properties of the kinetic terms in $\mathbf H$ and of $\mathbf V^\text{BH}$ as discussed in Section \ref{sec:diabatic}. 
The evolution of $\mathbf A$ along $u_1$ is determined by a skew-Hermitian matrix $\mathbf S$ via $\dot{\mathbf{ A}} \mathbf{ = A S}$, and the induced transformation of $\mathbf D$, $\mathbf V$ and $\mathbf C$ is given by simple conjugation with $\mathbf A$. For the generalized derivative coupling matrix $\mathbf F$, it was found that
 $\tilde{\mathbf F} = \mathbf {A F A^\dagger } - \frac{1}{2} \mathbf{ A \{ D, S\}  A^\dagger},$
which implies
\begin{equation*}
\mathbf A \left(\partial_1 + \mathbf F \right) \mathbf A^\dagger = \mathbf {A F A^\dagger } + \partial_1 + \mathbf{A}\dot{\mathbf{ A}}^\dagger = \partial_1 + \tilde{\mathbf F} - \mathbf{ASA^\dagger} +\frac{1}{2} \mathbf A \{\mathbf D, \mathbf S\} \mathbf A^\dagger,
\end{equation*}
and consequently
\begin{eqnarray*}
 && \mathbf A \left(\partial_1 + \mathbf F \right)^2 \mathbf A^\dagger 
=\left( \partial_1 + \tilde{\mathbf F} \right)^2 + \mathbf A \left( -\mathbf S + \frac{1}{2}  \{\mathbf D, \mathbf S\} \right)^2\mathbf A^\dagger + \{ \partial_1 + \tilde{\mathbf F}, - \mathbf{ASA^\dagger} +\frac{1}{2} \mathbf A \{\mathbf D, \mathbf S\} \mathbf A^\dagger\} \\
&&=\left( \partial_1 + \tilde{\mathbf F} \right)^2 + \mathbf A \left( \mathbf S^2 - \frac{1}{4}  \{\mathbf D, \mathbf S\}^2  - \{\mathbf F, \mathbf S \}  + \frac{1}{2} \{\mathbf F, \{\mathbf D, \mathbf S\}\right)\mathbf A^\dagger  +  \frac{1}{2} \{ \partial_1, \mathbf A  \{\mathbf D - \mathbf 1, \mathbf S\}   \mathbf A^\dagger \}.
\end{eqnarray*}
For the other kinetic term, a similar calculation yields
\begin{eqnarray*}
  &&\partial_1 (\tilde{\mathbf D} - \mathbf 1) \partial_1 
= \mathbf A \partial_1  (\mathbf D - \mathbf 1) \partial_1  \mathbf A^\dagger 
+\mathbf A \left( \frac{1}{2} \left(\partial_1 [ \mathbf D ,\mathbf{S}]\right) - \frac{1}{2} \{ \mathbf {D},\mathbf{ S^2}\}  + \mathbf S^2 \right) \mathbf{A^\dagger} + \frac{1}{2}\{ \partial_1, {\mathbf A} \{ \mathbf D - \mathbf 1, \mathbf{S} \}  \mathbf{A^\dagger} \}.
\end{eqnarray*}
So we find the result of Eq. (\ref{eq:trafo-kinetic}) for the sum of the transformed kinetic terms:
\begin{eqnarray}
&& \partial_1 (\tilde{\mathbf D} - \mathbf 1) \partial_1 +\left(\partial_1 + \tilde{\mathbf F} \right)^2 \nonumber\\
&&=  \mathbf A \left[ \partial_1  (\mathbf D - \mathbf 1) \partial_1  +\left( \partial_1 + \mathbf F \right)^2 
+ \frac{1}{2} \left(\partial_1 [ \mathbf D ,\mathbf{S}]\right) - \frac{1}{2} \{ \mathbf {D},\mathbf{ S^2}\}  + \frac{1}{4} \{\mathbf D, \mathbf S\}^2 + \{\mathbf F, \mathbf S\} - \frac{1}{2} \{\mathbf F, \{\mathbf D, \mathbf S\}\} 
 \right] \mathbf A^\dagger.
\label{eq:trafokinetic}
\end{eqnarray}
Now we proceed to the Born-Huang potential, which in the transformed basis reads
\begin{eqnarray*}
 \tilde V_{mn}^\text{BH} &=& -\frac{\hbar^2}{2M} \left[ - \sum_{k} \tilde F_{mk}\tilde F_{kn} + \frac{1}{2} \left( \langle \partial_1 \tilde\phi_m | \dot D | \tilde \phi_n \rangle + \langle \tilde \phi_m | \dot D | \partial_1 \tilde \phi_n \rangle + \langle \partial_1 ^2 \tilde\phi_m | D | \tilde\phi_n \rangle + \langle \tilde\phi_m | D | \partial_1 ^2 \tilde\phi_n \rangle\right) \right] 
\end{eqnarray*}
We consider the terms in $\tilde{ \mathbf V}^\text{BH}$ individually. First, we find after a straightforward calculation
\begin{eqnarray*}
 &&\tilde V_{mn}^\text{BH,0} := - \sum_{k}  \tilde F_{mk}\tilde F_{kn} 
=\left[ \mathbf A \left(\mathbf V^\text{BH,0}  - \frac{1}{4} \{\mathbf D, \mathbf S \}^2  +  \frac{1}{2} \{ \{\mathbf D, \mathbf S \}, \mathbf F \} \right) \mathbf{A^\dagger} \right]_{mn} .
\end{eqnarray*}
Similarly, for the second term
\begin{eqnarray*}
 \tilde V_{mn}^\text{BH,1} &:=& \frac{1}{2} \left(\langle \partial_1 \tilde\phi_m | \dot D | \tilde \phi_n \rangle + \langle \tilde \phi_m | \dot D | \partial_1 \tilde \phi_n \rangle \right)
= \left[ \mathbf{A}\left( \mathbf{V^\text{BH,1}} + \frac{1}{2} \left[\mathbf S, \mathring{ \mathbf {D} } \right]  \right) \mathbf{A^\dagger} \right]_{mn},
\end{eqnarray*}
where the matrix $\mathring{ \mathbf {D} }$ with entries $\mathring{D}_{mn} := \langle \phi_m | \dot D | \phi_n \rangle$ was introduced. Finally, the third contribution reads
\begin{eqnarray*}
 \tilde V_{mn}^\text{BH,2} &:=& \frac{1}{2} \left( \langle \partial_1^2 \tilde\phi_m | D | \tilde \phi_n \rangle + \langle \tilde \phi_m | D | \partial_1^2 \tilde \phi_n \rangle  \right) 
= \left[ \mathbf{A}\left(\mathbf{V^\text{BH,2}} + \frac{1}{2} \left[\dot{\mathbf {S}}, \mathbf D \right] +\frac{1}{2}\left\{\mathbf {S^2}, \mathbf D \right\} +\mathbf{ SL -L^\dagger S}  \right) \mathbf{A^\dagger} \right]_{mn},
\end{eqnarray*}
where we have introduced the shorthand notation $L_{ij}:=\langle \partial_1 \phi_i | D | \phi_j \rangle$, such that $2 \mathbf F = \mathbf{L^\dagger -L}$.
We can simplify the result by noting that
\begin{eqnarray*}
 \left[\mathbf S, \dot{\mathbf D} \right] &=& \left[ \mathbf S, \mathbf{L} +\mathring{ \mathbf {D} } + \mathbf{L}^\dagger\right]
= \left[ \mathbf S, \mathring{ \mathbf {D} } \right] + 2(\mathbf{ SL} - \mathbf {L^\dagger S}) + 2 \{ \mathbf{S},\mathbf{F} \},
\end{eqnarray*}
such that the full transformed Born-Huang potential reads
\begin{equation}
  \tilde {\mathbf V}^\text{BH} = \mathbf{A} \mathbf{V^\text{BH}}\mathbf{A^\dagger} 
 -\frac{\hbar^2}{2M}  \left[ \mathbf{A}\left(  - \frac{1}{4} \{\mathbf D, \mathbf S \}^2  +  \frac{1}{2} \{ \{\mathbf D, \mathbf S \}, \mathbf F \}+ \frac{1}{2} \left(  \partial_1\left[\mathbf { S}, \mathbf D \right] \right)  - \{ \mathbf{S},\mathbf{F} \}
 + \frac{1}{2}\left\{\mathbf {S^2}, \mathbf D \right\}  \right) \mathbf{A^\dagger} \right]
\label{eq:trafoBH}
\end{equation}
as provided in Eq. (\ref{eq:BHtrafo2}). The extra terms arising in Eq. (\ref{eq:trafokinetic}) and Eq. (\ref{eq:trafoBH}) cancel when the kinetic and potential parts of the Hamiltonian are summed.

\bibliography{bib_waveguides_NAC}

\begin{thebibliography}{106}%
\makeatletter
\providecommand \@ifxundefined [1]{%
 \@ifx{#1\undefined}
}%
\providecommand \@ifnum [1]{%
 \ifnum #1\expandafter \@firstoftwo
 \else \expandafter \@secondoftwo
 \fi
}%
\providecommand \@ifx [1]{%
 \ifx #1\expandafter \@firstoftwo
 \else \expandafter \@secondoftwo
 \fi
}%
\providecommand \natexlab [1]{#1}%
\providecommand \enquote  [1]{``#1''}%
\providecommand \bibnamefont  [1]{#1}%
\providecommand \bibfnamefont [1]{#1}%
\providecommand \citenamefont [1]{#1}%
\providecommand \href@noop [0]{\@secondoftwo}%
\providecommand \href [0]{\begingroup \@sanitize@url \@href}%
\providecommand \@href[1]{\@@startlink{#1}\@@href}%
\providecommand \@@href[1]{\endgroup#1\@@endlink}%
\providecommand \@sanitize@url [0]{\catcode `\\12\catcode `\$12\catcode
  `\&12\catcode `\#12\catcode `\^12\catcode `\_12\catcode `\%12\relax}%
\providecommand \@@startlink[1]{}%
\providecommand \@@endlink[0]{}%
\providecommand \url  [0]{\begingroup\@sanitize@url \@url }%
\providecommand \@url [1]{\endgroup\@href {#1}{\urlprefix }}%
\providecommand \urlprefix  [0]{URL }%
\providecommand \Eprint [0]{\href }%
\providecommand \doibase [0]{http://dx.doi.org/}%
\providecommand \selectlanguage [0]{\@gobble}%
\providecommand \bibinfo  [0]{\@secondoftwo}%
\providecommand \bibfield  [0]{\@secondoftwo}%
\providecommand \translation [1]{[#1]}%
\providecommand \BibitemOpen [0]{}%
\providecommand \bibitemStop [0]{}%
\providecommand \bibitemNoStop [0]{.\EOS\space}%
\providecommand \EOS [0]{\spacefactor3000\relax}%
\providecommand \BibitemShut  [1]{\csname bibitem#1\endcsname}%
\let\auto@bib@innerbib\@empty
\bibitem [{\citenamefont {Londergan}\ \emph {et~al.}(1999)\citenamefont
  {Londergan}, \citenamefont {Carini},\ and\ \citenamefont
  {Murdock}}]{Londergan1999i}%
  \BibitemOpen
  \bibfield  {author} {\bibinfo {author} {\bibfnamefont {J.~T.}\ \bibnamefont
  {Londergan}}, \bibinfo {author} {\bibfnamefont {J.~P.}\ \bibnamefont
  {Carini}}, \ and\ \bibinfo {author} {\bibfnamefont {D.~P.}\ \bibnamefont
  {Murdock}},\ }\href@noop {} {\emph {\bibinfo {title} {Binding and Scattering
  in Two-Dimensional Systems - Applications to Quantum Wires, Waveguides and
  Photonic Crystals}}},\ \bibinfo {series} {Lecture Notes in Physics
  Monographs}, Vol.~\bibinfo {volume} {60}\ (\bibinfo  {publisher} {Springer
  Berlin/Heidelberg},\ \bibinfo {year} {1999})\BibitemShut {NoStop}%
\bibitem [{\citenamefont {Hoffrogge}\ \emph {et~al.}(2011)\citenamefont
  {Hoffrogge}, \citenamefont {Fr\"ohlich}, \citenamefont {Kasevich},\ and\
  \citenamefont {Hommelhoff}}]{Hoffrogge2011}%
  \BibitemOpen
  \bibfield  {author} {\bibinfo {author} {\bibfnamefont {J.}~\bibnamefont
  {Hoffrogge}}, \bibinfo {author} {\bibfnamefont {R.}~\bibnamefont
  {Fr\"ohlich}}, \bibinfo {author} {\bibfnamefont {M.~A.}\ \bibnamefont
  {Kasevich}}, \ and\ \bibinfo {author} {\bibfnamefont {P.}~\bibnamefont
  {Hommelhoff}},\ }\href {\doibase 10.1103/PhysRevLett.106.193001} {\bibfield
  {journal} {\bibinfo  {journal} {Phys. Rev. Lett.}\ }\textbf {\bibinfo
  {volume} {106}},\ \bibinfo {pages} {193001} (\bibinfo {year}
  {2011})}\BibitemShut {NoStop}%
\bibitem [{\citenamefont {Klein}\ and\ \citenamefont
  {Werner}(1983)}]{Klein1983}%
  \BibitemOpen
  \bibfield  {author} {\bibinfo {author} {\bibfnamefont {A.~G.}\ \bibnamefont
  {Klein}}\ and\ \bibinfo {author} {\bibfnamefont {S.~A.}\ \bibnamefont
  {Werner}},\ }\href {http://stacks.iop.org/0034-4885/46/i=3/a=001} {\bibfield
  {journal} {\bibinfo  {journal} {Rep. Prog. Phys.}\ }\textbf {\bibinfo
  {volume} {46}},\ \bibinfo {pages} {259} (\bibinfo {year} {1983})}\BibitemShut
  {NoStop}%
\bibitem [{\citenamefont {Alvarez-Estrada}\ and\ \citenamefont
  {Calvo}(1984)}]{Alvarez-Estrada1984}%
  \BibitemOpen
  \bibfield  {author} {\bibinfo {author} {\bibfnamefont {R.~F.}\ \bibnamefont
  {Alvarez-Estrada}}\ and\ \bibinfo {author} {\bibfnamefont {M.~L.}\
  \bibnamefont {Calvo}},\ }\href {http://stacks.iop.org/0022-3727/17/i=3/a=007}
  {\bibfield  {journal} {\bibinfo  {journal} {J. Phys. D: Appl. Phys.}\
  }\textbf {\bibinfo {volume} {17}},\ \bibinfo {pages} {475} (\bibinfo {year}
  {1984})}\BibitemShut {NoStop}%
\bibitem [{\citenamefont {Szameit}\ \emph {et~al.}(2010)\citenamefont
  {Szameit}, \citenamefont {Dreisow}, \citenamefont {Heinrich}, \citenamefont
  {Keil}, \citenamefont {Nolte}, \citenamefont {T\"unnermann},\ and\
  \citenamefont {Longhi}}]{Szameit2010}%
  \BibitemOpen
  \bibfield  {author} {\bibinfo {author} {\bibfnamefont {A.}~\bibnamefont
  {Szameit}}, \bibinfo {author} {\bibfnamefont {F.}~\bibnamefont {Dreisow}},
  \bibinfo {author} {\bibfnamefont {M.}~\bibnamefont {Heinrich}}, \bibinfo
  {author} {\bibfnamefont {R.}~\bibnamefont {Keil}}, \bibinfo {author}
  {\bibfnamefont {S.}~\bibnamefont {Nolte}}, \bibinfo {author} {\bibfnamefont
  {A.}~\bibnamefont {T\"unnermann}}, \ and\ \bibinfo {author} {\bibfnamefont
  {S.}~\bibnamefont {Longhi}},\ }\href {\doibase
  10.1103/PhysRevLett.104.150403} {\bibfield  {journal} {\bibinfo  {journal}
  {Phys. Rev. Lett.}\ }\textbf {\bibinfo {volume} {104}},\ \bibinfo {pages}
  {150403} (\bibinfo {year} {2010})}\BibitemShut {NoStop}%
\bibitem [{\citenamefont {Kartashov}\ \emph {et~al.}(2011)\citenamefont
  {Kartashov}, \citenamefont {Szameit}, \citenamefont {Keil}, \citenamefont
  {Vysloukh},\ and\ \citenamefont {Torner}}]{Kartashov2011}%
  \BibitemOpen
  \bibfield  {author} {\bibinfo {author} {\bibfnamefont {Y.~V.}\ \bibnamefont
  {Kartashov}}, \bibinfo {author} {\bibfnamefont {A.}~\bibnamefont {Szameit}},
  \bibinfo {author} {\bibfnamefont {R.}~\bibnamefont {Keil}}, \bibinfo {author}
  {\bibfnamefont {V.~A.}\ \bibnamefont {Vysloukh}}, \ and\ \bibinfo {author}
  {\bibfnamefont {L.}~\bibnamefont {Torner}},\ }\href {\doibase
  10.1364/OL.36.003470} {\bibfield  {journal} {\bibinfo  {journal} {Opt.
  Lett.}\ }\textbf {\bibinfo {volume} {36}},\ \bibinfo {pages} {3470} (\bibinfo
  {year} {2011})}\BibitemShut {NoStop}%
\bibitem [{\citenamefont {Marcus}(1966)}]{Marcus1966}%
  \BibitemOpen
  \bibfield  {author} {\bibinfo {author} {\bibfnamefont {R.~A.}\ \bibnamefont
  {Marcus}},\ }\href {\doibase 10.1063/1.1727528} {\bibfield  {journal}
  {\bibinfo  {journal} {J. Chem. Phys.}\ }\textbf {\bibinfo {volume} {45}},\
  \bibinfo {pages} {4493} (\bibinfo {year} {1966})}\BibitemShut {NoStop}%
\bibitem [{\citenamefont {Torrontegui}\ \emph {et~al.}(2011)\citenamefont
  {Torrontegui}, \citenamefont {Ruschhaupt}, \citenamefont {Guéry-Odelin},\
  and\ \citenamefont {Muga}}]{Torrontegui2011}%
  \BibitemOpen
  \bibfield  {author} {\bibinfo {author} {\bibfnamefont {E.}~\bibnamefont
  {Torrontegui}}, \bibinfo {author} {\bibfnamefont {A.}~\bibnamefont
  {Ruschhaupt}}, \bibinfo {author} {\bibfnamefont {D.}~\bibnamefont
  {Guéry-Odelin}}, \ and\ \bibinfo {author} {\bibfnamefont {J.~G.}\
  \bibnamefont {Muga}},\ }\href
  {http://stacks.iop.org/0953-4075/44/i=19/a=195302} {\bibfield  {journal}
  {\bibinfo  {journal} {J. Phys. B: At. Mol. Opt. Phys.}\ }\textbf {\bibinfo
  {volume} {44}},\ \bibinfo {pages} {195302} (\bibinfo {year}
  {2011})}\BibitemShut {NoStop}%
\bibitem [{\citenamefont {Bittner}\ \emph {et~al.}(2013)\citenamefont
  {Bittner}, \citenamefont {Dietz}, \citenamefont {Miski-Oglu}, \citenamefont
  {Richter}, \citenamefont {Ripp}, \citenamefont {Sadurn\'\i},\ and\
  \citenamefont {Schleich}}]{Bittner2013}%
  \BibitemOpen
  \bibfield  {author} {\bibinfo {author} {\bibfnamefont {S.}~\bibnamefont
  {Bittner}}, \bibinfo {author} {\bibfnamefont {B.}~\bibnamefont {Dietz}},
  \bibinfo {author} {\bibfnamefont {M.}~\bibnamefont {Miski-Oglu}}, \bibinfo
  {author} {\bibfnamefont {A.}~\bibnamefont {Richter}}, \bibinfo {author}
  {\bibfnamefont {C.}~\bibnamefont {Ripp}}, \bibinfo {author} {\bibfnamefont
  {E.}~\bibnamefont {Sadurn\'\i}}, \ and\ \bibinfo {author} {\bibfnamefont
  {W.~P.}\ \bibnamefont {Schleich}},\ }\href {\doibase
  10.1103/PhysRevE.87.042912} {\bibfield  {journal} {\bibinfo  {journal} {Phys.
  Rev. E}\ }\textbf {\bibinfo {volume} {87}},\ \bibinfo {pages} {042912}
  (\bibinfo {year} {2013})}\BibitemShut {NoStop}%
\bibitem [{\citenamefont {Timp}\ \emph {et~al.}(1988)\citenamefont {Timp},
  \citenamefont {Baranger}, \citenamefont {deVegvar}, \citenamefont
  {Cunningham}, \citenamefont {Howard}, \citenamefont {Behringer},\ and\
  \citenamefont {Mankiewich}}]{Timp1988}%
  \BibitemOpen
  \bibfield  {author} {\bibinfo {author} {\bibfnamefont {G.}~\bibnamefont
  {Timp}}, \bibinfo {author} {\bibfnamefont {H.~U.}\ \bibnamefont {Baranger}},
  \bibinfo {author} {\bibfnamefont {P.}~\bibnamefont {deVegvar}}, \bibinfo
  {author} {\bibfnamefont {J.~E.}\ \bibnamefont {Cunningham}}, \bibinfo
  {author} {\bibfnamefont {R.~E.}\ \bibnamefont {Howard}}, \bibinfo {author}
  {\bibfnamefont {R.}~\bibnamefont {Behringer}}, \ and\ \bibinfo {author}
  {\bibfnamefont {P.~M.}\ \bibnamefont {Mankiewich}},\ }\href {\doibase
  10.1103/PhysRevLett.60.2081} {\bibfield  {journal} {\bibinfo  {journal}
  {Phys. Rev. Lett.}\ }\textbf {\bibinfo {volume} {60}},\ \bibinfo {pages}
  {2081} (\bibinfo {year} {1988})}\BibitemShut {NoStop}%
\bibitem [{\citenamefont {Exner}(1989)}]{Exner1989a}%
  \BibitemOpen
  \bibfield  {author} {\bibinfo {author} {\bibfnamefont {P.}~\bibnamefont
  {Exner}},\ }\href {\doibase 10.1016/0375-9601(89)90470-2} {\bibfield
  {journal} {\bibinfo  {journal} {Phys. Lett. A}\ }\textbf {\bibinfo {volume}
  {141}},\ \bibinfo {pages} {213 } (\bibinfo {year} {1989})}\BibitemShut
  {NoStop}%
\bibitem [{\citenamefont {Carini}\ \emph {et~al.}(1992)\citenamefont {Carini},
  \citenamefont {Londergan}, \citenamefont {Mullen},\ and\ \citenamefont
  {Murdock}}]{Carini1992}%
  \BibitemOpen
  \bibfield  {author} {\bibinfo {author} {\bibfnamefont {J.~P.}\ \bibnamefont
  {Carini}}, \bibinfo {author} {\bibfnamefont {J.~T.}\ \bibnamefont
  {Londergan}}, \bibinfo {author} {\bibfnamefont {K.}~\bibnamefont {Mullen}}, \
  and\ \bibinfo {author} {\bibfnamefont {D.~P.}\ \bibnamefont {Murdock}},\
  }\href {\doibase 10.1103/PhysRevB.46.15538} {\bibfield  {journal} {\bibinfo
  {journal} {Phys. Rev. B}\ }\textbf {\bibinfo {volume} {46}},\ \bibinfo
  {pages} {15538} (\bibinfo {year} {1992})}\BibitemShut {NoStop}%
\bibitem [{\citenamefont {Carini}\ \emph {et~al.}(1993)\citenamefont {Carini},
  \citenamefont {Londergan}, \citenamefont {Mullen},\ and\ \citenamefont
  {Murdock}}]{Carini1993}%
  \BibitemOpen
  \bibfield  {author} {\bibinfo {author} {\bibfnamefont {J.~P.}\ \bibnamefont
  {Carini}}, \bibinfo {author} {\bibfnamefont {J.~T.}\ \bibnamefont
  {Londergan}}, \bibinfo {author} {\bibfnamefont {K.}~\bibnamefont {Mullen}}, \
  and\ \bibinfo {author} {\bibfnamefont {D.~P.}\ \bibnamefont {Murdock}},\
  }\href {\doibase 10.1103/PhysRevB.48.4503} {\bibfield  {journal} {\bibinfo
  {journal} {Phys. Rev. B}\ }\textbf {\bibinfo {volume} {48}},\ \bibinfo
  {pages} {4503} (\bibinfo {year} {1993})}\BibitemShut {NoStop}%
\bibitem [{\citenamefont {Carini}\ \emph
  {et~al.}(1997{\natexlab{a}})\citenamefont {Carini}, \citenamefont
  {Londergan},\ and\ \citenamefont {Murdock}}]{Carini1997}%
  \BibitemOpen
  \bibfield  {author} {\bibinfo {author} {\bibfnamefont {J.~P.}\ \bibnamefont
  {Carini}}, \bibinfo {author} {\bibfnamefont {J.~T.}\ \bibnamefont
  {Londergan}}, \ and\ \bibinfo {author} {\bibfnamefont {D.~P.}\ \bibnamefont
  {Murdock}},\ }\href {\doibase 10.1103/PhysRevB.55.9852} {\bibfield  {journal}
  {\bibinfo  {journal} {Phys. Rev. B}\ }\textbf {\bibinfo {volume} {55}},\
  \bibinfo {pages} {9852} (\bibinfo {year} {1997}{\natexlab{a}})}\BibitemShut
  {NoStop}%
\bibitem [{\citenamefont {Carini}\ \emph
  {et~al.}(1997{\natexlab{b}})\citenamefont {Carini}, \citenamefont
  {Londergan}, \citenamefont {Murdock}, \citenamefont {Trinkle},\ and\
  \citenamefont {Yung}}]{Carini1997a}%
  \BibitemOpen
  \bibfield  {author} {\bibinfo {author} {\bibfnamefont {J.~P.}\ \bibnamefont
  {Carini}}, \bibinfo {author} {\bibfnamefont {J.~T.}\ \bibnamefont
  {Londergan}}, \bibinfo {author} {\bibfnamefont {D.~P.}\ \bibnamefont
  {Murdock}}, \bibinfo {author} {\bibfnamefont {D.}~\bibnamefont {Trinkle}}, \
  and\ \bibinfo {author} {\bibfnamefont {C.~S.}\ \bibnamefont {Yung}},\ }\href
  {\doibase 10.1103/PhysRevB.55.9842} {\bibfield  {journal} {\bibinfo
  {journal} {Phys. Rev. B}\ }\textbf {\bibinfo {volume} {55}},\ \bibinfo
  {pages} {9842} (\bibinfo {year} {1997}{\natexlab{b}})}\BibitemShut {NoStop}%
\bibitem [{\citenamefont {Reichel}(2002)}]{Reichel2002}%
  \BibitemOpen
  \bibfield  {author} {\bibinfo {author} {\bibfnamefont {J.}~\bibnamefont
  {Reichel}},\ }\href {\doibase 10.1007/s003400200861} {\bibfield  {journal}
  {\bibinfo  {journal} {App. Phys. B}\ }\textbf {\bibinfo {volume} {74}},\
  \bibinfo {pages} {469} (\bibinfo {year} {2002})}\BibitemShut {NoStop}%
\bibitem [{\citenamefont {Folman}\ \emph {et~al.}(2002)\citenamefont {Folman},
  \citenamefont {Kruger}, \citenamefont {Schmiedmayer}, \citenamefont
  {Denschlag},\ and\ \citenamefont {Henkel}}]{Folman2002}%
  \BibitemOpen
  \bibfield  {author} {\bibinfo {author} {\bibfnamefont {R.}~\bibnamefont
  {Folman}}, \bibinfo {author} {\bibfnamefont {P.}~\bibnamefont {Kruger}},
  \bibinfo {author} {\bibfnamefont {J.}~\bibnamefont {Schmiedmayer}}, \bibinfo
  {author} {\bibfnamefont {J.}~\bibnamefont {Denschlag}}, \ and\ \bibinfo
  {author} {\bibfnamefont {C.}~\bibnamefont {Henkel}},\ }\href
  {http://dx.doi.org/10.1016/S1049-250X(02)80011-8} {\bibfield  {journal}
  {\bibinfo  {journal} {Adv. At. Mol. Opt. Phys.}\ }\textbf {\bibinfo {volume}
  {48}},\ \bibinfo {pages} {263} (\bibinfo {year} {2002})}\BibitemShut
  {NoStop}%
\bibitem [{\citenamefont {Adams}\ \emph {et~al.}(1994)\citenamefont {Adams},
  \citenamefont {Sigel},\ and\ \citenamefont {Mlynek}}]{Adams1994}%
  \BibitemOpen
  \bibfield  {author} {\bibinfo {author} {\bibfnamefont {C.}~\bibnamefont
  {Adams}}, \bibinfo {author} {\bibfnamefont {M.}~\bibnamefont {Sigel}}, \ and\
  \bibinfo {author} {\bibfnamefont {J.}~\bibnamefont {Mlynek}},\ }\href
  {\doibase http://dx.doi.org/10.1016/0370-1573(94)90066-3} {\bibfield
  {journal} {\bibinfo  {journal} {Phys. Rep.}\ }\textbf {\bibinfo {volume}
  {240}},\ \bibinfo {pages} {143 } (\bibinfo {year} {1994})}\BibitemShut
  {NoStop}%
\bibitem [{\citenamefont {Bongs}\ \emph {et~al.}(2001)\citenamefont {Bongs},
  \citenamefont {Burger}, \citenamefont {Dettmer}, \citenamefont {Hellweg},
  \citenamefont {Arlt}, \citenamefont {Ertmer},\ and\ \citenamefont
  {Sengstock}}]{Bongs2001}%
  \BibitemOpen
  \bibfield  {author} {\bibinfo {author} {\bibfnamefont {K.}~\bibnamefont
  {Bongs}}, \bibinfo {author} {\bibfnamefont {S.}~\bibnamefont {Burger}},
  \bibinfo {author} {\bibfnamefont {S.}~\bibnamefont {Dettmer}}, \bibinfo
  {author} {\bibfnamefont {D.}~\bibnamefont {Hellweg}}, \bibinfo {author}
  {\bibfnamefont {J.}~\bibnamefont {Arlt}}, \bibinfo {author} {\bibfnamefont
  {W.}~\bibnamefont {Ertmer}}, \ and\ \bibinfo {author} {\bibfnamefont
  {K.}~\bibnamefont {Sengstock}},\ }\href {\doibase 10.1103/PhysRevA.63.031602}
  {\bibfield  {journal} {\bibinfo  {journal} {Phys. Rev. A}\ }\textbf {\bibinfo
  {volume} {63}},\ \bibinfo {pages} {031602} (\bibinfo {year}
  {2001})}\BibitemShut {NoStop}%
\bibitem [{\citenamefont {Leanhardt}\ \emph {et~al.}(2002)\citenamefont
  {Leanhardt}, \citenamefont {Chikkatur}, \citenamefont {Kielpinski},
  \citenamefont {Shin}, \citenamefont {Gustavson}, \citenamefont {Ketterle},\
  and\ \citenamefont {Pritchard}}]{Leanhardt2002}%
  \BibitemOpen
  \bibfield  {author} {\bibinfo {author} {\bibfnamefont {A.~E.}\ \bibnamefont
  {Leanhardt}}, \bibinfo {author} {\bibfnamefont {A.~P.}\ \bibnamefont
  {Chikkatur}}, \bibinfo {author} {\bibfnamefont {D.}~\bibnamefont
  {Kielpinski}}, \bibinfo {author} {\bibfnamefont {Y.}~\bibnamefont {Shin}},
  \bibinfo {author} {\bibfnamefont {T.~L.}\ \bibnamefont {Gustavson}}, \bibinfo
  {author} {\bibfnamefont {W.}~\bibnamefont {Ketterle}}, \ and\ \bibinfo
  {author} {\bibfnamefont {D.~E.}\ \bibnamefont {Pritchard}},\ }\href {\doibase
  10.1103/PhysRevLett.89.040401} {\bibfield  {journal} {\bibinfo  {journal}
  {Phys. Rev. Lett.}\ }\textbf {\bibinfo {volume} {89}},\ \bibinfo {pages}
  {040401} (\bibinfo {year} {2002})}\BibitemShut {NoStop}%
\bibitem [{\citenamefont {Guerin}\ \emph {et~al.}(2006)\citenamefont {Guerin},
  \citenamefont {Riou}, \citenamefont {Gaebler}, \citenamefont {Josse},
  \citenamefont {Bouyer},\ and\ \citenamefont {Aspect}}]{Guerin2006}%
  \BibitemOpen
  \bibfield  {author} {\bibinfo {author} {\bibfnamefont {W.}~\bibnamefont
  {Guerin}}, \bibinfo {author} {\bibfnamefont {J.-F.}\ \bibnamefont {Riou}},
  \bibinfo {author} {\bibfnamefont {J.~P.}\ \bibnamefont {Gaebler}}, \bibinfo
  {author} {\bibfnamefont {V.}~\bibnamefont {Josse}}, \bibinfo {author}
  {\bibfnamefont {P.}~\bibnamefont {Bouyer}}, \ and\ \bibinfo {author}
  {\bibfnamefont {A.}~\bibnamefont {Aspect}},\ }\href {\doibase
  10.1103/PhysRevLett.97.200402} {\bibfield  {journal} {\bibinfo  {journal}
  {Phys. Rev. Lett.}\ }\textbf {\bibinfo {volume} {97}},\ \bibinfo {pages}
  {200402} (\bibinfo {year} {2006})}\BibitemShut {NoStop}%
\bibitem [{\citenamefont {Cassettari}\ \emph {et~al.}(2000)\citenamefont
  {Cassettari}, \citenamefont {Hessmo}, \citenamefont {Folman}, \citenamefont
  {Maier},\ and\ \citenamefont {Schmiedmayer}}]{Cassettari2000}%
  \BibitemOpen
  \bibfield  {author} {\bibinfo {author} {\bibfnamefont {D.}~\bibnamefont
  {Cassettari}}, \bibinfo {author} {\bibfnamefont {B.}~\bibnamefont {Hessmo}},
  \bibinfo {author} {\bibfnamefont {R.}~\bibnamefont {Folman}}, \bibinfo
  {author} {\bibfnamefont {T.}~\bibnamefont {Maier}}, \ and\ \bibinfo {author}
  {\bibfnamefont {J.}~\bibnamefont {Schmiedmayer}},\ }\href {\doibase
  10.1103/PhysRevLett.85.5483} {\bibfield  {journal} {\bibinfo  {journal}
  {Phys. Rev. Lett.}\ }\textbf {\bibinfo {volume} {85}},\ \bibinfo {pages}
  {5483} (\bibinfo {year} {2000})}\BibitemShut {NoStop}%
\bibitem [{\citenamefont {Gattobigio}\ \emph {et~al.}(2012)\citenamefont
  {Gattobigio}, \citenamefont {Couvert}, \citenamefont {Reinaudi},
  \citenamefont {Georgeot},\ and\ \citenamefont
  {Gu\'ery-Odelin}}]{Gattobigio2012}%
  \BibitemOpen
  \bibfield  {author} {\bibinfo {author} {\bibfnamefont {G.~L.}\ \bibnamefont
  {Gattobigio}}, \bibinfo {author} {\bibfnamefont {A.}~\bibnamefont {Couvert}},
  \bibinfo {author} {\bibfnamefont {G.}~\bibnamefont {Reinaudi}}, \bibinfo
  {author} {\bibfnamefont {B.}~\bibnamefont {Georgeot}}, \ and\ \bibinfo
  {author} {\bibfnamefont {D.}~\bibnamefont {Gu\'ery-Odelin}},\ }\href
  {\doibase 10.1103/PhysRevLett.109.030403} {\bibfield  {journal} {\bibinfo
  {journal} {Phys. Rev. Lett.}\ }\textbf {\bibinfo {volume} {109}},\ \bibinfo
  {pages} {030403} (\bibinfo {year} {2012})}\BibitemShut {NoStop}%
\bibitem [{\citenamefont {Sauer}\ \emph {et~al.}(2001)\citenamefont {Sauer},
  \citenamefont {Barrett},\ and\ \citenamefont {Chapman}}]{Sauer2001}%
  \BibitemOpen
  \bibfield  {author} {\bibinfo {author} {\bibfnamefont {J.~A.}\ \bibnamefont
  {Sauer}}, \bibinfo {author} {\bibfnamefont {M.~D.}\ \bibnamefont {Barrett}},
  \ and\ \bibinfo {author} {\bibfnamefont {M.~S.}\ \bibnamefont {Chapman}},\
  }\href {\doibase 10.1103/PhysRevLett.87.270401} {\bibfield  {journal}
  {\bibinfo  {journal} {Phys. Rev. Lett.}\ }\textbf {\bibinfo {volume} {87}},\
  \bibinfo {pages} {270401} (\bibinfo {year} {2001})}\BibitemShut {NoStop}%
\bibitem [{\citenamefont {Gupta}\ \emph {et~al.}(2005)\citenamefont {Gupta},
  \citenamefont {Murch}, \citenamefont {Moore}, \citenamefont {Purdy},\ and\
  \citenamefont {Stamper-Kurn}}]{Gupta2005}%
  \BibitemOpen
  \bibfield  {author} {\bibinfo {author} {\bibfnamefont {S.}~\bibnamefont
  {Gupta}}, \bibinfo {author} {\bibfnamefont {K.~W.}\ \bibnamefont {Murch}},
  \bibinfo {author} {\bibfnamefont {K.~L.}\ \bibnamefont {Moore}}, \bibinfo
  {author} {\bibfnamefont {T.~P.}\ \bibnamefont {Purdy}}, \ and\ \bibinfo
  {author} {\bibfnamefont {D.~M.}\ \bibnamefont {Stamper-Kurn}},\ }\href
  {\doibase 10.1103/PhysRevLett.95.143201} {\bibfield  {journal} {\bibinfo
  {journal} {Phys. Rev. Lett.}\ }\textbf {\bibinfo {volume} {95}},\ \bibinfo
  {pages} {143201} (\bibinfo {year} {2005})}\BibitemShut {NoStop}%
\bibitem [{\citenamefont {Arnold}\ \emph {et~al.}(2006)\citenamefont {Arnold},
  \citenamefont {Garvie},\ and\ \citenamefont {Riis}}]{Arnold2006}%
  \BibitemOpen
  \bibfield  {author} {\bibinfo {author} {\bibfnamefont {A.~S.}\ \bibnamefont
  {Arnold}}, \bibinfo {author} {\bibfnamefont {C.~S.}\ \bibnamefont {Garvie}},
  \ and\ \bibinfo {author} {\bibfnamefont {E.}~\bibnamefont {Riis}},\ }\href
  {\doibase 10.1103/PhysRevA.73.041606} {\bibfield  {journal} {\bibinfo
  {journal} {Phys. Rev. A}\ }\textbf {\bibinfo {volume} {73}},\ \bibinfo
  {pages} {041606} (\bibinfo {year} {2006})}\BibitemShut {NoStop}%
\bibitem [{\citenamefont {Henderson}\ \emph {et~al.}(2009)\citenamefont
  {Henderson}, \citenamefont {Ryu}, \citenamefont {MacCormick},\ and\
  \citenamefont {Boshier}}]{Henderson2009}%
  \BibitemOpen
  \bibfield  {author} {\bibinfo {author} {\bibfnamefont {K.}~\bibnamefont
  {Henderson}}, \bibinfo {author} {\bibfnamefont {C.}~\bibnamefont {Ryu}},
  \bibinfo {author} {\bibfnamefont {C.}~\bibnamefont {MacCormick}}, \ and\
  \bibinfo {author} {\bibfnamefont {M.~G.}\ \bibnamefont {Boshier}},\ }\href
  {http://stacks.iop.org/1367-2630/11/i=4/a=043030} {\bibfield  {journal}
  {\bibinfo  {journal} {New J. Phys.}\ }\textbf {\bibinfo {volume} {11}},\
  \bibinfo {pages} {043030} (\bibinfo {year} {2009})}\BibitemShut {NoStop}%
\bibitem [{\citenamefont {Sagu\'e}\ \emph {et~al.}(2007)\citenamefont
  {Sagu\'e}, \citenamefont {Vetsch}, \citenamefont {Alt}, \citenamefont
  {Meschede},\ and\ \citenamefont {Rauschenbeutel}}]{Sagu'e2007}%
  \BibitemOpen
  \bibfield  {author} {\bibinfo {author} {\bibfnamefont {G.}~\bibnamefont
  {Sagu\'e}}, \bibinfo {author} {\bibfnamefont {E.}~\bibnamefont {Vetsch}},
  \bibinfo {author} {\bibfnamefont {W.}~\bibnamefont {Alt}}, \bibinfo {author}
  {\bibfnamefont {D.}~\bibnamefont {Meschede}}, \ and\ \bibinfo {author}
  {\bibfnamefont {A.}~\bibnamefont {Rauschenbeutel}},\ }\href {\doibase
  10.1103/PhysRevLett.99.163602} {\bibfield  {journal} {\bibinfo  {journal}
  {Phys. Rev. Lett.}\ }\textbf {\bibinfo {volume} {99}},\ \bibinfo {pages}
  {163602} (\bibinfo {year} {2007})}\BibitemShut {NoStop}%
\bibitem [{\citenamefont {Sagu\'e}\ \emph {et~al.}(2008)\citenamefont
  {Sagu\'e}, \citenamefont {Baade},\ and\ \citenamefont
  {Rauschenbeutel}}]{Sagu'e2008}%
  \BibitemOpen
  \bibfield  {author} {\bibinfo {author} {\bibfnamefont {G.}~\bibnamefont
  {Sagu\'e}}, \bibinfo {author} {\bibfnamefont {A.}~\bibnamefont {Baade}}, \
  and\ \bibinfo {author} {\bibfnamefont {A.}~\bibnamefont {Rauschenbeutel}},\
  }\href {http://stacks.iop.org/1367-2630/10/i=11/a=113008} {\bibfield
  {journal} {\bibinfo  {journal} {New J. Phys.}\ }\textbf {\bibinfo {volume}
  {10}},\ \bibinfo {pages} {113008} (\bibinfo {year} {2008})}\BibitemShut
  {NoStop}%
\bibitem [{\citenamefont {Reitz}\ and\ \citenamefont
  {Rauschenbeutel}(2012)}]{Reitz2012}%
  \BibitemOpen
  \bibfield  {author} {\bibinfo {author} {\bibfnamefont {D.}~\bibnamefont
  {Reitz}}\ and\ \bibinfo {author} {\bibfnamefont {A.}~\bibnamefont
  {Rauschenbeutel}},\ }\href {\doibase 10.1016/j.optcom.2012.06.034} {\bibfield
   {journal} {\bibinfo  {journal} {Opt. Comm.}\ }\textbf {\bibinfo {volume}
  {285}},\ \bibinfo {pages} {4705 } (\bibinfo {year} {2012})}\BibitemShut
  {NoStop}%
\bibitem [{\citenamefont {Kreutzmann}\ \emph {et~al.}(2004)\citenamefont
  {Kreutzmann}, \citenamefont {Poulsen}, \citenamefont {Lewenstein},
  \citenamefont {Dumke}, \citenamefont {Ertmer}, \citenamefont {Birkl},\ and\
  \citenamefont {Sanpera}}]{Kreutzmann2004}%
  \BibitemOpen
  \bibfield  {author} {\bibinfo {author} {\bibfnamefont {H.}~\bibnamefont
  {Kreutzmann}}, \bibinfo {author} {\bibfnamefont {U.~V.}\ \bibnamefont
  {Poulsen}}, \bibinfo {author} {\bibfnamefont {M.}~\bibnamefont {Lewenstein}},
  \bibinfo {author} {\bibfnamefont {R.}~\bibnamefont {Dumke}}, \bibinfo
  {author} {\bibfnamefont {W.}~\bibnamefont {Ertmer}}, \bibinfo {author}
  {\bibfnamefont {G.}~\bibnamefont {Birkl}}, \ and\ \bibinfo {author}
  {\bibfnamefont {A.}~\bibnamefont {Sanpera}},\ }\href {\doibase
  10.1103/PhysRevLett.92.163201} {\bibfield  {journal} {\bibinfo  {journal}
  {Phys. Rev. Lett.}\ }\textbf {\bibinfo {volume} {92}},\ \bibinfo {pages}
  {163201} (\bibinfo {year} {2004})}\BibitemShut {NoStop}%
\bibitem [{\citenamefont {Cronin}\ \emph {et~al.}(2009)\citenamefont {Cronin},
  \citenamefont {Schmiedmayer},\ and\ \citenamefont {Pritchard}}]{Cronin2009}%
  \BibitemOpen
  \bibfield  {author} {\bibinfo {author} {\bibfnamefont {A.~D.}\ \bibnamefont
  {Cronin}}, \bibinfo {author} {\bibfnamefont {J.}~\bibnamefont
  {Schmiedmayer}}, \ and\ \bibinfo {author} {\bibfnamefont {D.~E.}\
  \bibnamefont {Pritchard}},\ }\href {\doibase 10.1103/RevModPhys.81.1051}
  {\bibfield  {journal} {\bibinfo  {journal} {Rev. Mod. Phys.}\ }\textbf
  {\bibinfo {volume} {81}},\ \bibinfo {pages} {1051} (\bibinfo {year}
  {2009})}\BibitemShut {NoStop}%
\bibitem [{\citenamefont {Schmiedmayer}\ \emph {et~al.}(2002)\citenamefont
  {Schmiedmayer}, \citenamefont {Folman},\ and\ \citenamefont
  {Calarco}}]{Schmiedmayer2002}%
  \BibitemOpen
  \bibfield  {author} {\bibinfo {author} {\bibfnamefont {J.}~\bibnamefont
  {Schmiedmayer}}, \bibinfo {author} {\bibfnamefont {R.}~\bibnamefont
  {Folman}}, \ and\ \bibinfo {author} {\bibfnamefont {T.}~\bibnamefont
  {Calarco}},\ }\href {\doibase 10.1080/09500340110111077} {\bibfield
  {journal} {\bibinfo  {journal} {J. Mod. Opt.}\ }\textbf {\bibinfo {volume}
  {49}},\ \bibinfo {pages} {1375} (\bibinfo {year} {2002})}\BibitemShut
  {NoStop}%
\bibitem [{\citenamefont {Wright}\ \emph {et~al.}(2013)\citenamefont {Wright},
  \citenamefont {Blakestad}, \citenamefont {Lobb}, \citenamefont {Phillips},\
  and\ \citenamefont {Campbell}}]{Wright2013}%
  \BibitemOpen
  \bibfield  {author} {\bibinfo {author} {\bibfnamefont {K.~C.}\ \bibnamefont
  {Wright}}, \bibinfo {author} {\bibfnamefont {R.~B.}\ \bibnamefont
  {Blakestad}}, \bibinfo {author} {\bibfnamefont {C.~J.}\ \bibnamefont {Lobb}},
  \bibinfo {author} {\bibfnamefont {W.~D.}\ \bibnamefont {Phillips}}, \ and\
  \bibinfo {author} {\bibfnamefont {G.~K.}\ \bibnamefont {Campbell}},\ }\href
  {\doibase 10.1103/PhysRevLett.110.025302} {\bibfield  {journal} {\bibinfo
  {journal} {Phys. Rev. Lett.}\ }\textbf {\bibinfo {volume} {110}},\ \bibinfo
  {pages} {025302} (\bibinfo {year} {2013})}\BibitemShut {NoStop}%
\bibitem [{\citenamefont {Leboeuf}\ and\ \citenamefont
  {Pavloff}(2001)}]{Leboeuf2001}%
  \BibitemOpen
  \bibfield  {author} {\bibinfo {author} {\bibfnamefont {P.}~\bibnamefont
  {Leboeuf}}\ and\ \bibinfo {author} {\bibfnamefont {N.}~\bibnamefont
  {Pavloff}},\ }\href {\doibase 10.1103/PhysRevA.64.033602} {\bibfield
  {journal} {\bibinfo  {journal} {Phys. Rev. A}\ }\textbf {\bibinfo {volume}
  {64}},\ \bibinfo {pages} {033602} (\bibinfo {year} {2001})}\BibitemShut
  {NoStop}%
\bibitem [{\citenamefont {J\"a\"askel\"ainen}\ and\ \citenamefont
  {Stenholm}(2002{\natexlab{a}})}]{Jaaskelainen2002a}%
  \BibitemOpen
  \bibfield  {author} {\bibinfo {author} {\bibfnamefont {M.}~\bibnamefont
  {J\"a\"askel\"ainen}}\ and\ \bibinfo {author} {\bibfnamefont
  {S.}~\bibnamefont {Stenholm}},\ }\href {\doibase 10.1103/PhysRevA.66.023608}
  {\bibfield  {journal} {\bibinfo  {journal} {Phys. Rev. A}\ }\textbf {\bibinfo
  {volume} {66}},\ \bibinfo {pages} {023608} (\bibinfo {year}
  {2002}{\natexlab{a}})}\BibitemShut {NoStop}%
\bibitem [{\citenamefont {Bromley}\ and\ \citenamefont
  {Esry}(2003)}]{Bromley2003}%
  \BibitemOpen
  \bibfield  {author} {\bibinfo {author} {\bibfnamefont {M.~W.~J.}\
  \bibnamefont {Bromley}}\ and\ \bibinfo {author} {\bibfnamefont {B.~D.}\
  \bibnamefont {Esry}},\ }\href {\doibase 10.1103/PhysRevA.68.043609}
  {\bibfield  {journal} {\bibinfo  {journal} {Phys. Rev. A}\ }\textbf {\bibinfo
  {volume} {68}},\ \bibinfo {pages} {043609} (\bibinfo {year}
  {2003})}\BibitemShut {NoStop}%
\bibitem [{\citenamefont {Koehler}\ \emph {et~al.}(2005)\citenamefont
  {Koehler}, \citenamefont {Bromley},\ and\ \citenamefont
  {Esry}}]{Koehler2005}%
  \BibitemOpen
  \bibfield  {author} {\bibinfo {author} {\bibfnamefont {M.}~\bibnamefont
  {Koehler}}, \bibinfo {author} {\bibfnamefont {M.~W.~J.}\ \bibnamefont
  {Bromley}}, \ and\ \bibinfo {author} {\bibfnamefont {B.~D.}\ \bibnamefont
  {Esry}},\ }\href {\doibase 10.1103/PhysRevA.72.023603} {\bibfield  {journal}
  {\bibinfo  {journal} {Phys. Rev. A}\ }\textbf {\bibinfo {volume} {72}},\
  \bibinfo {pages} {023603} (\bibinfo {year} {2005})}\BibitemShut {NoStop}%
\bibitem [{\citenamefont {Bromley}\ and\ \citenamefont
  {Esry}(2004)}]{Bromley2004a}%
  \BibitemOpen
  \bibfield  {author} {\bibinfo {author} {\bibfnamefont {M.~W.~J.}\
  \bibnamefont {Bromley}}\ and\ \bibinfo {author} {\bibfnamefont {B.~D.}\
  \bibnamefont {Esry}},\ }\href {\doibase 10.1103/PhysRevA.69.053620}
  {\bibfield  {journal} {\bibinfo  {journal} {Phys. Rev. A}\ }\textbf {\bibinfo
  {volume} {69}},\ \bibinfo {pages} {053620} (\bibinfo {year}
  {2004})}\BibitemShut {NoStop}%
\bibitem [{\citenamefont {Gaididei}\ \emph {et~al.}(2005)\citenamefont
  {Gaididei}, \citenamefont {Christiansen}, \citenamefont {Kevrekidis},
  \citenamefont {Büttner},\ and\ \citenamefont {Bishop}}]{Gaididei2005}%
  \BibitemOpen
  \bibfield  {author} {\bibinfo {author} {\bibfnamefont {Y.~B.}\ \bibnamefont
  {Gaididei}}, \bibinfo {author} {\bibfnamefont {P.~L.}\ \bibnamefont
  {Christiansen}}, \bibinfo {author} {\bibfnamefont {P.~G.}\ \bibnamefont
  {Kevrekidis}}, \bibinfo {author} {\bibfnamefont {H.}~\bibnamefont
  {Büttner}}, \ and\ \bibinfo {author} {\bibfnamefont {A.~R.}\ \bibnamefont
  {Bishop}},\ }\href {http://stacks.iop.org/1367-2630/7/i=1/a=052} {\bibfield
  {journal} {\bibinfo  {journal} {New J. Phys.}\ }\textbf {\bibinfo {volume}
  {7}},\ \bibinfo {pages} {52} (\bibinfo {year} {2005})}\BibitemShut {NoStop}%
\bibitem [{\citenamefont {Schwartz}\ \emph {et~al.}(2006)\citenamefont
  {Schwartz}, \citenamefont {Cozzini}, \citenamefont {Menotti}, \citenamefont
  {Carusotto}, \citenamefont {Bouyer},\ and\ \citenamefont
  {Stringari}}]{Schwartz2006}%
  \BibitemOpen
  \bibfield  {author} {\bibinfo {author} {\bibfnamefont {S.}~\bibnamefont
  {Schwartz}}, \bibinfo {author} {\bibfnamefont {M.}~\bibnamefont {Cozzini}},
  \bibinfo {author} {\bibfnamefont {C.}~\bibnamefont {Menotti}}, \bibinfo
  {author} {\bibfnamefont {I.}~\bibnamefont {Carusotto}}, \bibinfo {author}
  {\bibfnamefont {P.}~\bibnamefont {Bouyer}}, \ and\ \bibinfo {author}
  {\bibfnamefont {S.}~\bibnamefont {Stringari}},\ }\href
  {http://stacks.iop.org/1367-2630/8/i=8/a=162} {\bibfield  {journal} {\bibinfo
   {journal} {New J. Phys.}\ }\textbf {\bibinfo {volume} {8}},\ \bibinfo
  {pages} {162} (\bibinfo {year} {2006})}\BibitemShut {NoStop}%
\bibitem [{\citenamefont {Paul}\ \emph {et~al.}(2007)\citenamefont {Paul},
  \citenamefont {Hartung}, \citenamefont {Richter},\ and\ \citenamefont
  {Schlagheck}}]{Paul2007}%
  \BibitemOpen
  \bibfield  {author} {\bibinfo {author} {\bibfnamefont {T.}~\bibnamefont
  {Paul}}, \bibinfo {author} {\bibfnamefont {M.}~\bibnamefont {Hartung}},
  \bibinfo {author} {\bibfnamefont {K.}~\bibnamefont {Richter}}, \ and\
  \bibinfo {author} {\bibfnamefont {P.}~\bibnamefont {Schlagheck}},\ }\href
  {\doibase 10.1103/PhysRevA.76.063605} {\bibfield  {journal} {\bibinfo
  {journal} {Phys. Rev. A}\ }\textbf {\bibinfo {volume} {76}},\ \bibinfo
  {pages} {063605} (\bibinfo {year} {2007})}\BibitemShut {NoStop}%
\bibitem [{\citenamefont {Gattobigio}\ \emph {et~al.}(2010)\citenamefont
  {Gattobigio}, \citenamefont {Couvert}, \citenamefont {Georgeot},\ and\
  \citenamefont {Guéry-Odelin}}]{Gattobigio2010}%
  \BibitemOpen
  \bibfield  {author} {\bibinfo {author} {\bibfnamefont {G.~L.}\ \bibnamefont
  {Gattobigio}}, \bibinfo {author} {\bibfnamefont {A.}~\bibnamefont {Couvert}},
  \bibinfo {author} {\bibfnamefont {B.}~\bibnamefont {Georgeot}}, \ and\
  \bibinfo {author} {\bibfnamefont {D.}~\bibnamefont {Guéry-Odelin}},\ }\href
  {http://stacks.iop.org/1367-2630/12/i=8/a=085013} {\bibfield  {journal}
  {\bibinfo  {journal} {New J. Phys.}\ }\textbf {\bibinfo {volume} {12}},\
  \bibinfo {pages} {085013} (\bibinfo {year} {2010})}\BibitemShut {NoStop}%
\bibitem [{\citenamefont {Ernst}\ \emph {et~al.}(2010)\citenamefont {Ernst},
  \citenamefont {Paul},\ and\ \citenamefont {Schlagheck}}]{Ernst2010}%
  \BibitemOpen
  \bibfield  {author} {\bibinfo {author} {\bibfnamefont {T.}~\bibnamefont
  {Ernst}}, \bibinfo {author} {\bibfnamefont {T.}~\bibnamefont {Paul}}, \ and\
  \bibinfo {author} {\bibfnamefont {P.}~\bibnamefont {Schlagheck}},\ }\href
  {\doibase 10.1103/PhysRevA.81.013631} {\bibfield  {journal} {\bibinfo
  {journal} {Phys. Rev. A}\ }\textbf {\bibinfo {volume} {81}},\ \bibinfo
  {pages} {013631} (\bibinfo {year} {2010})}\BibitemShut {NoStop}%
\bibitem [{\citenamefont {Tacla}\ and\ \citenamefont
  {Caves}(2011)}]{Tacla2011}%
  \BibitemOpen
  \bibfield  {author} {\bibinfo {author} {\bibfnamefont {A.~B.}\ \bibnamefont
  {Tacla}}\ and\ \bibinfo {author} {\bibfnamefont {C.~M.}\ \bibnamefont
  {Caves}},\ }\href {\doibase 10.1103/PhysRevA.84.053606} {\bibfield  {journal}
  {\bibinfo  {journal} {Phys. Rev. A}\ }\textbf {\bibinfo {volume} {84}},\
  \bibinfo {pages} {053606} (\bibinfo {year} {2011})}\BibitemShut {NoStop}%
\bibitem [{\citenamefont {Conti}()}]{Conti2013}%
  \BibitemOpen
  \bibfield  {author} {\bibinfo {author} {\bibfnamefont {C.}~\bibnamefont
  {Conti}},\ }\href {http://arxiv.org/abs/1302.3806} {}\Eprint
  {http://arxiv.org/abs/1302.3806} {arXiv:1302.3806} \BibitemShut {NoStop}%
\bibitem [{\citenamefont {del Campo}\ \emph {et~al.}()\citenamefont {del
  Campo}, \citenamefont {Boshier},\ and\ \citenamefont {Saxena}}]{Campo2013}%
  \BibitemOpen
  \bibfield  {author} {\bibinfo {author} {\bibfnamefont {A.}~\bibnamefont {del
  Campo}}, \bibinfo {author} {\bibfnamefont {M.~G.}\ \bibnamefont {Boshier}}, \
  and\ \bibinfo {author} {\bibfnamefont {A.}~\bibnamefont {Saxena}},\ }\href
  {http://arxiv.org/abs/1311.2062} {}\Eprint {http://arxiv.org/abs/1311.2062}
  {arXiv:1311.2062} \BibitemShut {NoStop}%
\bibitem [{\citenamefont {Loiko}\ \emph {et~al.}(2011)\citenamefont {Loiko},
  \citenamefont {Ahufinger}, \citenamefont {Corbal\'an}, \citenamefont
  {Birkl},\ and\ \citenamefont {Mompart}}]{Loiko2011}%
  \BibitemOpen
  \bibfield  {author} {\bibinfo {author} {\bibfnamefont {Y.}~\bibnamefont
  {Loiko}}, \bibinfo {author} {\bibfnamefont {V.}~\bibnamefont {Ahufinger}},
  \bibinfo {author} {\bibfnamefont {R.}~\bibnamefont {Corbal\'an}}, \bibinfo
  {author} {\bibfnamefont {G.}~\bibnamefont {Birkl}}, \ and\ \bibinfo {author}
  {\bibfnamefont {J.}~\bibnamefont {Mompart}},\ }\href {\doibase
  10.1103/PhysRevA.83.033629} {\bibfield  {journal} {\bibinfo  {journal} {Phys.
  Rev. A}\ }\textbf {\bibinfo {volume} {83}},\ \bibinfo {pages} {033629}
  (\bibinfo {year} {2011})}\BibitemShut {NoStop}%
\bibitem [{\citenamefont {Bücker}\ \emph {et~al.}(2013)\citenamefont
  {Bücker}, \citenamefont {Berrada}, \citenamefont {van Frank}, \citenamefont
  {Schaff}, \citenamefont {Schumm}, \citenamefont {Schmiedmayer}, \citenamefont
  {Jäger}, \citenamefont {Grond},\ and\ \citenamefont
  {Hohenester}}]{Buecker2013}%
  \BibitemOpen
  \bibfield  {author} {\bibinfo {author} {\bibfnamefont {R.}~\bibnamefont
  {Bücker}}, \bibinfo {author} {\bibfnamefont {T.}~\bibnamefont {Berrada}},
  \bibinfo {author} {\bibfnamefont {S.}~\bibnamefont {van Frank}}, \bibinfo
  {author} {\bibfnamefont {J.-F.}\ \bibnamefont {Schaff}}, \bibinfo {author}
  {\bibfnamefont {T.}~\bibnamefont {Schumm}}, \bibinfo {author} {\bibfnamefont
  {J.}~\bibnamefont {Schmiedmayer}}, \bibinfo {author} {\bibfnamefont
  {G.}~\bibnamefont {Jäger}}, \bibinfo {author} {\bibfnamefont
  {J.}~\bibnamefont {Grond}}, \ and\ \bibinfo {author} {\bibfnamefont
  {U.}~\bibnamefont {Hohenester}},\ }\href
  {http://stacks.iop.org/0953-4075/46/i=10/a=104012} {\bibfield  {journal}
  {\bibinfo  {journal} {J. Phys. B: At. Mol. Opt. Phys.}\ }\textbf {\bibinfo
  {volume} {46}},\ \bibinfo {pages} {104012} (\bibinfo {year}
  {2013})}\BibitemShut {NoStop}%
\bibitem [{\citenamefont {Mart\'{i}nez-Garaot}\ \emph
  {et~al.}(2013)\citenamefont {Mart\'{i}nez-Garaot}, \citenamefont
  {Torrontegui}, \citenamefont {Chen}, \citenamefont {Modugno}, \citenamefont
  {Guéry-Odelin}, \citenamefont {Tseng},\ and\ \citenamefont
  {Muga}}]{Martinez-Garaot2013}%
  \BibitemOpen
  \bibfield  {author} {\bibinfo {author} {\bibfnamefont {S.}~\bibnamefont
  {Mart\'{i}nez-Garaot}}, \bibinfo {author} {\bibfnamefont {E.}~\bibnamefont
  {Torrontegui}}, \bibinfo {author} {\bibfnamefont {X.}~\bibnamefont {Chen}},
  \bibinfo {author} {\bibfnamefont {M.}~\bibnamefont {Modugno}}, \bibinfo
  {author} {\bibfnamefont {D.}~\bibnamefont {Guéry-Odelin}}, \bibinfo {author}
  {\bibfnamefont {S.-Y.}\ \bibnamefont {Tseng}}, \ and\ \bibinfo {author}
  {\bibfnamefont {J.~G.}\ \bibnamefont {Muga}},\ }\href {\doibase
  10.1103/PhysRevLett.111.213001} {\bibfield  {journal} {\bibinfo  {journal}
  {Phys. Rev. Lett.}\ }\textbf {\bibinfo {volume} {111}},\ \bibinfo {pages}
  {213001} (\bibinfo {year} {2013})}\BibitemShut {NoStop}%
\bibitem [{\citenamefont {Kaplan}\ \emph {et~al.}(1997)\citenamefont {Kaplan},
  \citenamefont {Maitra},\ and\ \citenamefont {Heller}}]{Kaplan1997}%
  \BibitemOpen
  \bibfield  {author} {\bibinfo {author} {\bibfnamefont {L.}~\bibnamefont
  {Kaplan}}, \bibinfo {author} {\bibfnamefont {N.~T.}\ \bibnamefont {Maitra}},
  \ and\ \bibinfo {author} {\bibfnamefont {E.~J.}\ \bibnamefont {Heller}},\
  }\href {\doibase 10.1103/PhysRevA.56.2592} {\bibfield  {journal} {\bibinfo
  {journal} {Phys. Rev. A}\ }\textbf {\bibinfo {volume} {56}},\ \bibinfo
  {pages} {2592} (\bibinfo {year} {1997})}\BibitemShut {NoStop}%
\bibitem [{\citenamefont {da~Costa}(1981)}]{Costa1981}%
  \BibitemOpen
  \bibfield  {author} {\bibinfo {author} {\bibfnamefont {R.~C.~T.}\
  \bibnamefont {da~Costa}},\ }\href {\doibase 10.1103/PhysRevA.23.1982}
  {\bibfield  {journal} {\bibinfo  {journal} {Phys. Rev. A}\ }\textbf {\bibinfo
  {volume} {23}},\ \bibinfo {pages} {1982} (\bibinfo {year}
  {1981})}\BibitemShut {NoStop}%
\bibitem [{\citenamefont {da~Costa}(1982)}]{Costa1982}%
  \BibitemOpen
  \bibfield  {author} {\bibinfo {author} {\bibfnamefont {R.~C.~T.}\
  \bibnamefont {da~Costa}},\ }\href {\doibase 10.1103/PhysRevA.25.2893}
  {\bibfield  {journal} {\bibinfo  {journal} {Phys. Rev. A}\ }\textbf {\bibinfo
  {volume} {25}},\ \bibinfo {pages} {2893} (\bibinfo {year}
  {1982})}\BibitemShut {NoStop}%
\bibitem [{\citenamefont {Duclos}\ and\ \citenamefont
  {Exner}(1995)}]{Duclos1995}%
  \BibitemOpen
  \bibfield  {author} {\bibinfo {author} {\bibfnamefont {P.}~\bibnamefont
  {Duclos}}\ and\ \bibinfo {author} {\bibfnamefont {P.}~\bibnamefont {Exner}},\
  }\href {\doibase 10.1142/S0129055X95000062} {\bibfield  {journal} {\bibinfo
  {journal} {Rev. Math. Phys.}\ }\textbf {\bibinfo {volume} {7}},\ \bibinfo
  {pages} {73} (\bibinfo {year} {1995})}\BibitemShut {NoStop}%
\bibitem [{\citenamefont {Hurt}(2000)}]{Hurt2000}%
  \BibitemOpen
  \bibfield  {author} {\bibinfo {author} {\bibfnamefont {N.~E.}\ \bibnamefont
  {Hurt}},\ }\href@noop {} {\emph {\bibinfo {title} {Mathematical Physics of
  Quantum Wires and Devices: From Spectral Resonances to Anderson
  Localization}}}\ (\bibinfo  {publisher} {Kluwer Academic Publishers},\
  \bibinfo {year} {2000})\BibitemShut {NoStop}%
\bibitem [{\citenamefont {Exner}\ \emph {et~al.}(2008)\citenamefont {Exner},
  \citenamefont {Keating}, \citenamefont {Kuchment}, \citenamefont {Sunada},\
  and\ \citenamefont {Teplyaev}}]{Exner2008}%
  \BibitemOpen
  \bibinfo {editor} {\bibfnamefont {P.}~\bibnamefont {Exner}}, \bibinfo
  {editor} {\bibfnamefont {J.~P.}\ \bibnamefont {Keating}}, \bibinfo {editor}
  {\bibfnamefont {P.}~\bibnamefont {Kuchment}}, \bibinfo {editor}
  {\bibfnamefont {T.}~\bibnamefont {Sunada}}, \ and\ \bibinfo {editor}
  {\bibfnamefont {A.}~\bibnamefont {Teplyaev}},\ eds.,\ \href@noop {} {\emph
  {\bibinfo {title} {Analysis on Graphs and Its Applications}}},\ \bibinfo
  {series} {Proceedings of Symposia in Pure Mathematics}, Vol.~\bibinfo
  {volume} {77}\ (\bibinfo  {publisher} {American Mathematical Society},\
  \bibinfo {year} {2008})\BibitemShut {NoStop}%
\bibitem [{\citenamefont {Krejčiřík}\ and\ \citenamefont
  {Šediváková}(2012)}]{Krejcirik2012}%
  \BibitemOpen
  \bibfield  {author} {\bibinfo {author} {\bibfnamefont {D.}~\bibnamefont
  {Krejčiřík}}\ and\ \bibinfo {author} {\bibfnamefont {H.}~\bibnamefont
  {Šediváková}},\ }\href {\doibase 10.1142/S0129055X12500183} {\bibfield
  {journal} {\bibinfo  {journal} {Rev. Math. Phys.}\ }\textbf {\bibinfo
  {volume} {24}},\ \bibinfo {pages} {1250018} (\bibinfo {year}
  {2012})}\BibitemShut {NoStop}%
\bibitem [{\citenamefont {Bulla}\ \emph {et~al.}(1997)\citenamefont {Bulla},
  \citenamefont {Gesztesy}, \citenamefont {Renger},\ and\ \citenamefont
  {Simon}}]{Bulla1997}%
  \BibitemOpen
  \bibfield  {author} {\bibinfo {author} {\bibfnamefont {W.}~\bibnamefont
  {Bulla}}, \bibinfo {author} {\bibfnamefont {F.}~\bibnamefont {Gesztesy}},
  \bibinfo {author} {\bibfnamefont {W.}~\bibnamefont {Renger}}, \ and\ \bibinfo
  {author} {\bibfnamefont {B.}~\bibnamefont {Simon}},\ }\href
  {http://www.jstor.org/stable/2162096} {\bibfield  {journal} {\bibinfo
  {journal} {Proc. Am. Math. Soc}\ }\textbf {\bibinfo {volume} {125}},\
  \bibinfo {pages} {1487} (\bibinfo {year} {1997})}\BibitemShut {NoStop}%
\bibitem [{\citenamefont {Borisov}\ \emph {et~al.}(2001)\citenamefont
  {Borisov}, \citenamefont {Exner}, \citenamefont {Gadyl’shin},\ and\
  \citenamefont {Krejčiřík}}]{Borisov2001}%
  \BibitemOpen
  \bibfield  {author} {\bibinfo {author} {\bibfnamefont {D.}~\bibnamefont
  {Borisov}}, \bibinfo {author} {\bibfnamefont {P.}~\bibnamefont {Exner}},
  \bibinfo {author} {\bibfnamefont {R.}~\bibnamefont {Gadyl’shin}}, \ and\
  \bibinfo {author} {\bibfnamefont {D.}~\bibnamefont {Krejčiřík}},\ }\href
  {http://link.springer.com/article/10.1007%2FPL00001045} {\bibfield  {journal}
  {\bibinfo  {journal} {Ann. Henri Poincaré}\ }\textbf {\bibinfo {volume}
  {2}},\ \bibinfo {pages} {553 } (\bibinfo {year} {2001})}\BibitemShut
  {NoStop}%
\bibitem [{\citenamefont {Grushin}(2008)}]{Grushin2008}%
  \BibitemOpen
  \bibfield  {author} {\bibinfo {author} {\bibfnamefont {V.}~\bibnamefont
  {Grushin}},\ }\href {\doibase 10.1134/S000143460803019X} {\bibfield
  {journal} {\bibinfo  {journal} {Math. Notes}\ }\textbf {\bibinfo {volume}
  {83}},\ \bibinfo {pages} {463} (\bibinfo {year} {2008})}\BibitemShut
  {NoStop}%
\bibitem [{\citenamefont {Grushin}(2009)}]{Grushin2009}%
  \BibitemOpen
  \bibfield  {author} {\bibinfo {author} {\bibfnamefont {V.}~\bibnamefont
  {Grushin}},\ }\href {\doibase 10.1134/S000143460905006X} {\bibfield
  {journal} {\bibinfo  {journal} {Math. Notes}\ }\textbf {\bibinfo {volume}
  {85}},\ \bibinfo {pages} {661} (\bibinfo {year} {2009})}\BibitemShut
  {NoStop}%
\bibitem [{\citenamefont {Nakazato}\ and\ \citenamefont
  {Blaikie}(1991)}]{Nakazato1991}%
  \BibitemOpen
  \bibfield  {author} {\bibinfo {author} {\bibfnamefont {K.}~\bibnamefont
  {Nakazato}}\ and\ \bibinfo {author} {\bibfnamefont {R.~J.}\ \bibnamefont
  {Blaikie}},\ }\href {http://stacks.iop.org/0953-8984/3/i=30/a=006} {\bibfield
   {journal} {\bibinfo  {journal} {J. Phys.: Condens. Matter}\ }\textbf
  {\bibinfo {volume} {3}},\ \bibinfo {pages} {5729} (\bibinfo {year}
  {1991})}\BibitemShut {NoStop}%
\bibitem [{\citenamefont {J\"a\"askel\"ainen}\ and\ \citenamefont
  {Stenholm}(2002{\natexlab{b}})}]{Jaaskelainen2002}%
  \BibitemOpen
  \bibfield  {author} {\bibinfo {author} {\bibfnamefont {M.}~\bibnamefont
  {J\"a\"askel\"ainen}}\ and\ \bibinfo {author} {\bibfnamefont
  {S.}~\bibnamefont {Stenholm}},\ }\href {\doibase 10.1103/PhysRevA.66.053605}
  {\bibfield  {journal} {\bibinfo  {journal} {Phys. Rev. A}\ }\textbf {\bibinfo
  {volume} {66}},\ \bibinfo {pages} {053605} (\bibinfo {year}
  {2002}{\natexlab{b}})}\BibitemShut {NoStop}%
\bibitem [{\citenamefont {Bæk}\ and\ \citenamefont
  {Willatzen}(2008)}]{Baek2008}%
  \BibitemOpen
  \bibfield  {author} {\bibinfo {author} {\bibfnamefont {D.}~\bibnamefont
  {Bæk}}\ and\ \bibinfo {author} {\bibfnamefont {M.}~\bibnamefont
  {Willatzen}},\ }\href {\doibase doi:10.3813/AAA.918099} {\bibfield  {journal}
  {\bibinfo  {journal} {Acta Acust. United Ac.}\ }\textbf {\bibinfo {volume}
  {94}},\ \bibinfo {pages} {668} (\bibinfo {year} {2008})}\BibitemShut
  {NoStop}%
\bibitem [{\citenamefont {Gravesen}\ and\ \citenamefont
  {Willatzen}(2008)}]{Gravesen2008}%
  \BibitemOpen
  \bibfield  {author} {\bibinfo {author} {\bibfnamefont {J.}~\bibnamefont
  {Gravesen}}\ and\ \bibinfo {author} {\bibfnamefont {M.}~\bibnamefont
  {Willatzen}},\ }\href {\doibase 10.1016/j.spmi.2007.06.020} {\bibfield
  {journal} {\bibinfo  {journal} {Superlattice Microstruct.}\ }\textbf
  {\bibinfo {volume} {43}},\ \bibinfo {pages} {441 } (\bibinfo {year}
  {2008})}\BibitemShut {NoStop}%
\bibitem [{\citenamefont {Born}\ and\ \citenamefont
  {Oppenheimer}(1927)}]{Born1927}%
  \BibitemOpen
  \bibfield  {author} {\bibinfo {author} {\bibfnamefont {M.}~\bibnamefont
  {Born}}\ and\ \bibinfo {author} {\bibfnamefont {R.}~\bibnamefont
  {Oppenheimer}},\ }\href {\doibase 10.1002/andp.19273892002} {\bibfield
  {journal} {\bibinfo  {journal} {Ann. Phys.}\ }\textbf {\bibinfo {volume}
  {389}},\ \bibinfo {pages} {457} (\bibinfo {year} {1927})}\BibitemShut
  {NoStop}%
\bibitem [{\citenamefont {Born}\ and\ \citenamefont {Huang}(1954)}]{Born1954}%
  \BibitemOpen
  \bibfield  {author} {\bibinfo {author} {\bibfnamefont {M.}~\bibnamefont
  {Born}}\ and\ \bibinfo {author} {\bibfnamefont {K.}~\bibnamefont {Huang}},\
  }\href@noop {} {\emph {\bibinfo {title} {Dynamical theory of crystal
  lattices}}}\ (\bibinfo  {publisher} {Oxford University Press},\ \bibinfo
  {year} {1954})\BibitemShut {NoStop}%
\bibitem [{\citenamefont {Domcke}\ \emph {et~al.}(2004)\citenamefont {Domcke},
  \citenamefont {Yarkony},\ and\ \citenamefont {Köppel}}]{Domcke2004}%
  \BibitemOpen
  \bibinfo {editor} {\bibfnamefont {W.}~\bibnamefont {Domcke}}, \bibinfo
  {editor} {\bibfnamefont {D.~R.}\ \bibnamefont {Yarkony}}, \ and\ \bibinfo
  {editor} {\bibfnamefont {H.}~\bibnamefont {Köppel}},\ eds.,\ \href@noop {}
  {\emph {\bibinfo {title} {Conical Intersections: Electronic Structure,
  Dynamics \& Spectroscopy}}}\ (\bibinfo  {publisher} {World Scientific,
  Singapore},\ \bibinfo {year} {2004})\BibitemShut {NoStop}%
\bibitem [{\citenamefont {Belov}\ \emph {et~al.}(2006)\citenamefont {Belov},
  \citenamefont {Dobrokhotov},\ and\ \citenamefont {Tudorovskiy}}]{Belov2006}%
  \BibitemOpen
  \bibfield  {author} {\bibinfo {author} {\bibfnamefont {V.}~\bibnamefont
  {Belov}}, \bibinfo {author} {\bibfnamefont {S.}~\bibnamefont {Dobrokhotov}},
  \ and\ \bibinfo {author} {\bibfnamefont {T.}~\bibnamefont {Tudorovskiy}},\
  }\href {http://dx.doi.org/10.1007/s10665-006-9044-3} {\bibfield  {journal}
  {\bibinfo  {journal} {J. Eng. Math.}\ }\textbf {\bibinfo {volume} {55}},\
  \bibinfo {pages} {183} (\bibinfo {year} {2006})}\BibitemShut {NoStop}%
\bibitem [{\citenamefont {Jecko}()}]{Jecko2013}%
  \BibitemOpen
  \bibfield  {author} {\bibinfo {author} {\bibfnamefont {T.}~\bibnamefont
  {Jecko}},\ }\href {http://arxiv.org/abs/1303.5833} {}\Eprint
  {http://arxiv.org/abs/1303.5833} {arXiv:1303.5833} \BibitemShut {NoStop}%
\bibitem [{\citenamefont {Teufel}(2003)}]{Teufel2003}%
  \BibitemOpen
  \bibfield  {author} {\bibinfo {author} {\bibfnamefont {S.}~\bibnamefont
  {Teufel}},\ }\href@noop {} {\emph {\bibinfo {title} {Adiabatic Perturbation
  Theory in Quantum Dynamics}}},\ \bibinfo {series} {Lecture Notes in
  Mathematics}, Vol.\ \bibinfo {volume} {1821}\ (\bibinfo  {publisher}
  {Springer, Berlin},\ \bibinfo {year} {2003})\BibitemShut {NoStop}%
\bibitem [{\citenamefont {Panati}\ \emph {et~al.}(2003)\citenamefont {Panati},
  \citenamefont {Spohn},\ and\ \citenamefont {Teufel}}]{Panati2003}%
  \BibitemOpen
  \bibfield  {author} {\bibinfo {author} {\bibfnamefont {G.}~\bibnamefont
  {Panati}}, \bibinfo {author} {\bibfnamefont {H.}~\bibnamefont {Spohn}}, \
  and\ \bibinfo {author} {\bibfnamefont {S.}~\bibnamefont {Teufel}},\ }\href
  {http://dx.doi.org/10.4310/ATMP.2003.v7.n1.a6} {\bibfield  {journal}
  {\bibinfo  {journal} {Adv. Theor. Math. Phys.}\ }\textbf {\bibinfo {volume}
  {7}},\ \bibinfo {pages} {145} (\bibinfo {year} {2003})}\BibitemShut {NoStop}%
\bibitem [{\citenamefont {Panati}\ \emph {et~al.}(2007)\citenamefont {Panati},
  \citenamefont {Spohn},\ and\ \citenamefont {Teufel}}]{Panati2007}%
  \BibitemOpen
  \bibfield  {author} {\bibinfo {author} {\bibfnamefont {G.}~\bibnamefont
  {Panati}}, \bibinfo {author} {\bibfnamefont {H.}~\bibnamefont {Spohn}}, \
  and\ \bibinfo {author} {\bibfnamefont {S.}~\bibnamefont {Teufel}},\ }\href
  {\doibase 10.1051/m2an:2007023} {\bibfield  {journal} {\bibinfo  {journal}
  {ESAIM Math. Model. Num.}\ }\textbf {\bibinfo {volume} {41}},\ \bibinfo
  {pages} {297} (\bibinfo {year} {2007})}\BibitemShut {NoStop}%
\bibitem [{\citenamefont {Wachsmuth}\ and\ \citenamefont
  {Teufel}(2010)}]{Wachsmuth2010}%
  \BibitemOpen
  \bibfield  {author} {\bibinfo {author} {\bibfnamefont {J.}~\bibnamefont
  {Wachsmuth}}\ and\ \bibinfo {author} {\bibfnamefont {S.}~\bibnamefont
  {Teufel}},\ }\href {\doibase 10.1103/PhysRevA.82.022112} {\bibfield
  {journal} {\bibinfo  {journal} {Phys. Rev. A}\ }\textbf {\bibinfo {volume}
  {82}},\ \bibinfo {pages} {022112} (\bibinfo {year} {2010})}\BibitemShut
  {NoStop}%
\bibitem [{\citenamefont {Wachsmuth}\ and\ \citenamefont
  {Teufel}(2013)}]{Wachsmuth2013}%
  \BibitemOpen
  \bibfield  {author} {\bibinfo {author} {\bibfnamefont {J.}~\bibnamefont
  {Wachsmuth}}\ and\ \bibinfo {author} {\bibfnamefont {S.}~\bibnamefont
  {Teufel}},\ }\href {http://dx.doi.org/10.1090/memo/1083} {\bibfield
  {journal} {\bibinfo  {journal} {Memoirs of the American Mathematical
  Society}\ }\textbf {\bibinfo {volume} {230}},\ \bibinfo {pages} {1083}
  (\bibinfo {year} {2013})}\BibitemShut {NoStop}%
\bibitem [{\citenamefont {Haag}\ \emph {et~al.}(2014)\citenamefont {Haag},
  \citenamefont {Lampart},\ and\ \citenamefont {Teufel}}]{Haag2014}%
  \BibitemOpen
  \bibfield  {author} {\bibinfo {author} {\bibfnamefont {S.}~\bibnamefont
  {Haag}}, \bibinfo {author} {\bibfnamefont {J.}~\bibnamefont {Lampart}}, \
  and\ \bibinfo {author} {\bibfnamefont {S.}~\bibnamefont {Teufel}},\ }\href
  {http://arxiv.org/abs/1402.1067} {} (\bibinfo {year} {2014}),\ \Eprint
  {http://arxiv.org/abs/1402.1067} {arXiv:1402.1067} \BibitemShut {NoStop}%
\bibitem [{\citenamefont {J\"a\"askel\"ainen}\ and\ \citenamefont
  {Stenholm}(2002{\natexlab{c}})}]{Jaaskelainen2002b}%
  \BibitemOpen
  \bibfield  {author} {\bibinfo {author} {\bibfnamefont {M.}~\bibnamefont
  {J\"a\"askel\"ainen}}\ and\ \bibinfo {author} {\bibfnamefont
  {S.}~\bibnamefont {Stenholm}},\ }\href {\doibase 10.1103/PhysRevA.66.043612}
  {\bibfield  {journal} {\bibinfo  {journal} {Phys. Rev. A}\ }\textbf {\bibinfo
  {volume} {66}},\ \bibinfo {pages} {043612} (\bibinfo {year}
  {2002}{\natexlab{c}})}\BibitemShut {NoStop}%
\bibitem [{\citenamefont {Dall}\ \emph {et~al.}(2010)\citenamefont {Dall},
  \citenamefont {Hodgman}, \citenamefont {Johnsson}, \citenamefont {Baldwin},\
  and\ \citenamefont {Truscott}}]{Dall2010}%
  \BibitemOpen
  \bibfield  {author} {\bibinfo {author} {\bibfnamefont {R.~G.}\ \bibnamefont
  {Dall}}, \bibinfo {author} {\bibfnamefont {S.~S.}\ \bibnamefont {Hodgman}},
  \bibinfo {author} {\bibfnamefont {M.~T.}\ \bibnamefont {Johnsson}}, \bibinfo
  {author} {\bibfnamefont {K.~G.~H.}\ \bibnamefont {Baldwin}}, \ and\ \bibinfo
  {author} {\bibfnamefont {A.~G.}\ \bibnamefont {Truscott}},\ }\href {\doibase
  10.1103/PhysRevA.81.011602} {\bibfield  {journal} {\bibinfo  {journal} {Phys.
  Rev. A}\ }\textbf {\bibinfo {volume} {81}},\ \bibinfo {pages} {011602}
  (\bibinfo {year} {2010})}\BibitemShut {NoStop}%
\bibitem [{\citenamefont {Dall}\ \emph {et~al.}(2011)\citenamefont {Dall},
  \citenamefont {Hodgman}, \citenamefont {Manning},\ and\ \citenamefont
  {Truscott}}]{Dall2011}%
  \BibitemOpen
  \bibfield  {author} {\bibinfo {author} {\bibfnamefont {R.~G.}\ \bibnamefont
  {Dall}}, \bibinfo {author} {\bibfnamefont {S.~S.}\ \bibnamefont {Hodgman}},
  \bibinfo {author} {\bibfnamefont {A.~G.}\ \bibnamefont {Manning}}, \ and\
  \bibinfo {author} {\bibfnamefont {A.~G.}\ \bibnamefont {Truscott}},\ }\href
  {\doibase 10.1364/OL.36.001131} {\bibfield  {journal} {\bibinfo  {journal}
  {Opt. Lett.}\ }\textbf {\bibinfo {volume} {36}},\ \bibinfo {pages} {1131}
  (\bibinfo {year} {2011})}\BibitemShut {NoStop}%
\bibitem [{\citenamefont {Pacher}\ \emph {et~al.}(1989)\citenamefont {Pacher},
  \citenamefont {Mead}, \citenamefont {Cederbaum},\ and\ \citenamefont
  {K\"oppel}}]{Pacher1989}%
  \BibitemOpen
  \bibfield  {author} {\bibinfo {author} {\bibfnamefont {T.}~\bibnamefont
  {Pacher}}, \bibinfo {author} {\bibfnamefont {C.~A.}\ \bibnamefont {Mead}},
  \bibinfo {author} {\bibfnamefont {L.~S.}\ \bibnamefont {Cederbaum}}, \ and\
  \bibinfo {author} {\bibfnamefont {H.}~\bibnamefont {K\"oppel}},\ }\href
  {\doibase 10.1063/1.457323} {\bibfield  {journal} {\bibinfo  {journal} {J.
  Chem. Phys.}\ }\textbf {\bibinfo {volume} {91}},\ \bibinfo {pages} {7057}
  (\bibinfo {year} {1989})}\BibitemShut {NoStop}%
\bibitem [{\citenamefont {Pacher}\ \emph {et~al.}(1993)\citenamefont {Pacher},
  \citenamefont {Cederbaum},\ and\ \citenamefont {K\"oppel}}]{Pacher1993}%
  \BibitemOpen
  \bibfield  {author} {\bibinfo {author} {\bibfnamefont {T.}~\bibnamefont
  {Pacher}}, \bibinfo {author} {\bibfnamefont {L.~S.}\ \bibnamefont
  {Cederbaum}}, \ and\ \bibinfo {author} {\bibfnamefont {H.}~\bibnamefont
  {K\"oppel}},\ }\href {\doibase 10.1002/9780470141427.ch4} {\bibfield
  {journal} {\bibinfo  {journal} {Adv. Chem. Phys.}\ }\textbf {\bibinfo
  {volume} {84}},\ \bibinfo {pages} {293} (\bibinfo {year} {1993})}\BibitemShut
  {NoStop}%
\bibitem [{\citenamefont {Van~Voorhis}\ \emph {et~al.}(2010)\citenamefont
  {Van~Voorhis}, \citenamefont {Kowalczyk}, \citenamefont {Kaduk},
  \citenamefont {Wang}, \citenamefont {Cheng},\ and\ \citenamefont
  {Wu}}]{VanVoorhis2010}%
  \BibitemOpen
  \bibfield  {author} {\bibinfo {author} {\bibfnamefont {T.}~\bibnamefont
  {Van~Voorhis}}, \bibinfo {author} {\bibfnamefont {T.}~\bibnamefont
  {Kowalczyk}}, \bibinfo {author} {\bibfnamefont {B.}~\bibnamefont {Kaduk}},
  \bibinfo {author} {\bibfnamefont {L.-P.}\ \bibnamefont {Wang}}, \bibinfo
  {author} {\bibfnamefont {C.-L.}\ \bibnamefont {Cheng}}, \ and\ \bibinfo
  {author} {\bibfnamefont {Q.}~\bibnamefont {Wu}},\ }\href {\doibase
  10.1146/annurev.physchem.012809.103324} {\bibfield  {journal} {\bibinfo
  {journal} {Annu. Rev. Phys. Chem.}\ }\textbf {\bibinfo {volume} {61}},\
  \bibinfo {pages} {149 } (\bibinfo {year} {2010})}\BibitemShut {NoStop}%
\bibitem [{\citenamefont {Morgan}\ \emph {et~al.}(2013)\citenamefont {Morgan},
  \citenamefont {O'Riordan}, \citenamefont {Crowley}, \citenamefont
  {O'Sullivan},\ and\ \citenamefont {Busch}}]{Morgan2013}%
  \BibitemOpen
  \bibfield  {author} {\bibinfo {author} {\bibfnamefont {T.}~\bibnamefont
  {Morgan}}, \bibinfo {author} {\bibfnamefont {L.~J.}\ \bibnamefont
  {O'Riordan}}, \bibinfo {author} {\bibfnamefont {N.}~\bibnamefont {Crowley}},
  \bibinfo {author} {\bibfnamefont {B.}~\bibnamefont {O'Sullivan}}, \ and\
  \bibinfo {author} {\bibfnamefont {T.}~\bibnamefont {Busch}},\ }\href
  {\doibase 10.1103/PhysRevA.88.053618} {\bibfield  {journal} {\bibinfo
  {journal} {Phys. Rev. A}\ }\textbf {\bibinfo {volume} {88}},\ \bibinfo
  {pages} {053618} (\bibinfo {year} {2013})}\BibitemShut {NoStop}%
\bibitem [{\citenamefont {Pethick}\ and\ \citenamefont
  {Smith}(2008)}]{Pethick2008}%
  \BibitemOpen
  \bibfield  {author} {\bibinfo {author} {\bibfnamefont {C.~J.}\ \bibnamefont
  {Pethick}}\ and\ \bibinfo {author} {\bibfnamefont {H.}~\bibnamefont
  {Smith}},\ }\href@noop {} {\emph {\bibinfo {title} {Bose-Einstein
  Condensation in Dilute Gases}}}\ (\bibinfo  {publisher} {Cambridge University
  Press},\ \bibinfo {address} {Cambridge},\ \bibinfo {year} {2008})\BibitemShut
  {NoStop}%
\bibitem [{\citenamefont {Tang}(1970)}]{Tang1970}%
  \BibitemOpen
  \bibfield  {author} {\bibinfo {author} {\bibfnamefont {C.~H.}\ \bibnamefont
  {Tang}},\ }\href {http://dx.doi.org/10.1109/TMTT.1970.1127150} {\bibfield
  {journal} {\bibinfo  {journal} {IEEE Trans. Microw. Theory Tech.}\ }\textbf
  {\bibinfo {volume} {18}},\ \bibinfo {pages} {69} (\bibinfo {year}
  {1970})}\BibitemShut {NoStop}%
\bibitem [{\citenamefont {Clark}\ and\ \citenamefont
  {Bracken}(1996)}]{Clark1996}%
  \BibitemOpen
  \bibfield  {author} {\bibinfo {author} {\bibfnamefont {I.~J.}\ \bibnamefont
  {Clark}}\ and\ \bibinfo {author} {\bibfnamefont {A.~J.}\ \bibnamefont
  {Bracken}},\ }\href {http://stacks.iop.org/0305-4470/29/i=15/a=022}
  {\bibfield  {journal} {\bibinfo  {journal} {J. Phys. A: Math. Gen.}\ }\textbf
  {\bibinfo {volume} {29}},\ \bibinfo {pages} {4527} (\bibinfo {year}
  {1996})}\BibitemShut {NoStop}%
\bibitem [{\citenamefont {Exner}\ and\ \citenamefont {Seba}(1989)}]{Exner1989}%
  \BibitemOpen
  \bibfield  {author} {\bibinfo {author} {\bibfnamefont {P.}~\bibnamefont
  {Exner}}\ and\ \bibinfo {author} {\bibfnamefont {P.}~\bibnamefont {Seba}},\
  }\href {\doibase 10.1063/1.528538} {\bibfield  {journal} {\bibinfo  {journal}
  {J. Math. Phys.}\ }\textbf {\bibinfo {volume} {30}},\ \bibinfo {pages} {2574}
  (\bibinfo {year} {1989})}\BibitemShut {NoStop}%
\bibitem [{\citenamefont {Stickney}\ and\ \citenamefont
  {Zozulya}(2003)}]{Stickney2003}%
  \BibitemOpen
  \bibfield  {author} {\bibinfo {author} {\bibfnamefont {J.~A.}\ \bibnamefont
  {Stickney}}\ and\ \bibinfo {author} {\bibfnamefont {A.~A.}\ \bibnamefont
  {Zozulya}},\ }\href {\doibase 10.1103/PhysRevA.68.013611} {\bibfield
  {journal} {\bibinfo  {journal} {Phys. Rev. A}\ }\textbf {\bibinfo {volume}
  {68}},\ \bibinfo {pages} {013611} (\bibinfo {year} {2003})}\BibitemShut
  {NoStop}%
\bibitem [{\citenamefont {J\"a\"askel\"ainen}\ and\ \citenamefont
  {Stenholm}(2003)}]{Jaaskelainen2003}%
  \BibitemOpen
  \bibfield  {author} {\bibinfo {author} {\bibfnamefont {M.}~\bibnamefont
  {J\"a\"askel\"ainen}}\ and\ \bibinfo {author} {\bibfnamefont
  {S.}~\bibnamefont {Stenholm}},\ }\href {\doibase 10.1103/PhysRevA.68.033607}
  {\bibfield  {journal} {\bibinfo  {journal} {Phys. Rev. A}\ }\textbf {\bibinfo
  {volume} {68}},\ \bibinfo {pages} {033607} (\bibinfo {year}
  {2003})}\BibitemShut {NoStop}%
\bibitem [{\citenamefont {Bortolotti}\ and\ \citenamefont
  {Bohn}(2004)}]{Bortolotti2004}%
  \BibitemOpen
  \bibfield  {author} {\bibinfo {author} {\bibfnamefont {D.~C.~E.}\
  \bibnamefont {Bortolotti}}\ and\ \bibinfo {author} {\bibfnamefont {J.~L.}\
  \bibnamefont {Bohn}},\ }\href {\doibase 10.1103/PhysRevA.69.033607}
  {\bibfield  {journal} {\bibinfo  {journal} {Phys. Rev. A}\ }\textbf {\bibinfo
  {volume} {69}},\ \bibinfo {pages} {033607} (\bibinfo {year}
  {2004})}\BibitemShut {NoStop}%
\bibitem [{\citenamefont {Willatzen}\ \emph {et~al.}(2010)\citenamefont
  {Willatzen}, \citenamefont {Gravesen},\ and\ \citenamefont {Lew
  Yan~Voon}}]{Willatzen2010}%
  \BibitemOpen
  \bibfield  {author} {\bibinfo {author} {\bibfnamefont {M.}~\bibnamefont
  {Willatzen}}, \bibinfo {author} {\bibfnamefont {J.}~\bibnamefont {Gravesen}},
  \ and\ \bibinfo {author} {\bibfnamefont {L.~C.}\ \bibnamefont {Lew
  Yan~Voon}},\ }\href {\doibase 10.1103/PhysRevA.81.060102} {\bibfield
  {journal} {\bibinfo  {journal} {Phys. Rev. A}\ }\textbf {\bibinfo {volume}
  {81}},\ \bibinfo {pages} {060102} (\bibinfo {year} {2010})}\BibitemShut
  {NoStop}%
\bibitem [{Note1()}]{Note1}%
  \BibitemOpen
  \bibinfo {note} {We remark that strictly speaking a pure harmonic oscillator
  confinement is not consistent with the need to restrict the wavefunction to
  the region of well-defined Tang frame coordinates. However, adding Dirichlet
  boundary conditions at a distance much larger than the oscillator length only
  weakly deforms energetically low-lying harmonic oscillator modes, since it
  mainly affects their exponentially suppressed tail beyond the classical
  turning point. The important expectation values of $\protect \mathaccentV
  {hat}05En^l$, $\protect \mathaccentV {hat}05Eb^l$ etc. for small $l$ are
  almost insensitive and can be evaluated neglecting the Dirichlet boundary
  conditions.}\BibitemShut {Stop}%
\bibitem [{\citenamefont {Worth}\ and\ \citenamefont
  {Cederbaum}(2004)}]{Worth2004}%
  \BibitemOpen
  \bibfield  {author} {\bibinfo {author} {\bibfnamefont {G.~A.}\ \bibnamefont
  {Worth}}\ and\ \bibinfo {author} {\bibfnamefont {L.~S.}\ \bibnamefont
  {Cederbaum}},\ }\href
  {www.dx.doi.org/10.1146/annurev.physchem.55.091602.094335} {\bibfield
  {journal} {\bibinfo  {journal} {Annu. Rev. Phys. Chem.}\ }\textbf {\bibinfo
  {volume} {55}},\ \bibinfo {pages} {127} (\bibinfo {year} {2004})}\BibitemShut
  {NoStop}%
\bibitem [{\citenamefont {Mitchell}(2001)}]{Mitchell2001}%
  \BibitemOpen
  \bibfield  {author} {\bibinfo {author} {\bibfnamefont {K.~A.}\ \bibnamefont
  {Mitchell}},\ }\href {\doibase 10.1103/PhysRevA.63.042112} {\bibfield
  {journal} {\bibinfo  {journal} {Phys. Rev. A}\ }\textbf {\bibinfo {volume}
  {63}},\ \bibinfo {pages} {042112} (\bibinfo {year} {2001})}\BibitemShut
  {NoStop}%
\bibitem [{\citenamefont {Bouchitté}\ \emph {et~al.}(2007)\citenamefont
  {Bouchitté}, \citenamefont {Mascarenhas},\ and\ \citenamefont
  {Trabucho}}]{Bouchitte2007}%
  \BibitemOpen
  \bibfield  {author} {\bibinfo {author} {\bibfnamefont {G.}~\bibnamefont
  {Bouchitté}}, \bibinfo {author} {\bibfnamefont {M.~L.}\ \bibnamefont
  {Mascarenhas}}, \ and\ \bibinfo {author} {\bibfnamefont {L.}~\bibnamefont
  {Trabucho}},\ }\href {\doibase 10.1051/cocv:2007042} {\bibfield  {journal}
  {\bibinfo  {journal} {ESAIM Control Optim. Calc. Var.}\ }\textbf {\bibinfo
  {volume} {13}},\ \bibinfo {pages} {793} (\bibinfo {year} {2007})}\BibitemShut
  {NoStop}%
\bibitem [{\citenamefont {Exner}\ and\ \citenamefont
  {Kovařík}(2005)}]{Exner2005}%
  \BibitemOpen
  \bibfield  {author} {\bibinfo {author} {\bibfnamefont {P.}~\bibnamefont
  {Exner}}\ and\ \bibinfo {author} {\bibfnamefont {H.}~\bibnamefont
  {Kovařík}},\ }\href {\doibase 10.1007/s11005-005-0016-8} {\bibfield
  {journal} {\bibinfo  {journal} {Lett. Math. Phys.}\ }\textbf {\bibinfo
  {volume} {73}},\ \bibinfo {pages} {183} (\bibinfo {year} {2005})}\BibitemShut
  {NoStop}%
\bibitem [{\citenamefont {Maraner}(1996)}]{Maraner1996}%
  \BibitemOpen
  \bibfield  {author} {\bibinfo {author} {\bibfnamefont {P.}~\bibnamefont
  {Maraner}},\ }\href {\doibase 10.1006/aphy.1996.0029} {\bibfield  {journal}
  {\bibinfo  {journal} {Ann. Phys.}\ }\textbf {\bibinfo {volume} {246}},\
  \bibinfo {pages} {325 } (\bibinfo {year} {1996})}\BibitemShut {NoStop}%
\bibitem [{\citenamefont {Fujii}\ \emph {et~al.}(1997)\citenamefont {Fujii},
  \citenamefont {Ogawa}, \citenamefont {Uchiyama},\ and\ \citenamefont
  {Chepilko}}]{Fujii1997}%
  \BibitemOpen
  \bibfield  {author} {\bibinfo {author} {\bibfnamefont {K.}~\bibnamefont
  {Fujii}}, \bibinfo {author} {\bibfnamefont {N.}~\bibnamefont {Ogawa}},
  \bibinfo {author} {\bibfnamefont {S.}~\bibnamefont {Uchiyama}}, \ and\
  \bibinfo {author} {\bibfnamefont {N.~M.}\ \bibnamefont {Chepilko}},\ }\href
  {\doibase 10.1142/S0217751X97002814} {\bibfield  {journal} {\bibinfo
  {journal} {Int. J. Mod. Phys.}\ }\textbf {\bibinfo {volume} {A12}},\ \bibinfo
  {pages} {5235} (\bibinfo {year} {1997})}\BibitemShut {NoStop}%
\bibitem [{\citenamefont {Ulreich}\ and\ \citenamefont
  {Zwerger}(1998)}]{Ulreich1998}%
  \BibitemOpen
  \bibfield  {author} {\bibinfo {author} {\bibfnamefont {S.}~\bibnamefont
  {Ulreich}}\ and\ \bibinfo {author} {\bibfnamefont {W.}~\bibnamefont
  {Zwerger}},\ }\href {http://stacks.iop.org/0295-5075/41/i=2/a=117} {\bibfield
   {journal} {\bibinfo  {journal} {Europhys. Lett.}\ }\textbf {\bibinfo
  {volume} {41}},\ \bibinfo {pages} {117} (\bibinfo {year} {1998})}\BibitemShut
  {NoStop}%
\bibitem [{\citenamefont {Schuster}\ and\ \citenamefont
  {Jaffe}(2003)}]{Schuster2003}%
  \BibitemOpen
  \bibfield  {author} {\bibinfo {author} {\bibfnamefont {P.}~\bibnamefont
  {Schuster}}\ and\ \bibinfo {author} {\bibfnamefont {R.}~\bibnamefont
  {Jaffe}},\ }\href {\doibase 10.1016/S0003-4916(03)00080-0} {\bibfield
  {journal} {\bibinfo  {journal} {Ann. Phys.}\ }\textbf {\bibinfo {volume}
  {307}},\ \bibinfo {pages} {132 } (\bibinfo {year} {2003})}\BibitemShut
  {NoStop}%
\bibitem [{\citenamefont {von Neumann}\ and\ \citenamefont
  {Wigner}(1929)}]{Neumann1929}%
  \BibitemOpen
  \bibfield  {author} {\bibinfo {author} {\bibfnamefont {J.}~\bibnamefont {von
  Neumann}}\ and\ \bibinfo {author} {\bibfnamefont {E.}~\bibnamefont
  {Wigner}},\ }\href@noop {} {\bibfield  {journal} {\bibinfo  {journal} {Phys.
  Z.}\ }\textbf {\bibinfo {volume} {30}},\ \bibinfo {pages} {467} (\bibinfo
  {year} {1929})}\BibitemShut {NoStop}%
\bibitem [{Note2()}]{Note2}%
  \BibitemOpen
  \bibinfo {note} {After multiplication with $i$, Eq. (\ref {eq:sylv}) assumes
  the standard form of a Lyapunov matrix equation $ \protect \mathbf Q +
  \protect \mathbf {DX + X D^\dagger } = \protect \mathbf 0$, where $\protect
  \mathbf Q := -2i \protect \mathbf F$ is Hermitian and $\protect \mathbf X :=
  i \protect \mathbf S$ is to be determined.}\BibitemShut {Stop}%
\bibitem [{\citenamefont {Jameson}(1968)}]{Jameson1968}%
  \BibitemOpen
  \bibfield  {author} {\bibinfo {author} {\bibfnamefont {A.}~\bibnamefont
  {Jameson}},\ }\href {http://www.jstor.org/stable/2099227} {\bibfield
  {journal} {\bibinfo  {journal} {SIAM J. Appl. Math.}\ }\textbf {\bibinfo
  {volume} {16}},\ \bibinfo {pages} {1020} (\bibinfo {year}
  {1968})}\BibitemShut {NoStop}%
\bibitem [{\citenamefont {Bartels}\ and\ \citenamefont
  {Stewart}(1972)}]{Bartels1972}%
  \BibitemOpen
  \bibfield  {author} {\bibinfo {author} {\bibfnamefont {R.~H.}\ \bibnamefont
  {Bartels}}\ and\ \bibinfo {author} {\bibfnamefont {G.~W.}\ \bibnamefont
  {Stewart}},\ }\href {\doibase 10.1145/361573.361582} {\bibfield  {journal}
  {\bibinfo  {journal} {Commun. ACM}\ }\textbf {\bibinfo {volume} {15}},\
  \bibinfo {pages} {820} (\bibinfo {year} {1972})}\BibitemShut {NoStop}%
\bibitem [{\citenamefont {Baer}(2006)}]{Baer2006}%
  \BibitemOpen
  \bibfield  {author} {\bibinfo {author} {\bibfnamefont {M.}~\bibnamefont
  {Baer}},\ }\href@noop {} {\emph {\bibinfo {title} {Beyond Born-Oppenheimer:
  Electronic Nonadiabatic Coupling Terms and Conical Intersections}}}\
  (\bibinfo  {publisher} {Wiley-Interscience},\ \bibinfo {year}
  {2006})\BibitemShut {NoStop}%
\bibitem [{\citenamefont {van Dishoeck}\ \emph {et~al.}(1984)\citenamefont {van
  Dishoeck}, \citenamefont {van Hemert}, \citenamefont {Allison},\ and\
  \citenamefont {Dalgarno}}]{Dishoeck1984}%
  \BibitemOpen
  \bibfield  {author} {\bibinfo {author} {\bibfnamefont {E.~F.}\ \bibnamefont
  {van Dishoeck}}, \bibinfo {author} {\bibfnamefont {M.~C.}\ \bibnamefont {van
  Hemert}}, \bibinfo {author} {\bibfnamefont {A.~C.}\ \bibnamefont {Allison}},
  \ and\ \bibinfo {author} {\bibfnamefont {A.}~\bibnamefont {Dalgarno}},\
  }\href {\doibase 10.1063/1.447622} {\bibfield  {journal} {\bibinfo  {journal}
  {J. Chem. Phys.}\ }\textbf {\bibinfo {volume} {81}},\ \bibinfo {pages} {5709}
  (\bibinfo {year} {1984})}\BibitemShut {NoStop}%
\end{thebibliography}%
\end{document}